\documentclass{article}

\usepackage[margin=4cm]{geometry}
\usepackage{graphicx}
\usepackage{booktabs}
\usepackage{multirow}
\usepackage{subfig}
\usepackage[round]{natbib}
\usepackage{amsmath}
\usepackage{enumerate}
\newcommand{\degrees}{\ensuremath{^{\circ}}}
\newcommand{\mr}{\multirow}
\newcommand{\mc}{\multicolumn}
\newcommand{\tej}{Tropical Easterly Jet}
\newcommand{\tp}{Tibetan Plateau}
\newcommand{\cam}{CAM-3.1}

\newcommand{\cc}{Ctrl}
\newcommand{\cct}{Ctrl$\tau_{d12}$}

\newcommand{\nglo}{noGlOrog}
\newcommand{\nglot}{noGlOrog$\tau_{d12}$}
\newcommand{\rn}{reanalysis}
\newcommand{\nc}{NCEP}
\newcommand{\er}{ERA40}
\newcommand{\gp}{GPCP}
\newcommand{\cm}{CMAP}
\newcommand{\rp}{real-planet}
\newcommand{\ap}{aqua-planet}
\newcommand{\pc}{$P_c$}

\newcommand{\um}{$U_{max}$}
\newcommand{\md}{ mm day$^{-1}$}
\newcommand{\ms}{ m s$^{-1}$}

\newcommand{\dc}{\degrees{}C}
\newcommand{\de}{\degrees{}E}
\newcommand{\dw}{\degrees{}W}
\newcommand{\dn}{\degrees{}N}
\newcommand{\ds}{\degrees{}S}
\newcommand{\as}{AP\_S}
\newcommand{\am}{AP\_}

\begin{document}

\title{The impact of latent heating on the location, strength and structure of the \tej{} in the Community Atmosphere Model, version 3.1: Aqua-planet simulations}

\author{S. Rao$^{1,2}$ \\
           $^1$Department of Mechanical Engineering \\
               Indian Institute of Science \\
               Bangalore 560012, India \\
           $^2$Engineering Mechanics Unit \\
               Jawaharlal Centre for Advanced Scientific Research \\
               Bangalore 560064, India \\
               Email: {samrat.rao@gmail.com, samrat.rao@jncasr.ac.in} \\
               Mobile: +919916675221}

\maketitle

\begin{abstract}

The Tropical Easterly Jet (TEJ) is a prominent atmospheric circulation feature  observed during the Asian Summer Monsoon (ASM). The simulation of TEJ by the Community Atmosphere Model, version 3.1 (\cam{}) has been discussed in detail. Although the simulated TEJ replicates many observed features of the jet, the jet maximum is located too far to the west when compared to observation. Orography has minimal impact on the simulated TEJ hence indicating that latent heating is the crucial parameter. A series of \ap{} experiments with increasing complexity was undertaken to understand the reasons for the extreme westward shift of the TEJ.

The \ap{} simulations show that a single heat source in the deep tropics is inadequate to explain the structure of the observed TEJ. Equatorial heating is necessary to impart a baroclinic structure and a realistic meridional structure. Jet zonal wind speeds are directly related to the magnitude of deep tropical heating. The location of peak zonal wind is influenced by off-equatorial heating which is closest to it. Hence the presence of excess rainfall in Saudi Arabia has been shown to be the primary reason for the extreme westward shift of the TEJ maximum.



\end{abstract}

\section{Introduction}\label{intro}

The \tej{} (TEJ) is one of the most defining features of the Indian Summer Monsoon (ISM) which itself is a part of the Asian summer monsoon (ASM). The jet is observed mostly during the ISM that is in the months of June to September. It has a maximum between 50\de{}-80\de{}, Equator-15\dn{} and at 150hPa. The TEJ was first discovered by \citet{koteswaram-58}. It has a great influence on the rainfall in Africa (\citealp{hulme-89}). The correct simulation of the TEJ is important for accurate seasonal predictions and weather forecasting.

The \tej{} is believed to be influenced by the \tp{}. Previous studies (\citealp{flohn-68}, \citealp{krishnamurti-71}) have highlighted the presence of a huge upper tropospheric anticyclone above the \tp{} in summer. The origin of the Tibetan anticyclone itself has been attributed to the summertime insolation on the \tp{}. This sensible heating is widely regarded as the primary reason for this region to act as an elevated heat source. \cite{flohn-65} first suggested that in summer, southern and southeastern Tibet, i.e. south of 34\dn{}-35\dn{}, act as an elevated heat source which changes the meridional temperature and pressure gradients and contributes significantly to the reversal of high tropospheric flow during early June. \cite{flohn-68} showed that sensible heat source over the \tp{} as well as latent heat release due to monsoonal rains over central and eastern Himalayas and their southern approaches generates a warm core anticyclone in the upper troposphere at around 30\dn{}. This was a primary mechanism for establishing the south Asian monsoonal circulation over south Asia. According to Koteswaram these winds are a part of the Tibetan Anticyclone which forms during the summer monsoon over South Asia. He believed that the southern flank of the anticyclone preserved its angular momentum and became an easterly at approximately 15\dn{}. Contradicting Koteswaram, \cite{raghavan-73} opined that the upper tropospheric zonal component of the equatorward outflow from the Tibetan anticyclone does not agree with the law of conservation of angular momentum. However the thermal gradient balance was applicable and this was responsible for the origin of the TEJ.

The TEJ is not simply a passive atmospheric phenomena. \cite{hulme-89} studied the impact of the TEJ on Sudan rainfall. El Ni\~{n}o events were suggested as a direct control over Sahelian rainfall via the TEJ. The decelerating limb of the TEJ showed interannual variations in location in relation to Eastern Pacific warming. Fluctuations in the jet were shown to be responsible for altering precipitation patterns in Sudan. \cite{camberlin-95} also reported on the significant linkages between interannual variations of summer rainfall in Ethiopia-Sudan region and strength and latitudinal extent of upper-tropospheric easterlies. More recently \cite{nicholson-07} showed that the wave activity of the TEJ influences the African easterly jet (AEJ). The AEJ in turn influences the weather patterns over the Atlantic coast of Africa. \cite{bordoni-08} discussed the role of TEJ in modifying the monsoonal circulation. Upper-level easterlies shield the lower-level cross-equatorial monsoonal flow (which later becomes the Somali jet) from extratropical eddies. This made the angular momentum conservation principle applicable to overturning Hadley cell dynamics. In the larger picture this implies a strengthening of the monsoonal regime. This shows that the TEJ actively shapes the climate and weather pattern in regions not under the direct influence of the Indian Summer Monsoon.

In fact the TEJ also shows intraseasonal and interdecadal variations in its location. \cite{sathiyamoorthy-07} reported that the axis of the jet undergoes meridional movement in response to active and break phases of the Indian Monsoon. \cite{sathiyamoorthy-05} also observed a reduction in the spatial extent of the TEJ between 1960-1990. According to him the TEJ almost disappeared over the Atlantic and African regions. This reduction coincided with the 4-5 decade prolonged drought conditions over the Sahel region. He speculated that these two phenomena were associated with each other. This explanation also implies that latent heating significantly influences the characteristics of the TEJ.

\cite{ye-81} did laboratory experiments to simulate the heating effect of elevated land. He introduced heating in an ellipsoidal block resulting in vertical circulation, an anticyclone in the upper layer and a cyclonic flow in the lower layer. The pattern was found to be qualitatively similar to summer time atmospheric circulation in south Asia. \cite{raghavan-73} discussed the importance of \tp{} and explained that the jet owed its existence the the temperature gradients that was partly influenced by the Tibetan high. According to \cite{wang-06} the elevated heat source of the \tp{} was instrumental in providing an anchor to locate the Tibetan high. On the other hand \cite{hoskins-95} and \cite{liu-07} have argued that orography plays a secondary role in determining the position of the summertime upper tropospheric anticyclone. \cite{jingxi-89} studied the TEJ at 200 hPa and found that precipitation changes in the west coast of India led to changes in the jet structure. \cite{chen-87} observed that at 200 hPa the jet was weaker during El Ni\~{n}o and Indian drought events. The TEJ was weaker in the drought years of 1979, 1983 and 1987 and stronger in the excess monsoon years 1985 and 1988.

According to \cite{zhang-02}, the South Asian High had a bimodal structure. Its two modes -- one the Tibetan mode and the other the Iranian mode, both of which were fairly regular in their occurrence, were mostly influenced by heating effects. The former owed its existence to diabatic heating of the \tp{}, while the latter occurred due to adiabatic heating in the free atmosphere and diabatic heating near the surface.

Thus it can be seen that while there have been attempts in the past to explain some aspects of the TEJ and its effects, the issue about the importance of orography on the TEJ has not yet been settled. In the previous discussions variations in the jet were in the presence of the Himalayas and \tp{}. Although the importance of latent heating on the TEJ is quite clear now, it is necessary to confirm the importance of orography in deciding the location of the TEJ. It is necessary to study the jet in a GCM and understand the relative roles played by latent heating and orography on the TEJ. An Atmospheric General Circulation Model (AGCM), \cam{}, has been used to study the impact of orography and monsoonal heat sources on the TEJ. The jet location and its response to have been studied by:
\begin{enumerate}[(i)]
\item Modifying the deep-convective scheme and thereby changing the monsoonal heating pattern.
\item Removing orography.
\end{enumerate}
A series of \ap{} experiments have been conducted determine the factors influencing the location, structure and strength of the TEJ in \cam{}.

\section{The \tej{} in \rn{}}\label{tej-rean}

\begin{figure}[htbp]
   \begin{center}
   \includegraphics[trim = 5mm 00mm 15mm 45mm, clip, scale=0.4]{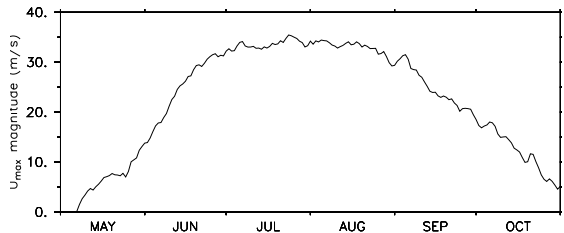}
   \end{center}
   \caption{Time series of 150 hPa zonal wind averaged between 50\de{}-80\de{}, Equator-15\dn{} showing TEJ peaking in July-August (\nc{}: 1980-2002 avg)}
   \label{fig: nc-U150xyavg-t}
\end{figure}

For wind data, \nc{} (\citealp{kalnay-96}) and \er{} (\citealp{uppala-05}) data have been used, and \gp{} (\citealp{adler-03}) and \cm{} (\citealp{xie-97}) data for precipitation. Unless otherwise mentioned the time period chosen spans years 1980 to 2002. The maximum zonal winds occur at 150hPa. From Fig. \ref{fig: nc-U150xyavg-t} it is seen that the TEJ peaks in July to August. Hence, in present work the focus is on the TEJ during the month of July when it first reaches its maximum value.

\begin{figure}[htbp]
   \begin{center}
   \subfloat[\nc{} (150 hPa)]{\label{nc-xy}\includegraphics[trim = 5mm 5mm 12mm 75mm, clip, scale=0.35]{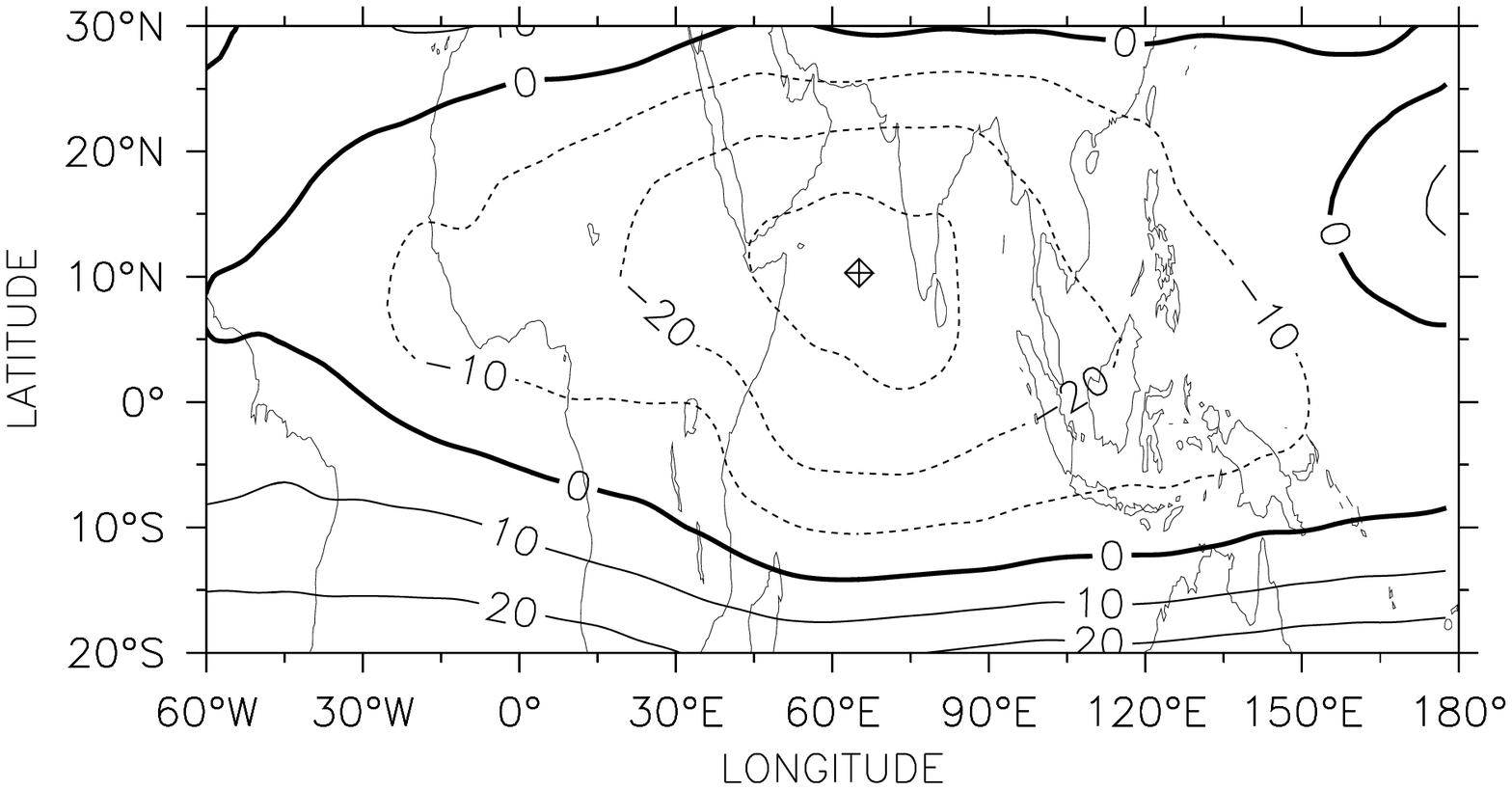}}
   \vskip 5mm
   \subfloat[\cc{} (125 hPa)]{\label{cc-xy}\includegraphics[trim = 5mm 5mm 12mm 75mm, clip, scale=0.35]{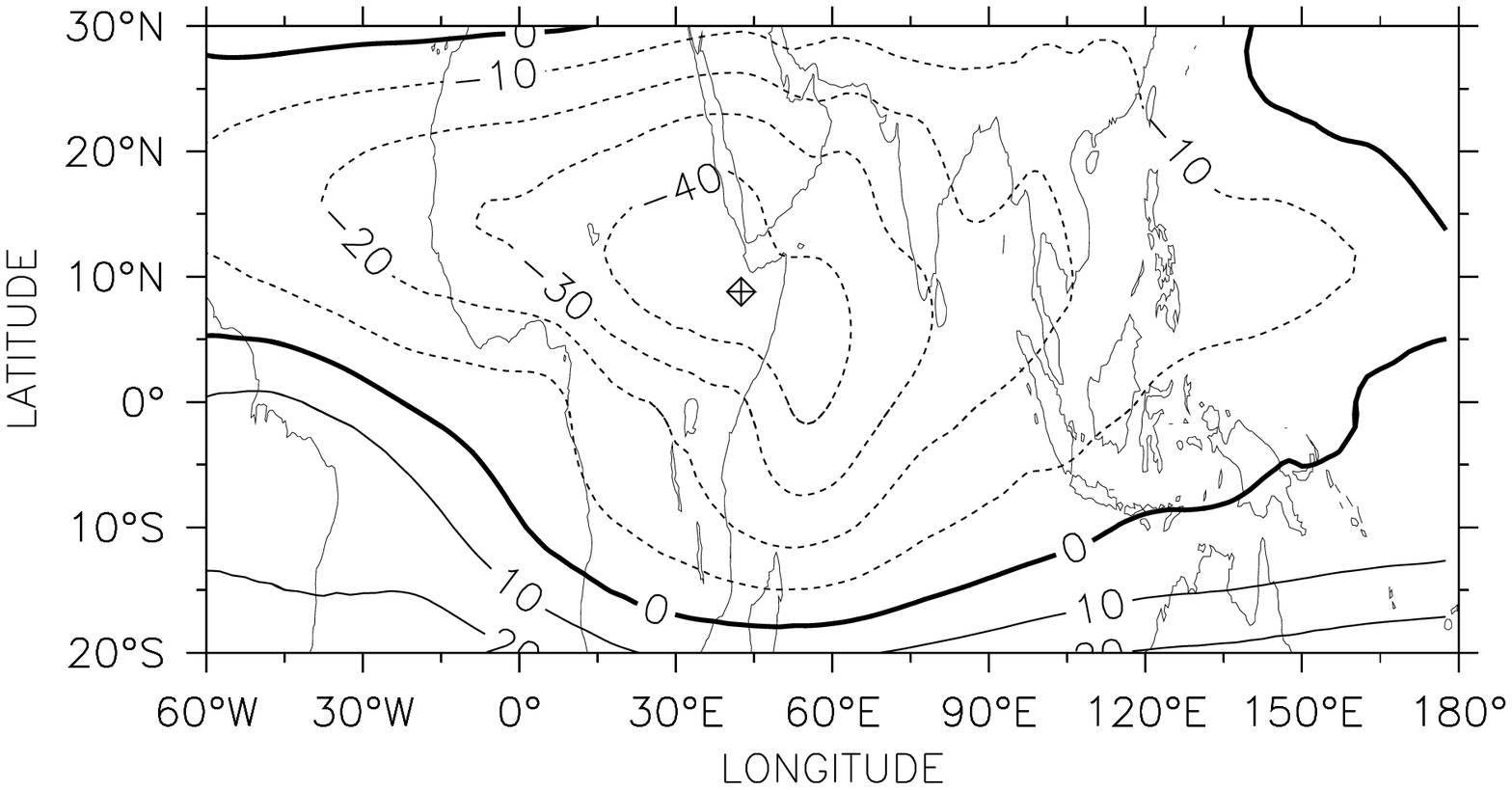}}
   \vskip 5mm
   \subfloat[\nglo{} (125 hPa)]{\label{nglo-xy}\includegraphics[trim = 5mm 5mm 12mm 75mm, clip, scale=0.35]{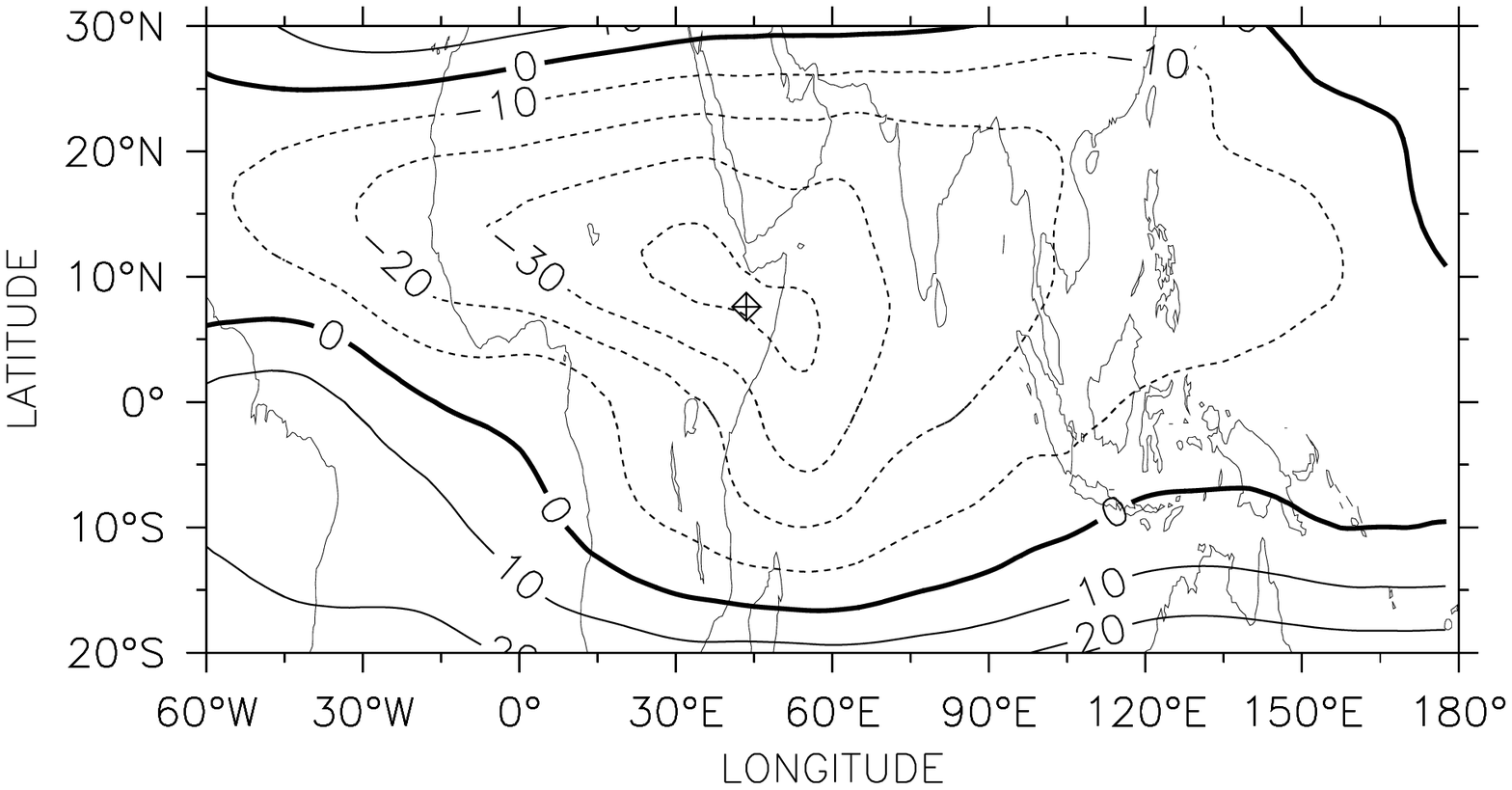}}
   \end{center}
   \caption[July horizontal zonal wind profile at pressure level where \um{} is attained: \nc{}, \cc{} and \nglo{}.]{July horizontal zonal wind profile at pressure level where maximum zonal wind is attained, `cross-diamond' is location of peak zonal wind: \nc{}, \cc{} and \nglo{}.}
   \label{fig: mix_jul_Umax_xy}
\end{figure}

\begin{figure}[htbp]
   \begin{center}
   \subfloat[\nc{}]{\label{nc-yz}\includegraphics[trim = 0mm 15mm 80mm 60mm, clip, scale=0.35]{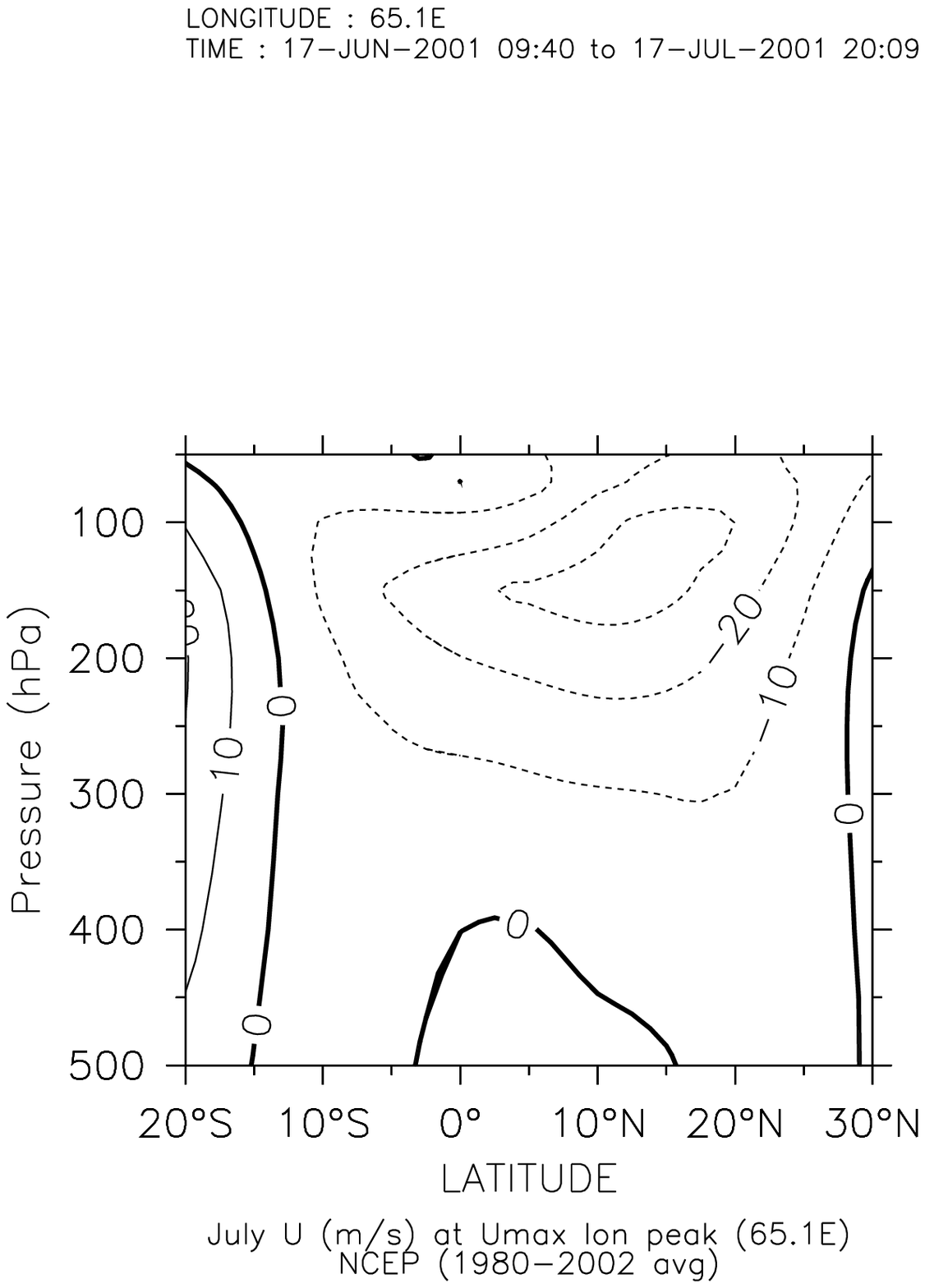}}
   \hspace{5mm}
   \subfloat[\nc{}]{\label{nc-zx}\includegraphics[trim = 0mm 15mm 10mm 60mm, clip, scale=0.35]{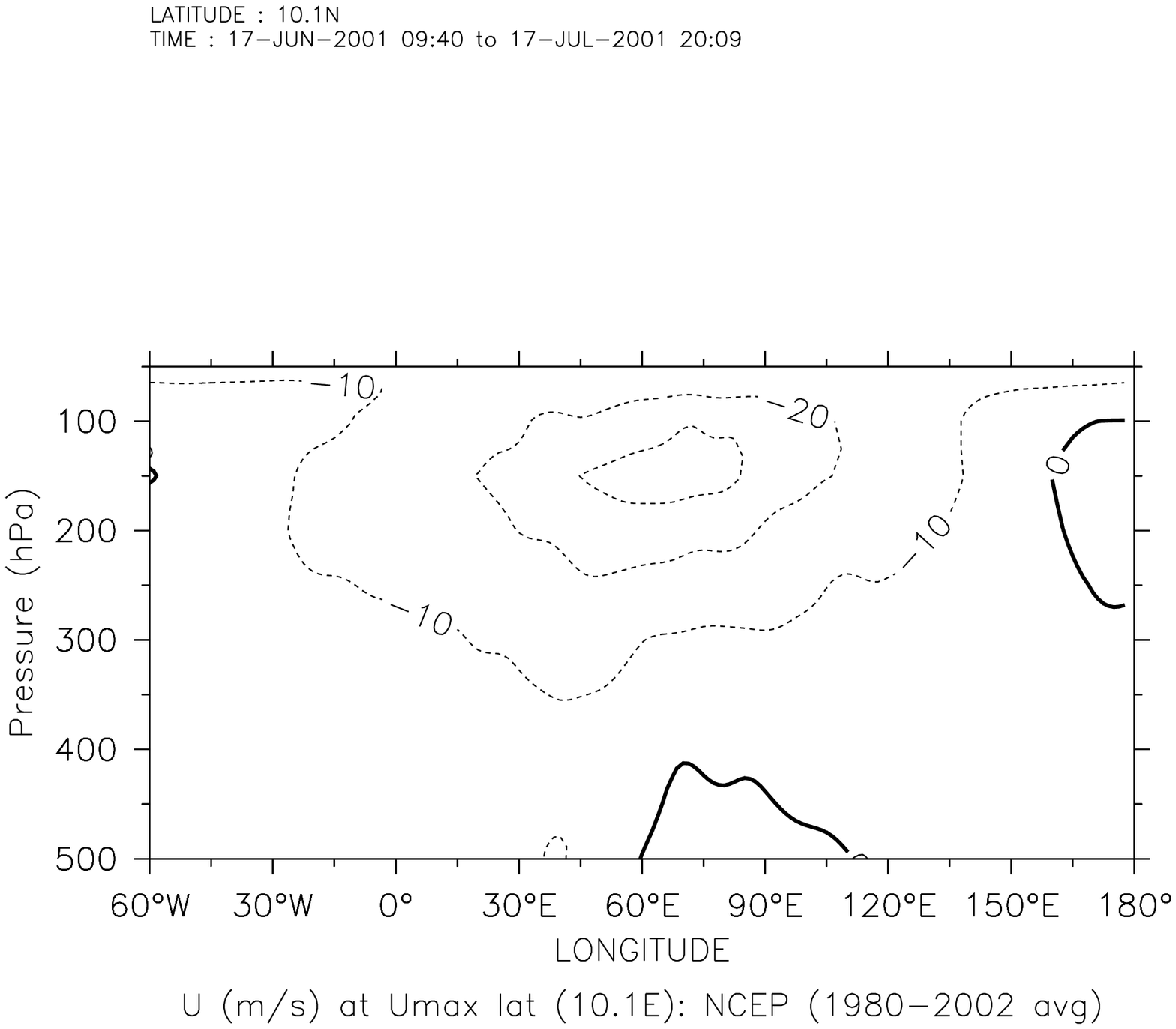}}
   \vskip 5mm
   \subfloat[\cc{}]{\label{cc-yz}\includegraphics[trim = 0mm 15mm 80mm 60mm, clip, scale=0.35]{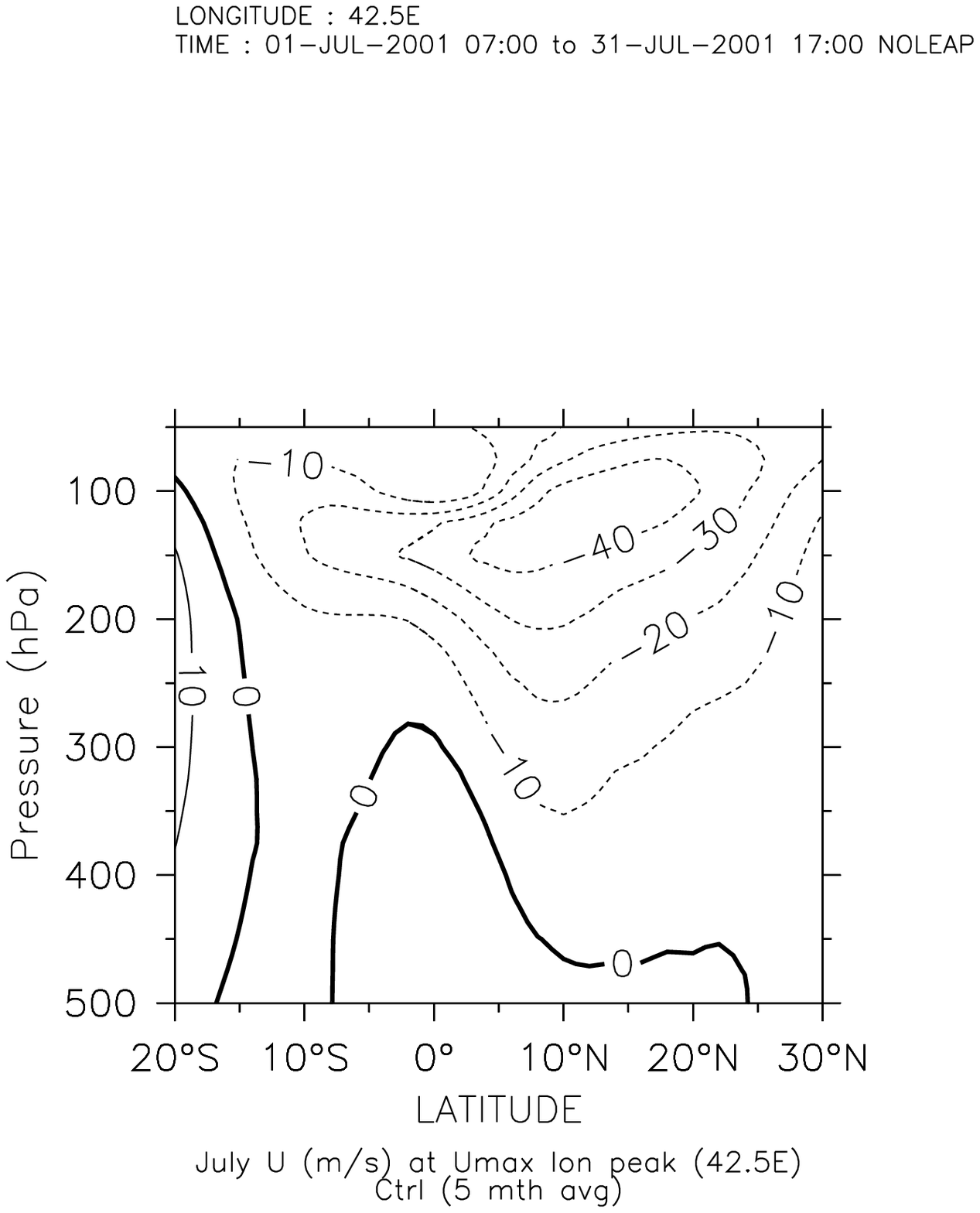}}
   \hspace{5mm}
   \subfloat[\cc{}]{\label{cc-zx}\includegraphics[trim = 0mm 15mm 10mm 60mm, clip, scale=0.35]{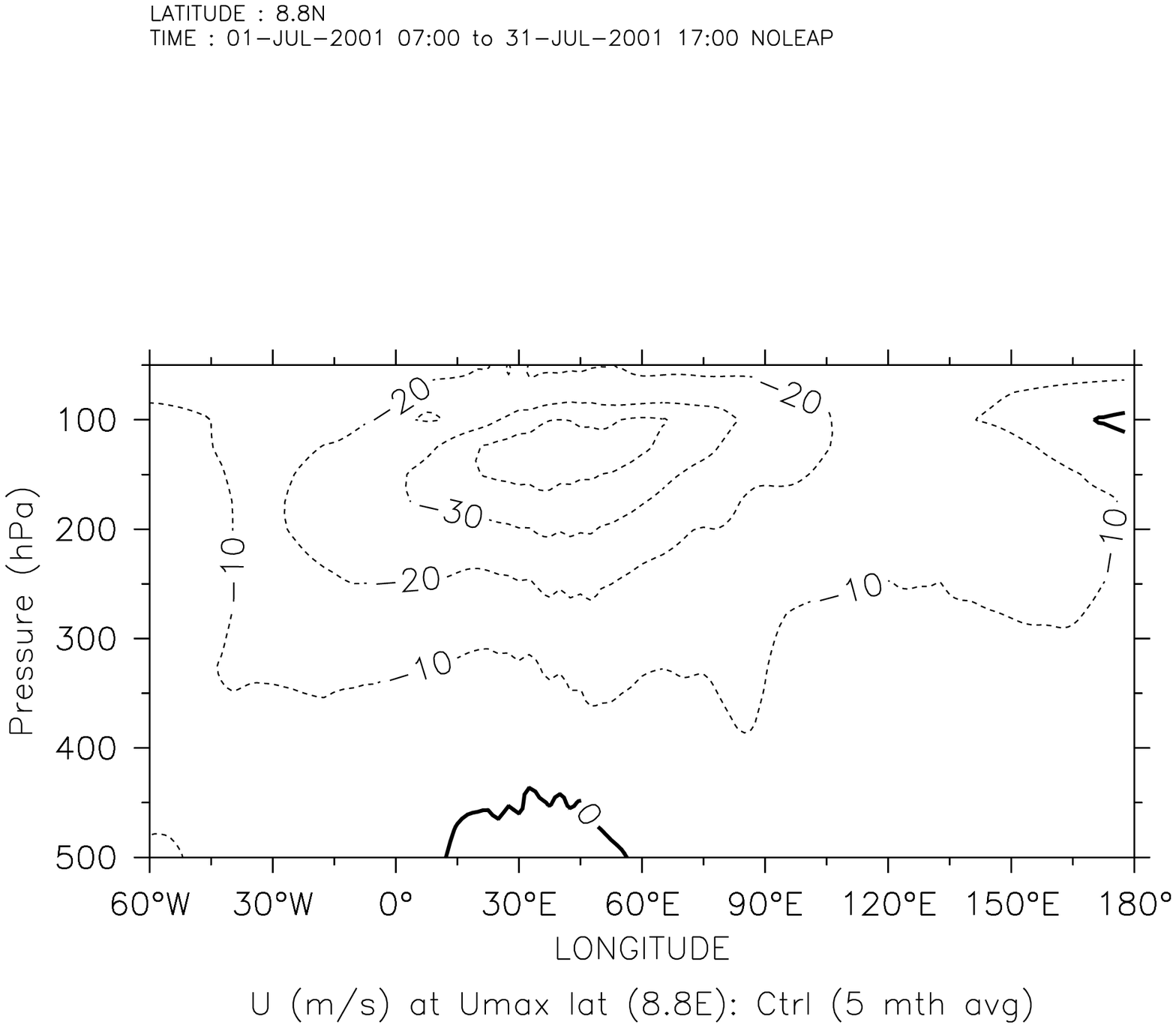}}
   \vskip 5mm
   \subfloat[\nglo{}]{\label{nglo-yz}\includegraphics[trim = 0mm 15mm 80mm 60mm, clip, scale=0.35]{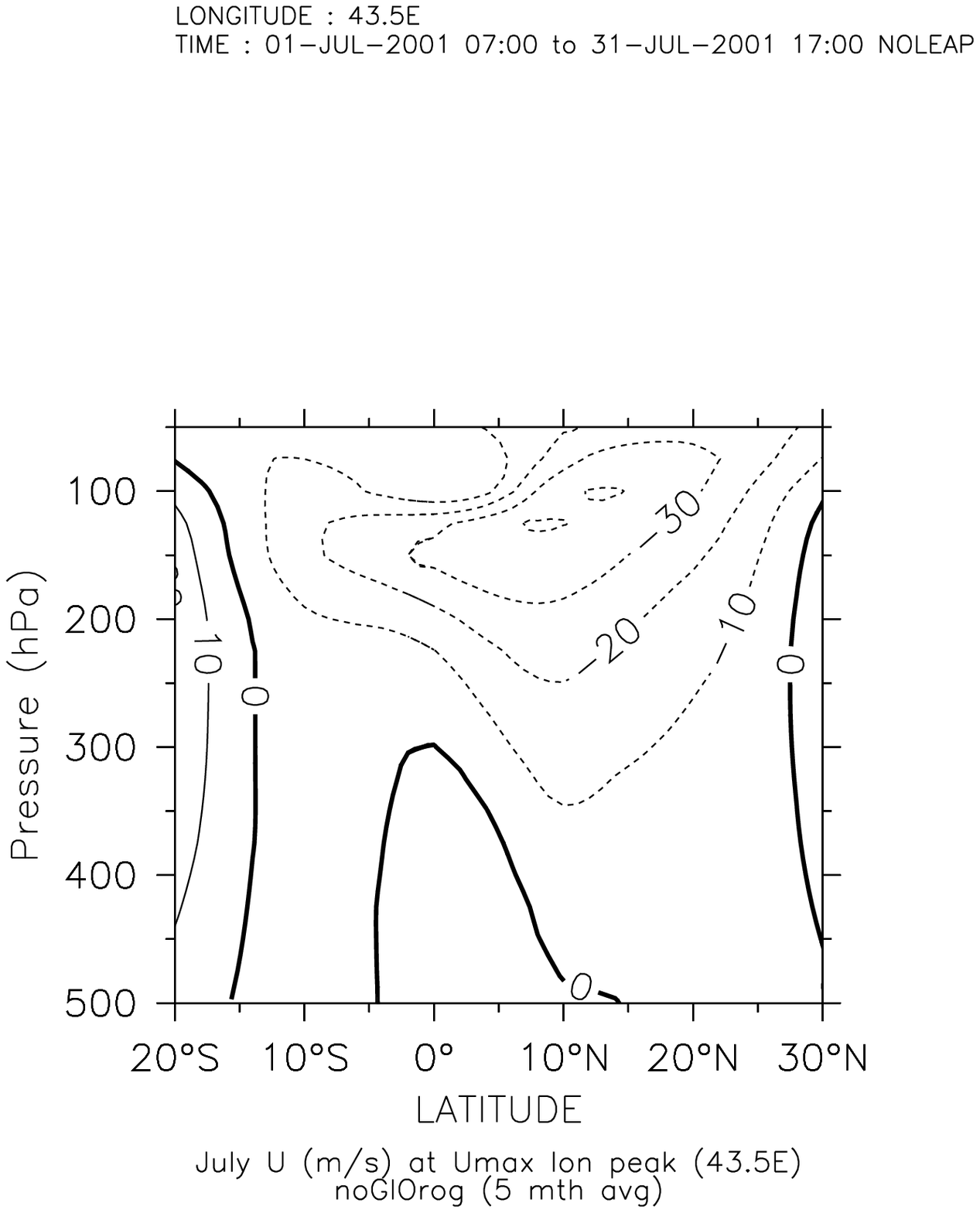}}
   \hspace{5mm}
   \subfloat[\nglo{}]{\label{nglo-zx}\includegraphics[trim = 0mm 15mm 10mm 60mm, clip, scale=0.35]{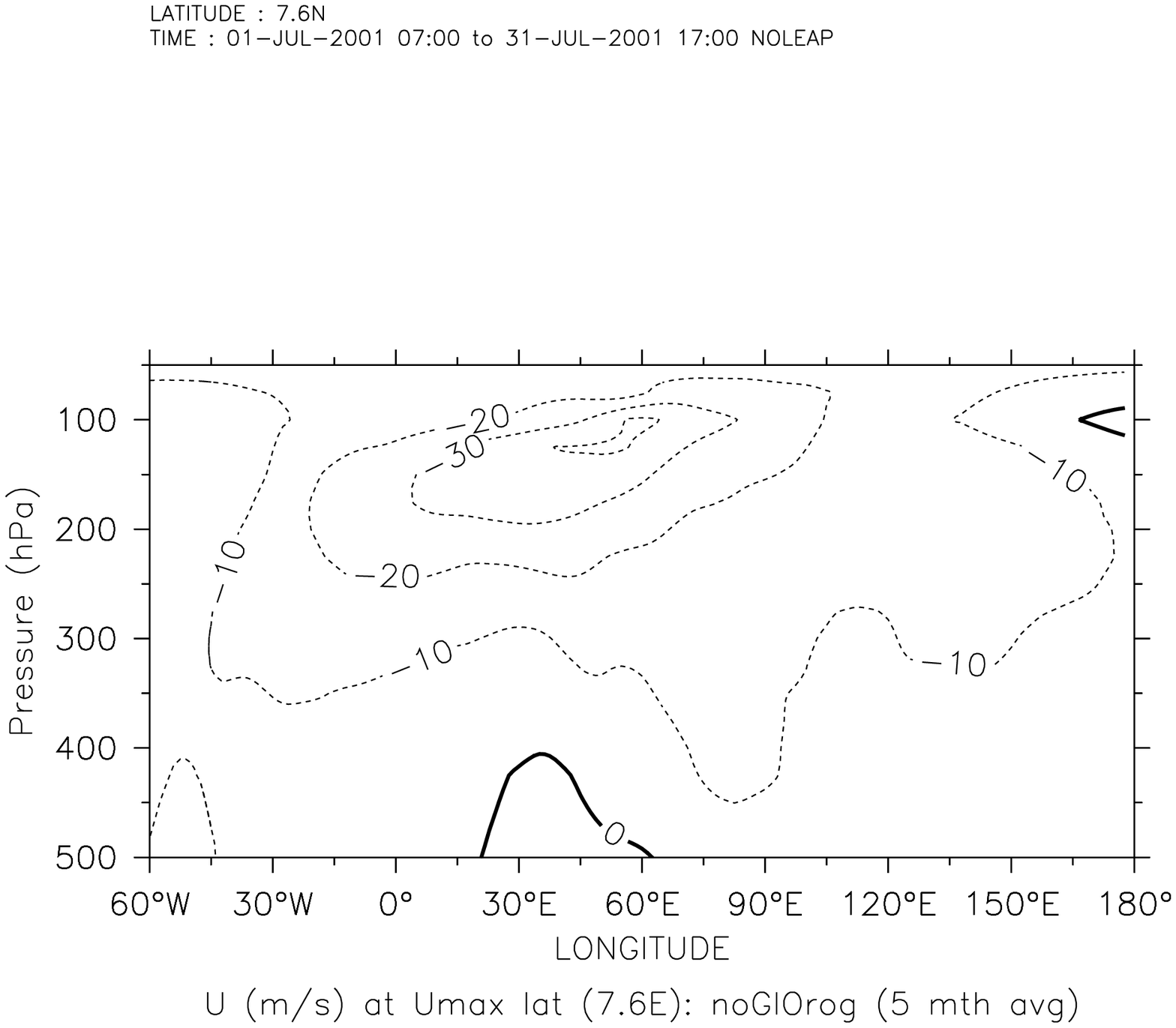}}
   \end{center}
   \caption[July zonal and meridional zonal wind profile: \rn{} and \rp{}.]{July cross-section of zonal wind profile at the longitude (left panel) and latitude (right panel) where \um{} is attained (see Table \ref{tab: mix-Umax,R} for details): \nc{}, \cc{} and \nglo{}.}
   \label{fig: mix_jul_Umax_yz-zx}
\end{figure}

\begin{figure}[htbp]
   \begin{center}
   \subfloat[\nc{}]{\label{nc-z3-xy}\includegraphics[trim = 5mm 5mm 10mm 70mm, clip, scale=0.35]{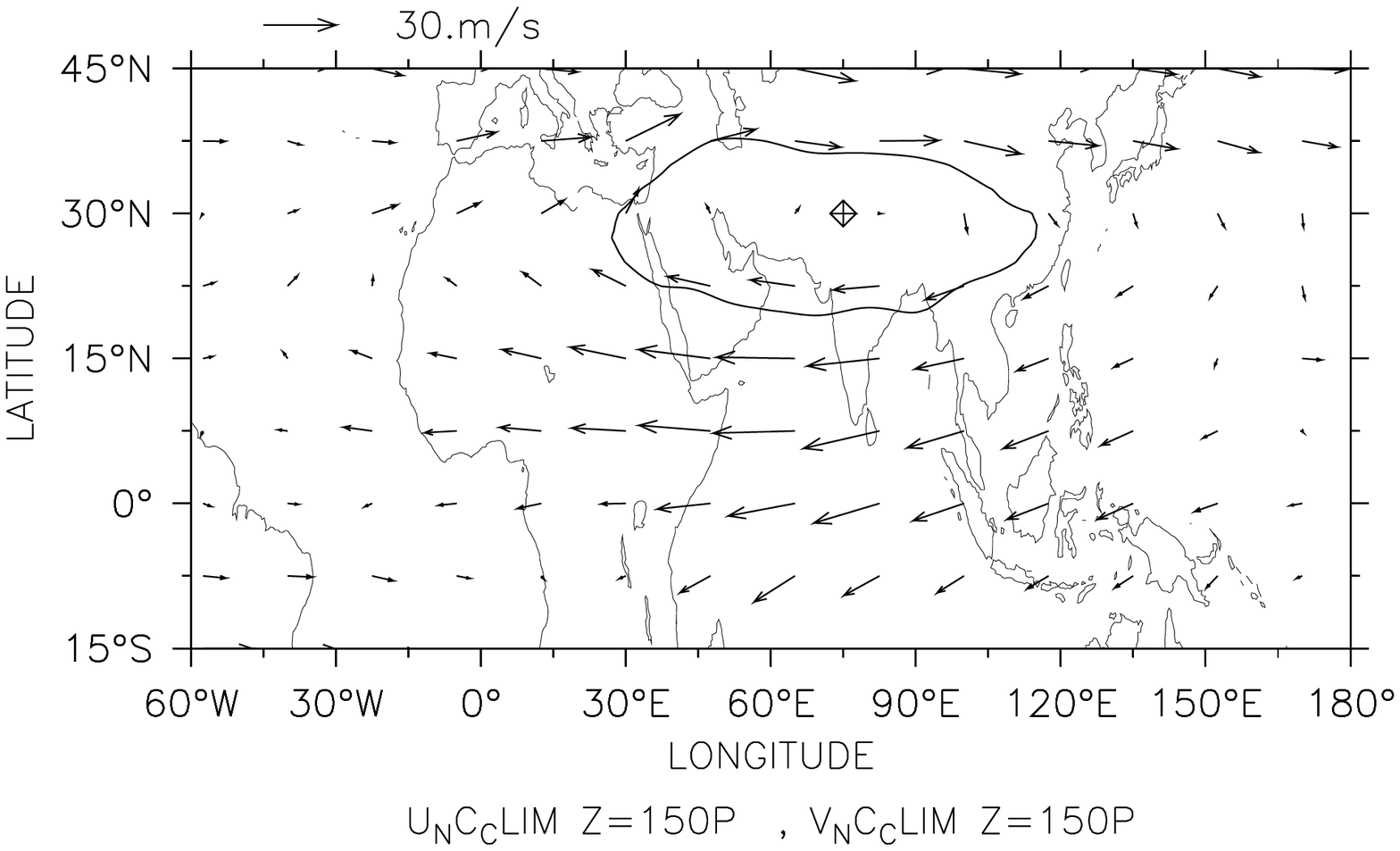}}
   \vskip 5mm
   \subfloat[\cc{}]{\label{cc-z3-xy}\includegraphics[trim = 5mm 5mm 10mm 70mm, clip, scale=0.35]{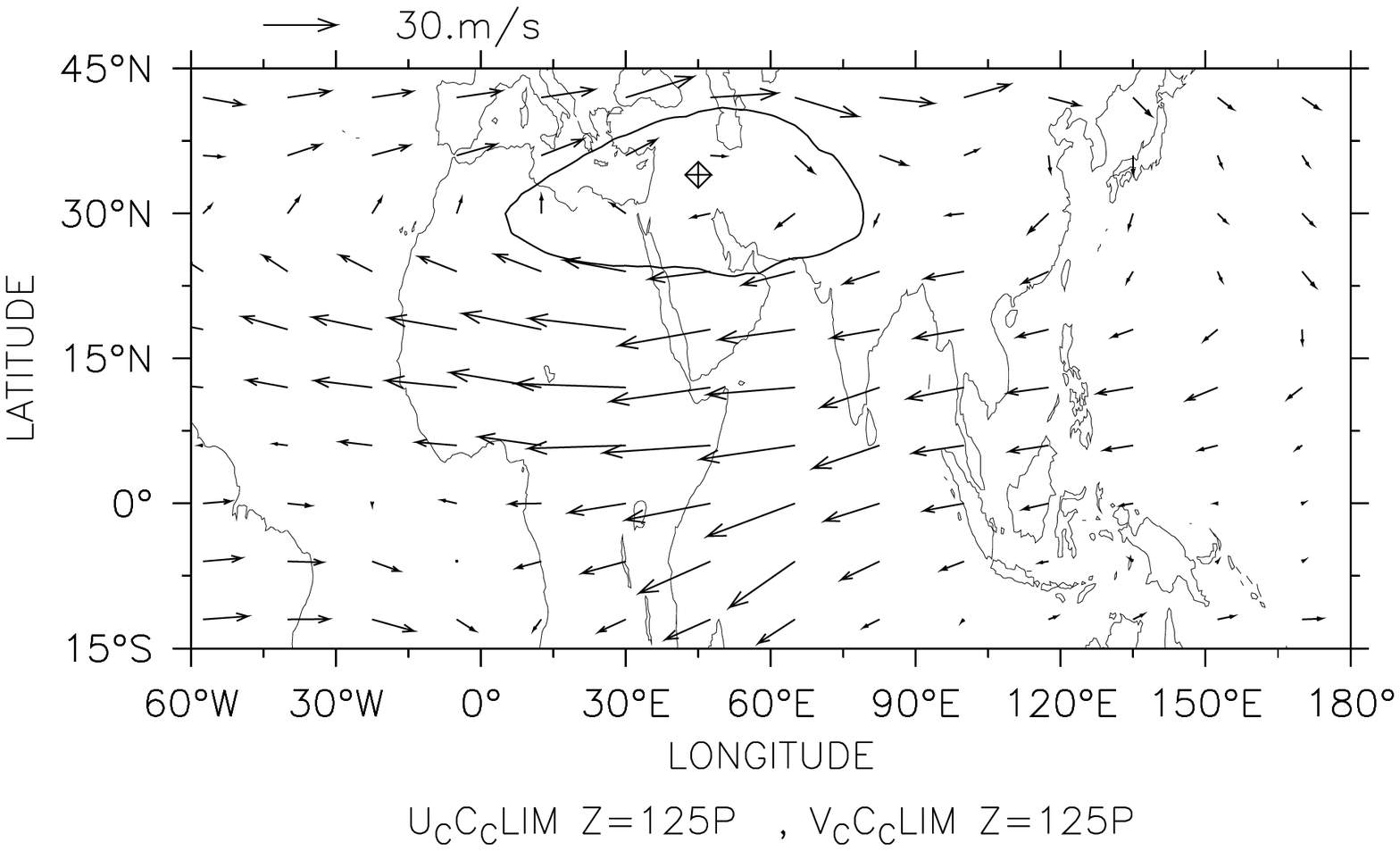}}
   \end{center}
   \caption[July velocity vectors and geopotential height at pressure level where \um{} is attained: \nc{} and \cc{}.]{July velocity vectors and geopotential height ($\times10^{3}$m) at 150 hPa. `Cross-diamond' is location of peak geopotential height. Contour for \nc{} and \cc{} is 14.36 and 14.43 respectively which is 99.5\% of individual peak.}
   \label{fig: mix-z3-xy}
\end{figure}

The magnitude of peak zonal wind and its location are listed in Table \ref{tab: mix-Umax,R}. Due to the good agreement between zonal winds of \nc{} and \er{}, only shown the former is shown. In Fig. \ref{nc-xy}, the zonal wind and location of peak zonal wind, henceforth \um{}, for \nc{} at 150hPa, is shown. The `cross-diamond' indicates the horizontal location of \um{}. The location of peak zonal wind of each individual year from 1980-2002 was found and then the means were calculated. The standard deviation for the zonal winds for \nc{} and \er{} was found to be 2\ms{} and 1.7\ms{} respectively. The meridional section of the zonal wind is shown in Fig. \ref{nc-yz}. This vertical cross-section corresponds to the longitude where the zonal wind is maximum. The meridional structure shows the jet peak lying between equator and $\sim$20\dn{}. A vertical equator to pole tilt towards higher pressure levels can be observed. The mean height of maximum easterly zonal wind speeds are higher at higher latitudes. In Fig. \ref{nc-z3-xy} the velocity vectors and geopotential high is shown. The peak is not over the \tp{} but to the west of it.

\begin{table}[htbp]
\caption{Magnitude (\ms{}) and location of peak zonal wind, and centroid of mean precipitation}
\label{tab: mix-Umax,R}
\begin{center}
\begin{tabular}{ccccccc} \toprule
\mc{7}{c}{\textbf{\rn{} and observation}} \\
\mr{2}{*}{Case} & \mc{4}{c}{Zonal wind}                 & \mc{2}{c}{Precipitation} \\ \cline{2-7}
                & Peak  & Lon       & Lat       & Press & Lon       & Lat \\ \midrule
\nc{}           & 34.48 & 65.1\de{} & 10.3\dn{} & 150 & \\
\er{}           & 33.61 & 68.7\de{} & 10.4\dn{} & 150 & \\
\gp{}           &       &           &           &       & 96.0\de{} & 11.9\dn{} \\
\cm{}           &       &           &           &       & 95.8\de{} & 10.0\dn{} \\ \midrule \midrule
\mc{7}{c}{\textbf{\cam{} simulations}} \\
\mr{2}{*}{Case} & \mc{4}{c}{Zonal wind}                 & \mc{2}{c}{Precipitation} \\ \cline{2-7}
                & Peak  & Lon       & Lat       & Press & Lon       & Lat \\ \midrule
\cc{}           & 47.19 & 42.5\de{} & 8.8\dn{}  & 125   & 76.7\de{} & 9.7\dn{} \\
\nglo{}         & 43.45 & 43.5\de{} & 7.6\dn{}  & 125   & 77.0\de{} & 7.4\dn{} \\ \midrule
\mc{7}{c}{\cc{}: default orography} \\
\mc{7}{c}{\nglo{}: no orography} \\ \bottomrule
\end{tabular}
\end{center}
\end{table}

\subsection{Impact of latent heating on the location of the TEJ}\label{tej-latent}

The TEJ is not a stationary entity. Variations in its position in different years can also be observed. This is most strikingly observed in July 1988 and 2002. The location of the TEJ in during these two years is radically different. This is seen in Fig. \ref{nc-jul-Umax-88,02-xy} where the 30\ms{} zonal wind contour and \um{} locations show significant differences. In fact the location of \um{} is shifted eastwards by $\sim$20\degrees{} in 2002. These differences are seen in \er{} data as well. The vertical structure in these two years is similar to Fig. \ref{nc-yz}.

\begin{figure}[htbp]
   \begin{center}
   \subfloat[\nc{} zonal wind, 1988 (red), 2002 (blue)]{\label{nc-jul-Umax-88,02-xy}\includegraphics[trim = 5mm 20mm 10mm 60mm, clip, scale=0.35]{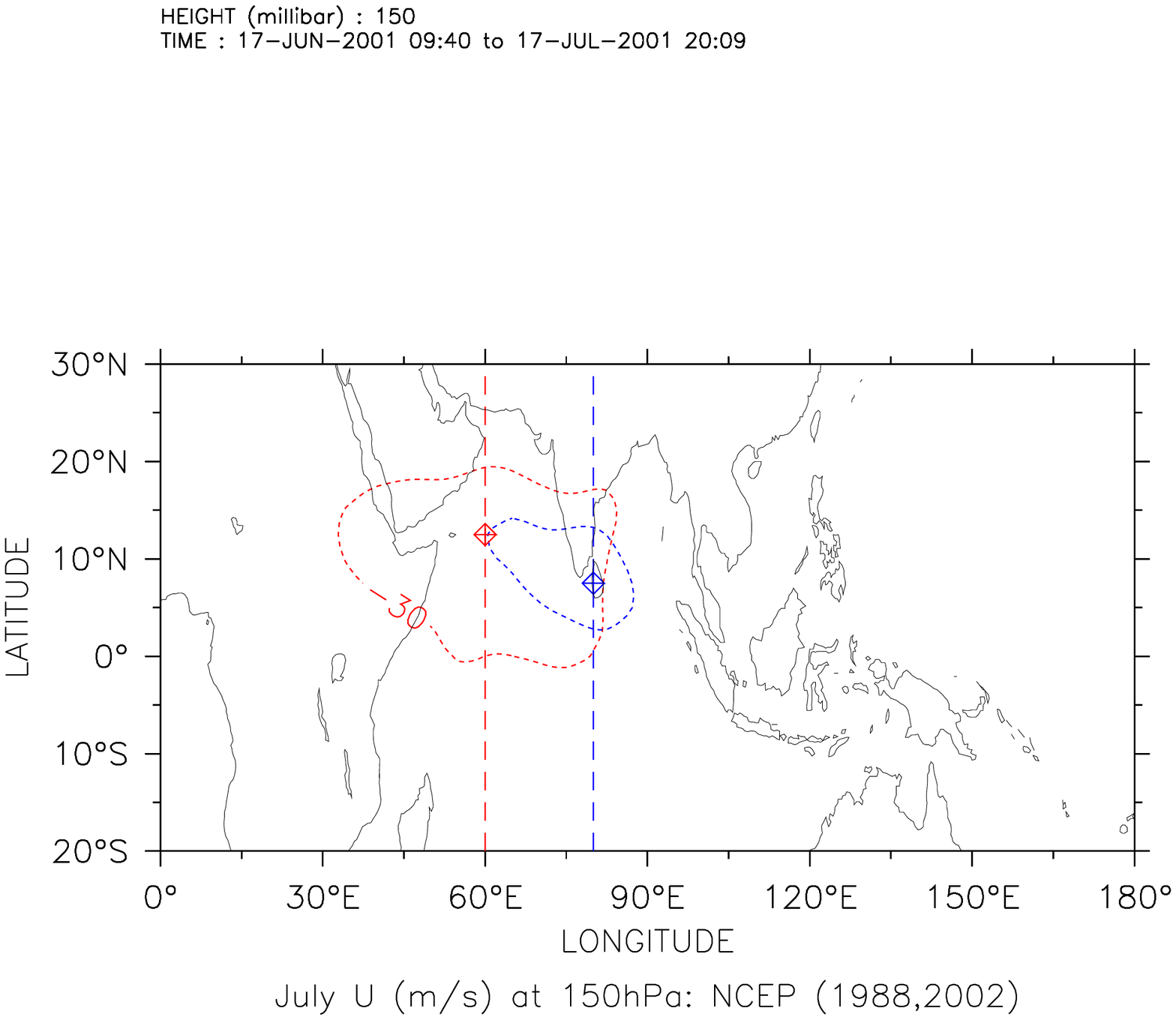}}
   \vskip 5mm
   \subfloat[\nc{} precipitation difference: 1988--2002]{\label{nc-jul-Rdiff-xy}\includegraphics[trim = 5mm 20mm 0mm 60mm, clip, scale=0.35]{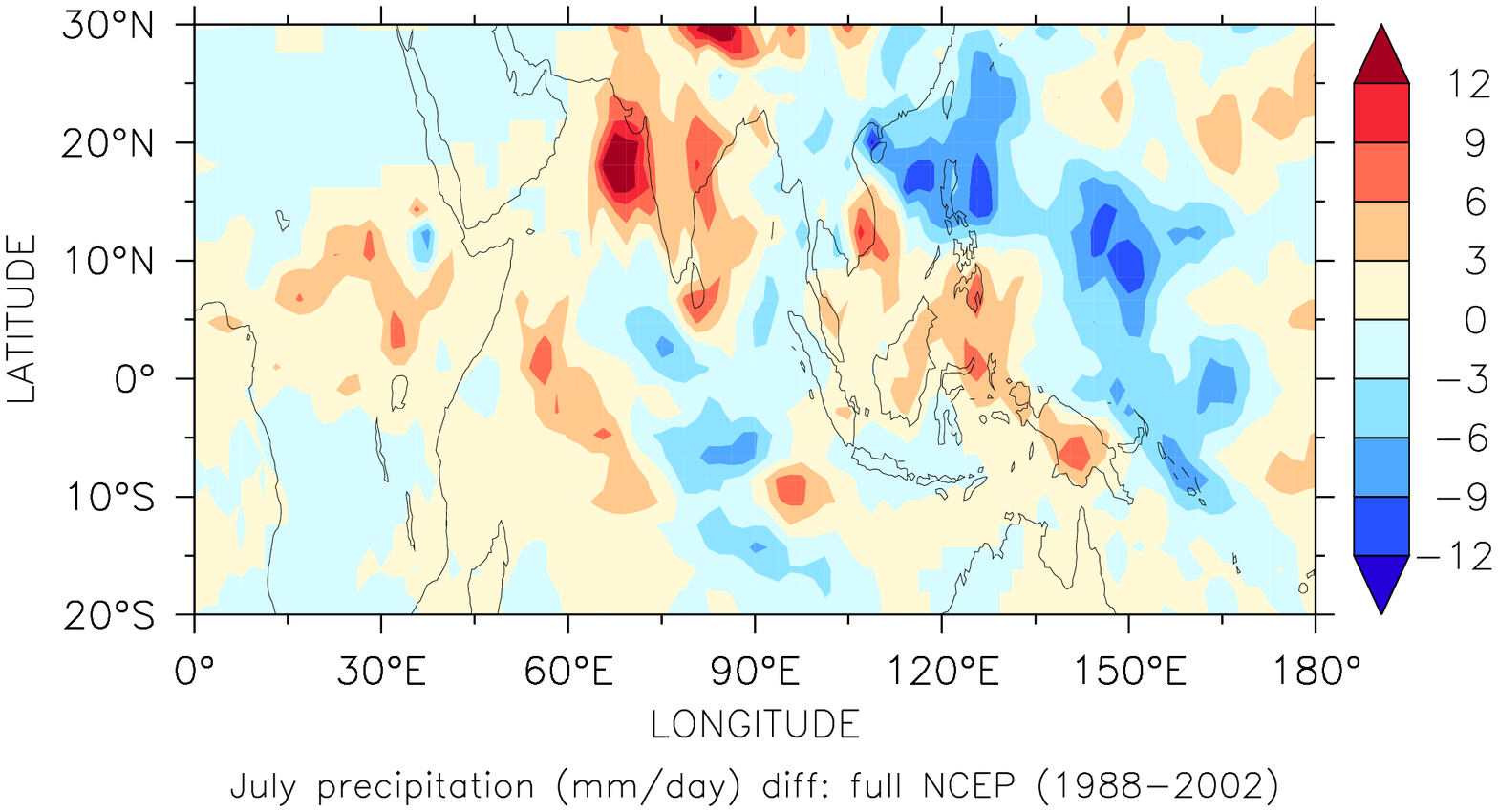}}
   \end{center}
   \caption[July precipitation and zonal wind: \nc{} in 1988 and 2002]{\protect\subref{nc-jul-Umax-88,02-xy} 30\ms{} zonal wind contour (dashed) in 1988 and 2002 showing shift in TEJ in response to heating, cross-sections are at \um{} pressure level (150hPa), `cross-diamond' shows the location of maximum zonal wind; \protect\subref{nc-jul-Rdiff-xy} precipitation difference (\md{}), year 2002 subtracted from 1988. All data from \nc{}.}
   \label{fig: nc-R,Umax-xy}
\end{figure}

It is well known that in India 2002 was a drought year (\citealp{sikka-03}, \citealp{bhat-06}) while 1988 was an excess monsoon year. The relationship between precipitation and the jet location in the month of July for these two years is now clarified. Here in order to be self-consistent,both precipitation and zonal winds from \nc{} data have been used. This is because \rn{} data is self-consistent and that the dynamic response of the atmosphere is entirely on account of forcing from the same data. If the TEJ is influenced largely by latent heating due to Indian Summer Monsoon then the July 2002 shift should be due to the mean rainfall being eastward shifted. This is clearly seen in Fig. \ref{nc-jul-Rdiff-xy} where precipitation differences between July 1988 and 2002 have been plotted. The differences are quite striking. In July 2002 there is a clear eastward shift in rainfall with the maximum being in the Pacific warm pool and relatively little in the Indian region. This is in contrast with July 1988 where Indian region received high amounts of precipitation. The jet in 1988 peaks at 60\de{}-65\de{} while in 2002 the maximum is in the southern Indian peninsula, implying a $\sim$20\degrees{} westward shift. The same behaviour was also found for \er{} data and hence is not shown.

\section{The \tej{} in \cam{}}\label{tej-cam}

\subsection{Description of \cam{} and experiment details}\label{cam-exp}

The AGCM that has been used for the present work is the Community Atmosphere Model, version 3.1 (\cam{}). The finite-volume dynamical core using the recommended 2\degrees{}$\times$2.5\degrees{} grid resolution has been used for all simulations. The time step is 30 minutes and 26 levels in the vertical are used. Deep convection is the \cite{zhang-95} scheme while shallow convection is the \cite{hack-94} scheme. Stratiform processes employs the \cite{rasch-98} scheme updated by \cite{zhang-03}. Cloud fraction is computed using a generalization of the scheme introduced by \cite{slingo-89}. The shortwave radiation scheme employed is described in \cite{briegleb-92}. The longwave radiation scheme is from \cite{ramanathan-86}. Land surface fluxes of momentum, sensible heat, and latent heat are calculated from Monin-Obukhov similarity theory applied to the surface. Climatological mean SST was specified as the boundary condition. Sea surface temperatures are the blended products that combine the global Hadley Centre Sea Ice and Sea Surface Temperature (HadISST) dataset (\citealp{rayner-03}) for years up to 1981 and \cite{reynolds-02} dataset after 1981.

The model was run in its default configuration for a five year period. This simulation is referred to as the control (\cc{}) simulation. Additionally another simulation (referred to as \nglo{}) has been conducted to check the influence of orography on the TEJ. This has also been run for five years with same boundary conditions but with orography all over the globe removed. This latter simulation is used to investigate the direct influence of topography on the TEJ. All the simulation results presented in this paper are based on five year means.

\subsection{Impact of orography on the simulated TEJ}\label{cam-orog}

The 5 year average of maximum zonal wind and its corresponding location have been computed. The horizontal (Figs. \ref{cc-xy},\subref*{nglo-xy}) and meridional (Figs. \ref{cc-yz},\subref*{nglo-yz}) profiles of the TEJ for \cc{} and \nglo{} simulations are shown. As with \rn{}, the vertical cross-section is at the location of zonal wind maximum. The existence of a \tej{} can be observed. In both the simulations the first noticeable feature is a $\sim$30\degrees{} westward shift of the simulated jet. This shift in the default \cc{} was also documented by \cite{hurrell-06} where they analyzed the 200hPa JJA zonal wind fields. The location of the peak zonal wind is virtually the same for \cc{} and \nglo{} simulations while the jet is weaker in the absence of orography. The zonal cross-section also shows that the zero line at 500 hPa is sandwiched between the peaks in the TEJ and low-level Somali jet. One major difference appears in the \cc{} and \nglo{} simulation - the maximum in the Somali jet is greater and restricted east of $\sim$50\de{} in the former. This was explained by \cite{chakraborty-08}. The absence of orography in the latter caused the low-level wind to spread out and thereby reduce the intensity without compromising on the total flow. This is also seen in \nc{} data (Fig. \ref{nc-zx}) where the Somali jet is also maximum to the east of $\sim$50\de{}. The meridional cross-section is very similar to \rp{} and shows the familiar vertical equator to pole tilt with the maximum lying between equator and $\sim$20\dn{}. The depth is also maximum on the poleward side.

The impact of orography in determining the location and spatial structure of the jet in \cam{} is thus minimal. The location of the peak zonal wind is virtually the same for \cc{} and \nglo{} simulations while the jet is weaker in the absence of orography. Although CAM-3.1 does show reasonable fidelity in determining the spatial features of the TEJ, the discrepancies in location and magnitude of the maximum velocity of the TEJ need to be understood. With the insight gained from section \ref{tej-latent} the spatial distribution of simulated and observed rainfall is now studied.

\section{Spatial distribution of heating in \rn{} and \cc{}}\label{precip}

The westward shift of the simulated TEJ indicates that sensible heating from the \tp{} may not play a major role in the existence and location of TEJ. During the monsoon season the major source of heating in the tropics is latent heating and hence it is necessary to look at the role of latent heating.

\begin{figure}[htbp]
   \begin{center}
   \subfloat[Precipitation: \gp{}]{\label{gpcp-jul-R-xy}\includegraphics[trim = 5mm 20mm 0mm 60mm, clip, scale=0.29]{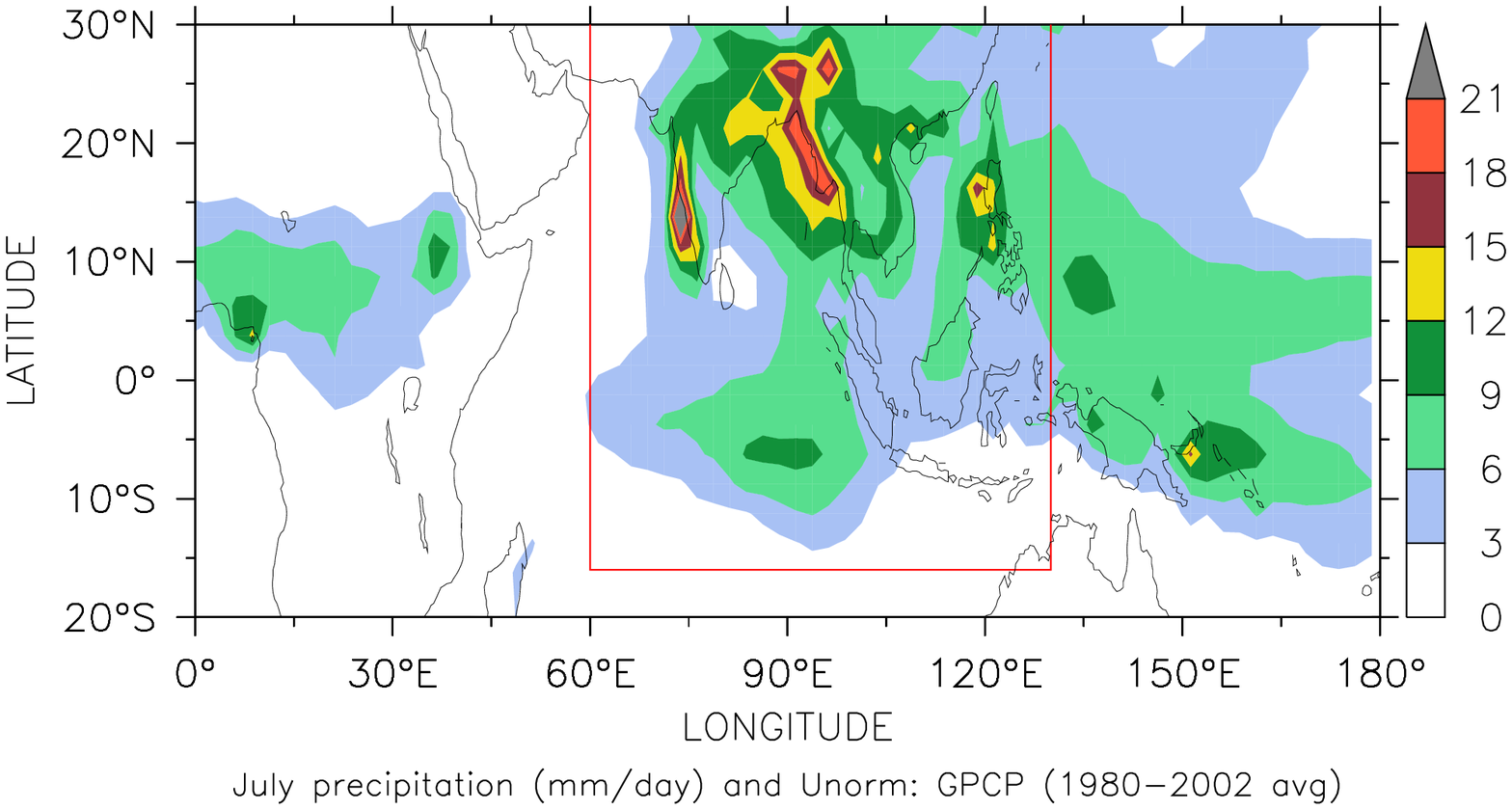}}
   \hspace{1mm}
   \subfloat[Precipitation: \cm{}]{\label{cmap-jul-R-xy}\includegraphics[trim = 5mm 20mm 0mm 60mm, clip, scale=0.29]{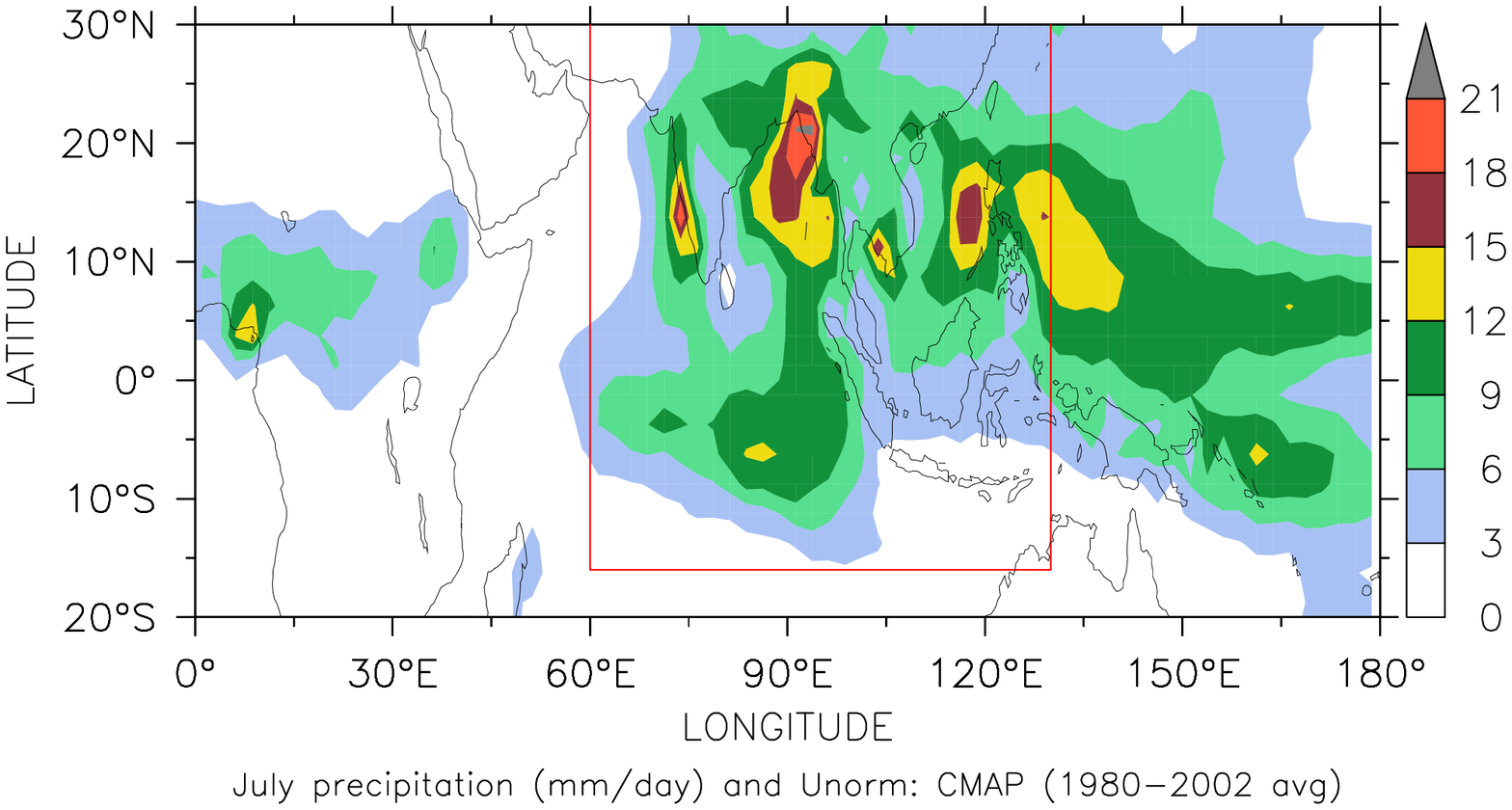}}
   \vskip 5mm
   \subfloat[Precipitation: \cc{}]{\label{cc-jul-R-xy}\includegraphics[trim = 5mm 20mm 0mm 60mm, clip, scale=0.29]{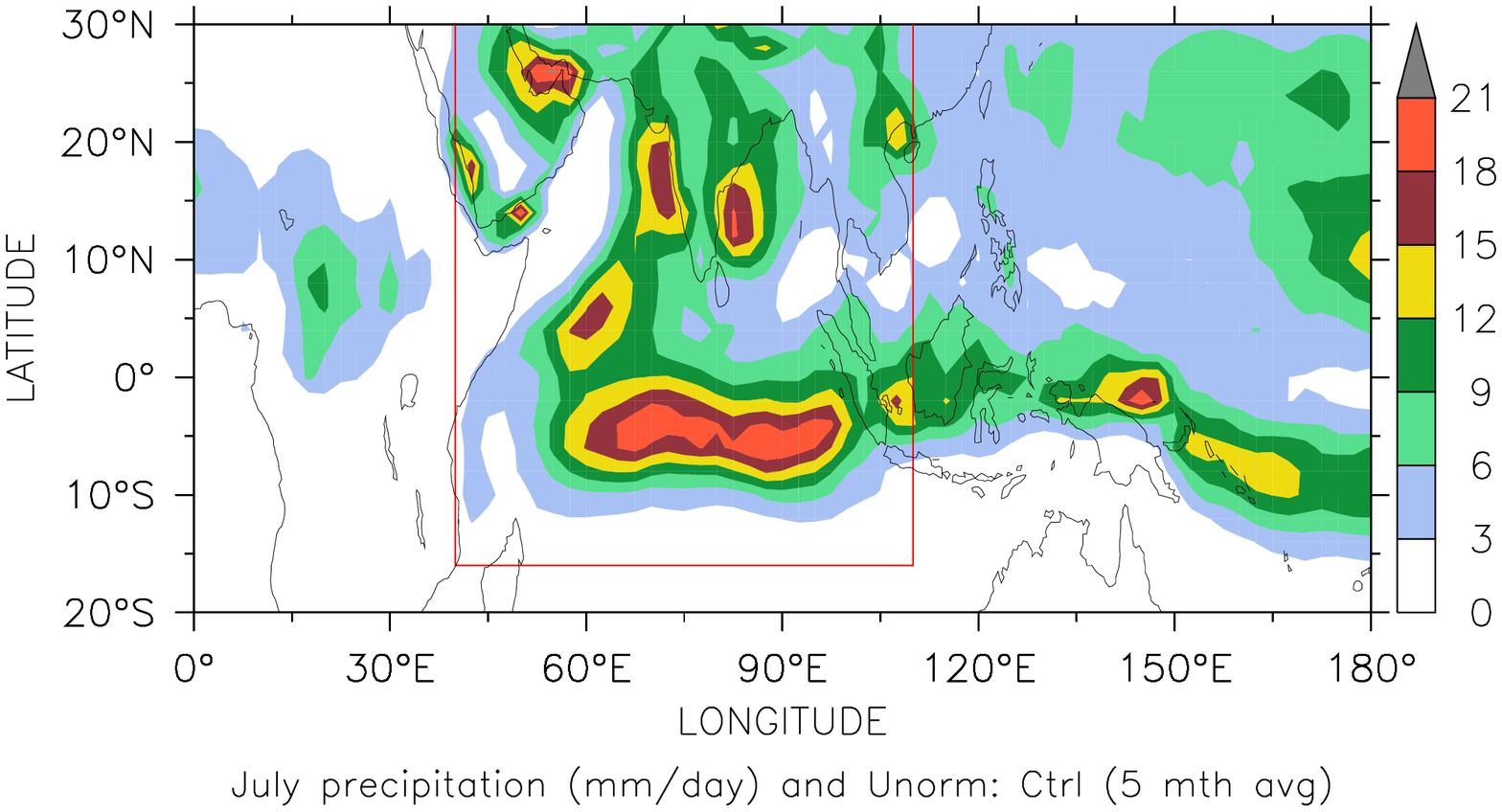}}
   \hspace{1mm}
   \subfloat[Precipitation difference]{\label{mix-Rdiff-xy}\includegraphics[trim = 5mm 20mm 0mm 60mm, clip, scale=0.29]{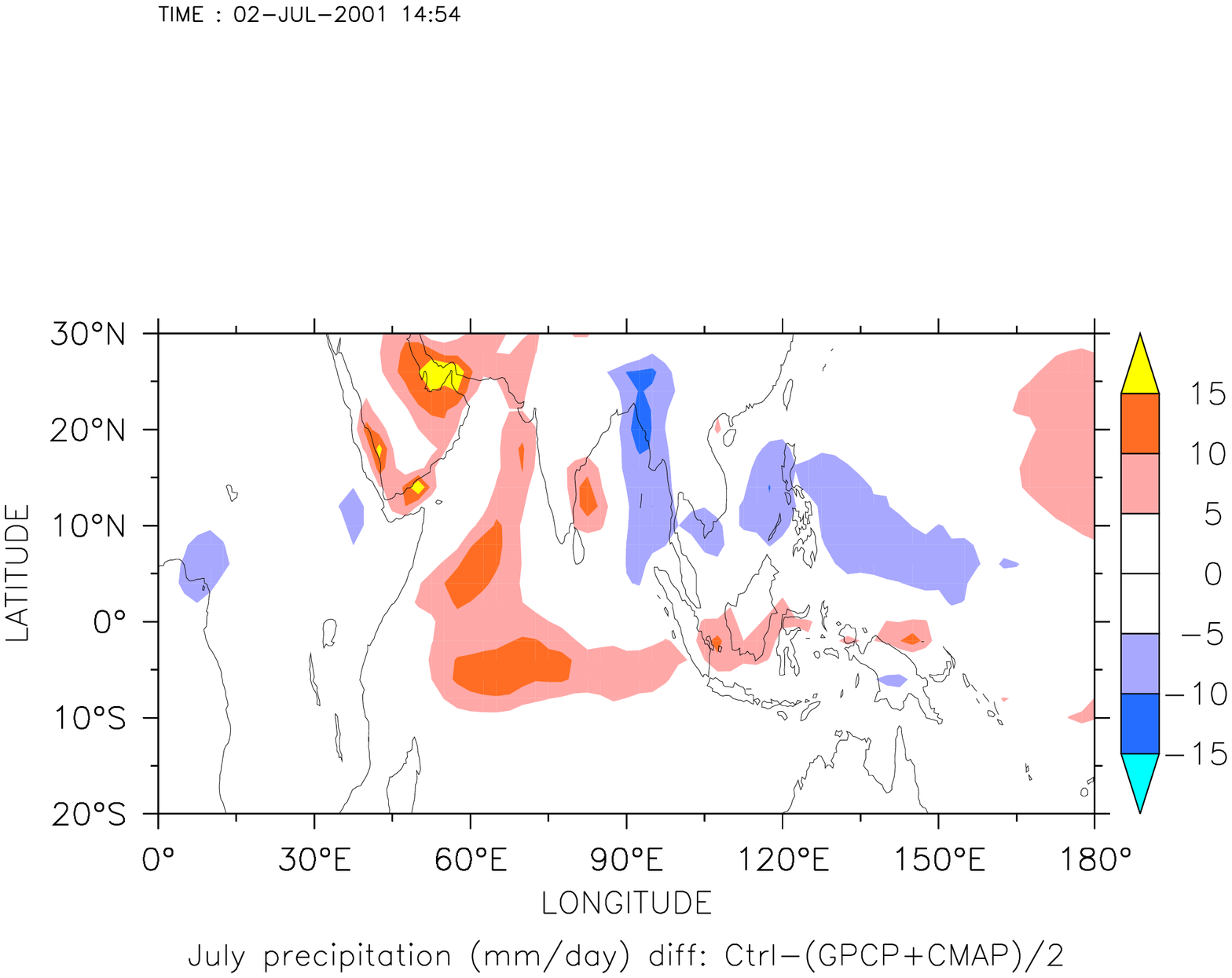}}
   \end{center}
   \caption[July precipitation: \gp{}, \cm{}, \cc{}]{\protect\subref{gpcp-jul-R-xy}, \protect\subref{cmap-jul-R-xy}, \protect\subref{cc-jul-R-xy} July precipitation (\md{}). \protect\subref{mix-Rdiff-xy} July precipitation difference (\md{}) (\cc{}--(\gp{}+\cm{})/2).}
   \label{fig: rean-cam3-Jul-R-xy}
\end{figure}

Figs. \ref{gpcp-jul-R-xy}-\subref*{cc-jul-R-xy} show the precipitation in the month of July of \gp{}, \cm{} and \cc{}. The difference between \cc{} and mean of \gp{} and \cm{} is shown in Fig. \ref{mix-Rdiff-xy}. The observational data have been averaged since the precipitation patterns are very similar. The contrast between \rn{} and model simulations is quite striking. Most noticeable discrepancies in model simulations are (i) significantly reduced precipitation in northern Bay of Bengal, East Asia, western Pacific warm pool (ii) a significant precipitation tongue just south of the equator between 50\de{}-100\de{} and (iii) Spurious precipitation in the Saudi Arabian region which is quite prominent and equal in magnitude to the precipitation peaks in central Arabian Sea and south-western Bay of Bengal. This implies a major realignment in the local heating pattern. This unrealistic precipitation has been discussed by \cite{hurrell-06}.

Thus from Table \ref{tab: mix-Umax,R} and Fig. \ref{fig: rean-cam3-Jul-R-xy} it appears that westward shift in the peak of the simulated precipitation is responsible for the TEJ to be centered east Africa.  The strongest impact seems to be from the significantly high anomalous precipitation in the Saudi Arabian region which could play a part in the extreme westward shift of the TEJ.

As in \cite{kucharski-09} and \cite{davis-12}, precipitation is used as a proxy for latent heating. The centroid of precipitation (\pc{}) has been computed for in a region that is spatially quite significant and covers the monsoon region. Different regions for \rn{} and \cam{} simulations have been chosen since the precipitation pattern is spatially different. For \rn{} the west Pacific warm pool is also considered, while it is ignored for model simulations. The region chosen for \rn{} is 60\de{}-130\de{}, 16\ds{}-36\dn{} and for the simulations it is 40\de{}-110\de{}, 16\ds{}-36\dn{} (region demarcated by red boxes in Figs. \ref{gpcp-jul-R-xy}-\subref*{cc-jul-R-xy}). From Table \ref{tab: mix-Umax,R} it is seen that the mean precipitation in simulations and \rn{} are in reasonable agreement although \cc{} experiment overestimated the precipitation by $\sim$15\%. The excess precipitation could be a cause for stronger jet speeds. The centroid of the precipitation, henceforth \pc{}, is computed using the following equation:

\begin{equation}\label{centroid}
x_c = \frac{\sum_i{P_ix_i}}{\sum_i{P_i}}, y_c =  \frac{\sum_i{P_iy_i}}{\sum_i{P_i}}
\end{equation}

\noindent{where,} \\
$x_c$ and $y_c$ are the zonal and meridional coordinates of the \pc{}, \\
$P_i$ is the precipitation at each grid point, \\
$x_i$ and $y_i$ are the zonal and meridional distances from a fixed coordinate system, in each case the grid point where peak precipitation occurs.

The distances are measured from a coordinate system centered at a point where the precipitation is maximum in the chosen region. There is a clear $\sim$20\degrees{} westward shift in the precipitation centroid in \cc{} with respect to \rn{}. Precipitation pattern of \nglo{} is similar to \cc{} and hence even in this case the shift is $\sim$20\degrees{} westward. The latitudinal differences are not significant. Choosing slightly different averaging regions does not significantly distort this relationship.

Consequent to the westward shift in the precipitation centroid and TEJ, the peak geopotential height at 150hPa in the simulations has also shifted westwards by $\sim$25\degrees{}. This is clearly seen in Fig. \ref{fig: mix-z3-xy} where the geopotential contour corresponds to 99.5\% of the peak, the location of which is indicated by the ‘cross-diamond’. The velocity vectors clearly show the anti-cyclonic flow around the geopotential high.

\begin{figure}[htbp]
   \begin{center}
   \subfloat[\cct{} (125 hPa)]{\label{cct-xy}\includegraphics[trim = 5mm 5mm 10mm 75mm, clip, scale=0.35]{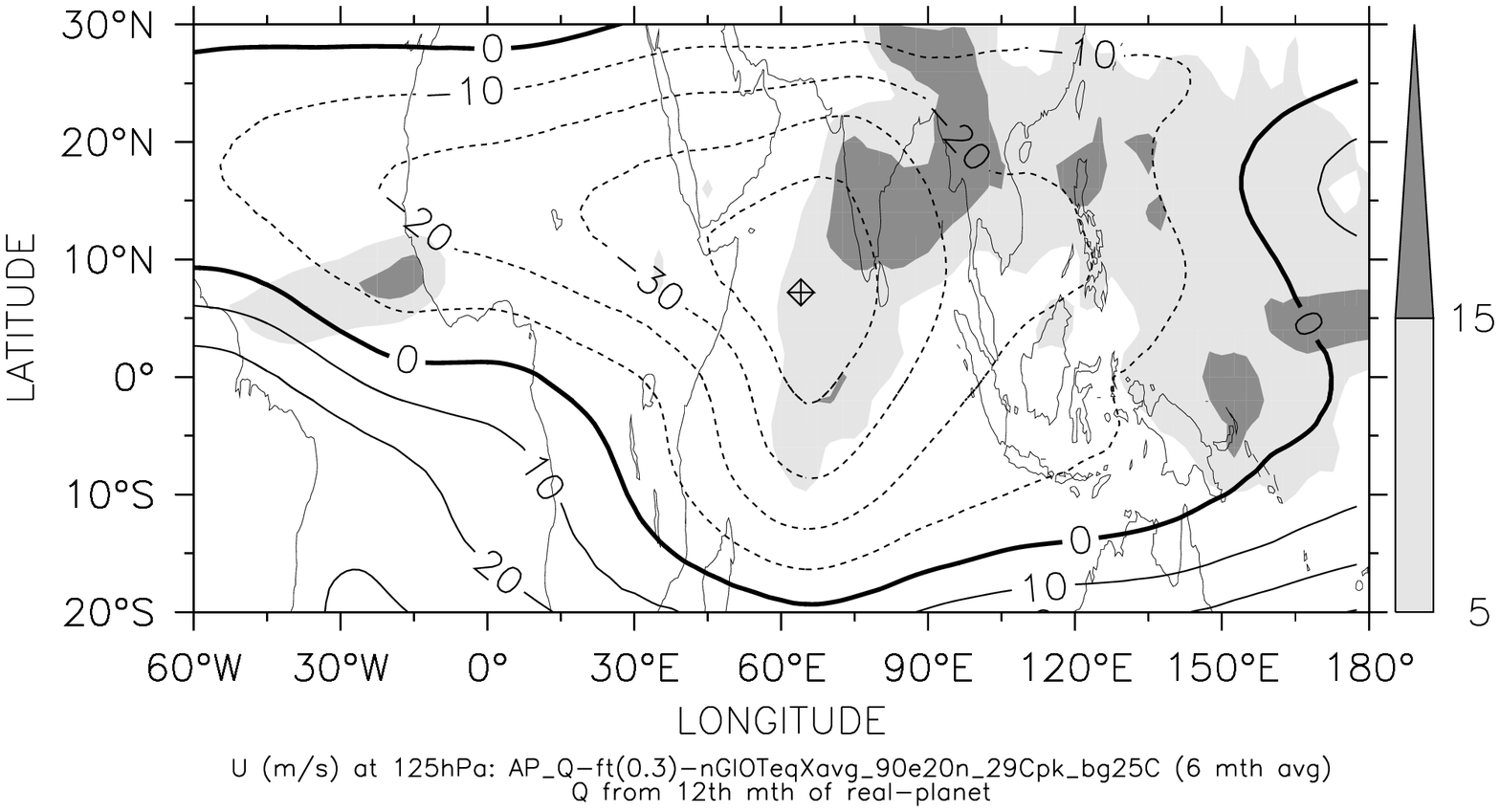}}
   \vskip 5mm
   \subfloat[\nglot{} (125 hPa)]{\label{nglot-xy}\includegraphics[trim = 5mm 5mm 10mm 75mm, clip, scale=0.35]{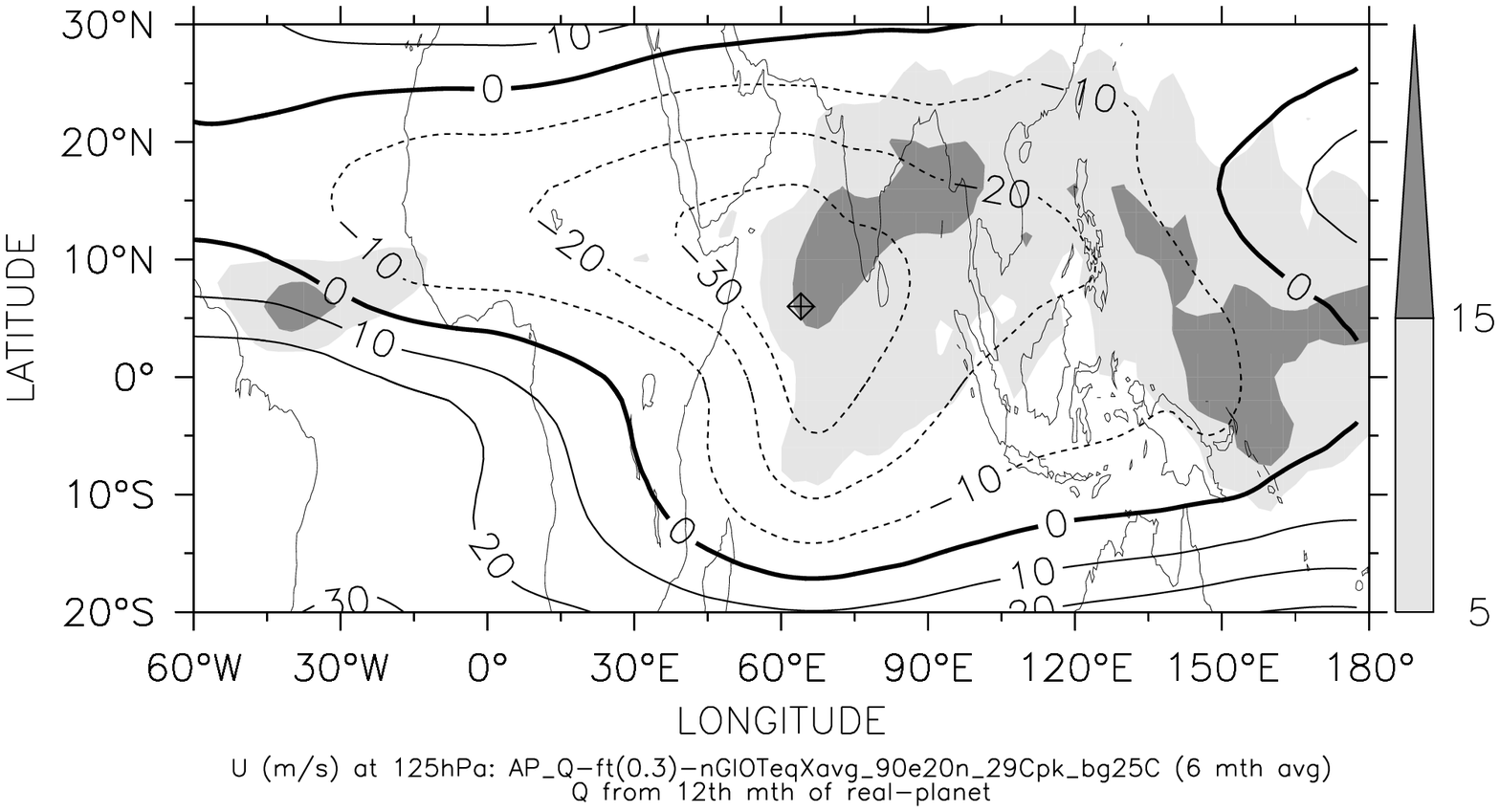}}
   \vskip 5mm
   \subfloat[\cct{} (150 hPa)]{\label{cct-z3-xy}\includegraphics[trim = 5mm 5mm 10mm 70mm, clip, scale=0.35]{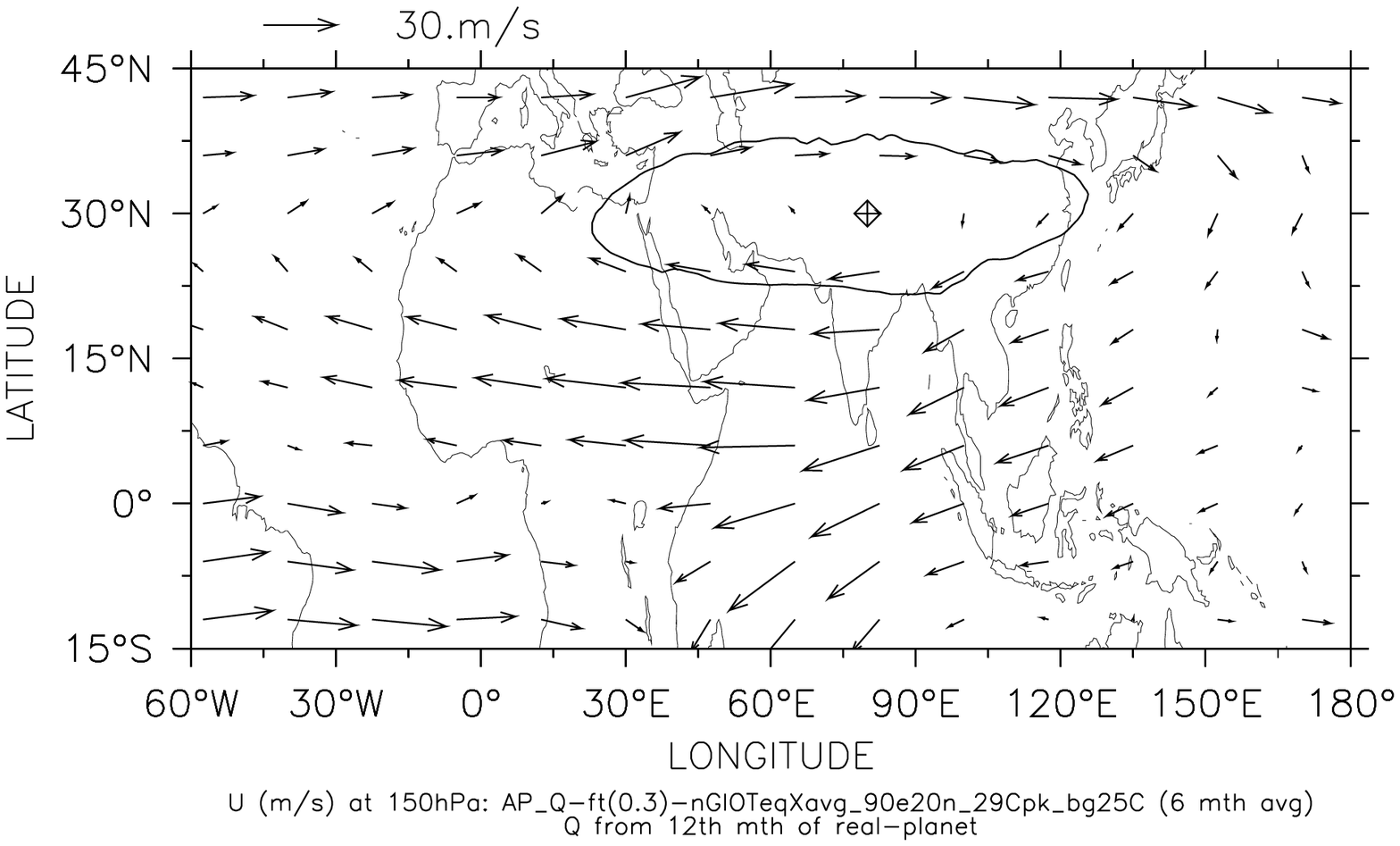}}
   \end{center}
   \caption[July precipitation and horizontal zonal wind profile, velocity vectors and geopotential height: \cct{} and \nglot{}.]{\protect\subref{cct-xy}, \protect\subref{nglot-xy} July precipitation (\md{}, shaded) and horizontal zonal wind (\ms{}) profile at pressure level where maximum zonal wind is attained; `cross-diamond' is location of peak zonal wind, \protect\subref{cct-z3-xy} velocity vectors and geopotential height ($\times10^{3}$m) at 150 hPa; `cross-diamond' is location of peak geopotential height; contour value is 14.46 which is 99.5\% of peak: \cct{} and \nglot{}.}
   \label{fig: tau_jul_xy}
\end{figure}

\begin{figure}[htbp]
   \begin{center}
   \includegraphics[trim = 0mm 0mm 0mm 70mm, clip, scale=0.35]{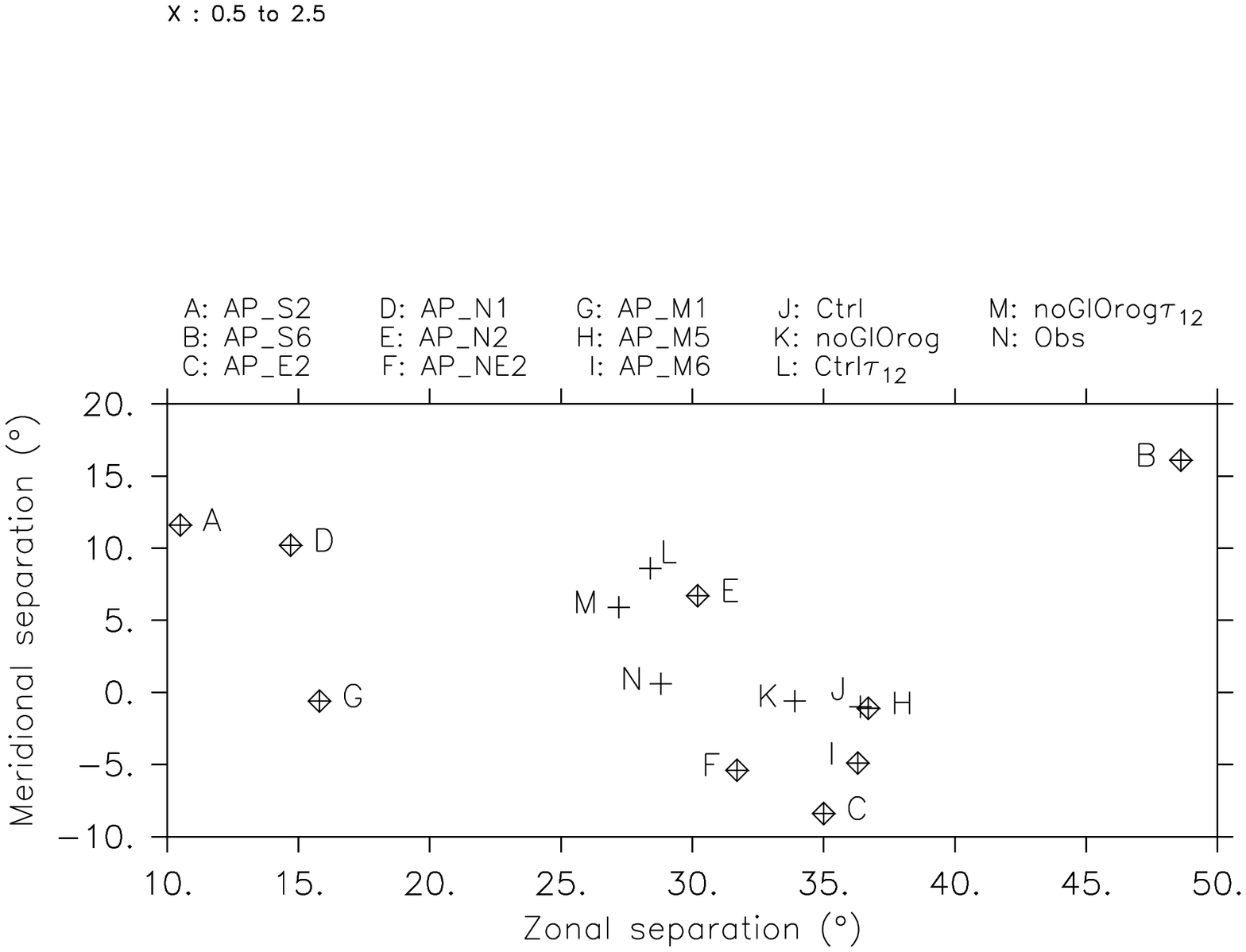}
   \end{center}
   \caption[Separation ($P_{c}$--$U_{l}$) between location of precipitation centroid and maximum zonal wind: \rn{}, \rp{} and \ap{}.]{Separation ($P_{c}$--$U_{l}$) between location of precipitation centroid and maximum zonal wind. `Cross-diamond' denotes \ap{} simulations, `plus' denotes \rp{} simulations and \rn{}.}
   \label{fig: mix-Pc-Umax-loc}
\end{figure}

Though this offers evidence of the primacy of latent heating in determining the location of the jet, the validity of this argument is further demonstrated by showing the precipitation patterns of the \cct{} and \nglot{} simulations. The mean precipitation and centroid are shown in Table \ref{tab: mix-Umax,R}. Since the precipitation pattern is now similar to \rp{} (not shown), the same region as in \rp{} has been used to compute the location of the precipitation centroid. Figs. \ref{cct-xy},\subref*{nglot-xy} show the precipitation and TEJ in these two simulations. The location of the TEJ agrees well with \rp{}. Although the mean precipitation is more in comparison to the previous simulations, the peak zonal wind speed is almost the same. The geopotential high (Fig. \ref{cct-z3-xy}) in the \cct{} simulation is now almost same as that in \nc{} (Fig. \ref{nc-z3-xy}). The spatial separation between the precipitation centroid and \um{} is included in Fig. \ref{fig: mix-Pc-Umax-loc}. The zonal separation is also in closer agreement to \rp{} and this is may be attributed to the increased similarity in spatial pattern of precipitation in \rp{} and these two simulations. The similarity in the precipitation patterns and TEJ structures in both these simulations as well as the previously discussed \cc{} and \nglo{} simulations suggests that it is the location and strength of the heat source and not orography that controls the TEJ. The only difference in precipitation between the \cct{} and \nglot{} simulations is that the former has additional heating to the north of 20\dn{} around 90\de{} longitude. The absence of this peak in the latter further suggests that the \tp{} is not the primary reason to set up the temperature gradients that lead to the formation of the TEJ. Sensible heating due to the presence of \tp{} may be less important when dominated by latent heat release due to convective processes.

Thus it has been demonstrated that there is a close correspondence between the source of heating, spatial location of TEJ and the geopotential high. The negligible influence of orography and strong effect of heating on upper level wind patterns has also been discussed by \cite{liu-07}. It is also worthwhile to note that the Somali jet in \nc{} and \cc{} simulation is roughly in the same location and hence the lower and upper-level winds (Figs. \ref{nc-zx},\subref*{cc-zx}) are not merely reflections of each other, not just in magnitude but also in location, as simple though insightful models, e.g. \cite{gill-80} suggest. However the spatial location of the low-level jet and upper-level TEJ is more closely correlated in the \nglo{} simulation and this is due to the absence of orography as will also be shown in the subsequent discussion on \ap{} simulations.

However this mean pattern does not present the full picture. Between 70\de{}-100\de{} the precipitation peak in \rp{} is more northwards in comparison to \cam{} while the opposite is true between 30\de{}-60\de{}. In fact \rp{} hardly shows rainfall in the latter region. Hence it is not clear how the spatial structure of the precipitation influences the positioning of the TEJ. This aspect will need to be clarified before further analysis and this will be the focus of the next section where a major simplification is adopted by running \cam{} in the \ap{} (AP) mode.

\section{Aqua-planet simulations}\label{sec: aqua}

Since orography has hardly any impact on the TEJ in the \rp{} simulations it is instructive to study the atmospheric response only due to heating. The multiplicity of heat sources both in \rn{} and \cc{} preclude any easy interpretation of the influence that each heat source has on the TEJ. A major simplification is conceivable if one removes orography, land, sea-ice totally and further remove seasonal cycles, and yet retain all the important physics that the AGCM offers. This also implies that one needs to prescribe SSTs as a boundary condition. All this is possible in the \ap{} configuration of \cam{}. The solar insolation is perpetually fixed at 21$^{st}$ March which is March Equinox.

In order to understand the role played by heating in determining the structure and location, \cam{} has been run in \ap{} configuration. The basic state consists of a uniform background SST on which additional heat sources are imposed. The heat sources are indirectly specified by setting an SST perturbation on this uniform SST background. Precipitation induced on account of this SST perturbation is representative of total atmospheric heating. The implicit assumption is that latent heating is the dominant effect in the region of precipitation.

The rationale for imposing heat sources on a uniform SST background in contrast to a zonally symmetric and meridionally varying SST profile is now explained. \cite{rajendran-09} showed that equatorial easterlies will be simulated even if a zonally symmetric but meridionally varying SST profile, symmetric about the equator, is used. The existence of an equatorial jet also depends on the presence of twin or single Inter-Tropical Convergence Zone (\cite{rajendran-09}, \cite{neale-01}). In such a meridionally varying SST profile it is more difficult to determine the role played by weak heat sources in the formation of a jet in \ap{}.

The \ap{} simulations with just uniform background SSTs of 20\dc{} and 25\dc{} had a weak equatorial easterly of about 10-15\ms{} peaking at $\sim$250hPa. In this simple configuration, it is implicitly assumed that SSTs beyond 60\dn{}/\ds{} do not significantly influence equatorial dynamics. This was also suggested by \cite{hoskins-95} when they conducted a series of experiments to understand the Asian summer monsoon. Thus a uniform background SST is deemed to be the simplest basic state on which additional heating may be imposed to study the influence of heating on the TEJ.

The reasons for the existence of these easterlies is as follows: these simulations had a band of weak precipitation near the equator. This would naturally imply that moist parcels arise aloft implying the presence of a weak Hadley Cell. This means that air is mixed latitudinally. Therefore there must be westerly motion at higher latitudes and easterly motion at lower latitudes owing to angular momentum conservation. The time-averaged torque on the whole atmosphere due to surface friction must be zero, which requires that there be both easterly and westerly winds. Thus easterlies must prevail near the equator. Drag on surface easterlies also transfers angular momentum from the surface to the atmosphere (\cite{schneider-06}). The issue of the existence of equatorial upper tropospheric easterlies has also been discussed by \cite{lee-99}. According to her, in the deep tropics the horizontal transient eddy momentum flux accelerates the zonal mean zonal wind. Transient eddies of intraannual and interannual timescales were important determining factors. Hadley cell dynamics was also important. According to \cite{bordoni-08} the Hadley cell approaches the angular momentum conservation limit because of these upper-level easterlies (upper-level easterlies strengthen when additional off-equatorial heating is present).

\begin{figure}[htbp]
   \begin{center}
   \includegraphics[trim = 5mm 5mm 10mm 75mm, clip, scale=0.35]{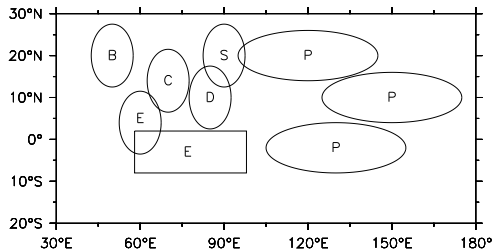}
   \end{center}
   \caption[Schematic showing locations of imposed SSTs for \ap{} simulations.]{Schematic showing locations of imposed SSTs for \ap{} simulations. Refer Table \ref{tab: AP-SSTloc} for further details. (Not drawn to scale).}
   \label{fig: AP-SSTloc}
\end{figure}

The details of all the \ap{} simulations are in Tables \ref{tab: AP-SSTloc} and \ref{tab: AP-Umax,R}. In Fig. \ref{fig: AP-SSTloc}, the different locations where SSTs are imposed are shown. The names of the \ap{} simulations start with `AP'. The magnitude of peak zonal wind and mean precipitation are listed in Table \ref{tab: AP-Umax,R}. The important point to note is that the SSTs have been chosen such that mean of precipitation exceeding 5\md{} compares well with \rn{} and \rp{} simulations. Fig. \ref{fig: mix-Pc-Umax-loc} shows the zonal and meridional distance between \pc{} and \um{}. Equation \eqref{centroid} has been used to compute the centroid. The zonal wind for each month has not been individually computed and then averaged; the zonal find fields have been added and then averaged for the six month period. Hence these locations correspond to grid point values of the model. This method is acceptable since the heat source is stationary and the response too will average out over a six month time scale. The coarse model resolution implies that minor fluctuations in location will hardly distort the main observations and inferences. The vertical sections at the longitude and latitude where \um{} is attained. Some representative cases are discussed.

\begin{table}[htbp]
\caption{Location and nomenclature of SST profiles for \ap{} simulations. See table \ref{tab: AP-Umax,R} for full nomenclature. Figure \ref{fig: AP-SSTloc} shows the schematic of the locations.}
\begin{center}
\begin{tabular}{cccc} \toprule
Region       & \mc{2}{c}{Shape of SST profile}                         & Location of SST peak \\ \hline
\mr{5}{*}{S} & Circle (C)           & 20\degrees{} diameter            & \mr{5}{*}{90\degrees{}E,20\degrees{}N} \\ \cline{2-3}
             & \mr{2}{*}{Oval (Oa)} & 50\degrees{} major axis          & \\
             &                      & 16\degrees{} minor axis          & \\ \cline{2-3}
             & \mr{2}{*}{Oval (Ob)} & 90\degrees{} major axis          & \\
             &                      & 16\degrees{} minor axis          & \\ \hline
\mr{2}{*}{E} & Rectangle            & 10\degrees{} slope               & 62\degrees{}E,4\degrees{}S to 94\degrees{}E,2\degrees{}S \\
             & Circle               & 20\degrees{} diameter            & 60\degrees{}E,4\degrees{}N  \\ \hline
B            & \mr{3}{*}{Circle}    & \mr{3}{*}{20\degrees{} diameter} & 50\degrees{}E,20\degrees{}N \\ \cline{1-1} \cline{4-4}
C            &                      &                                  & 70\degrees{}E,14\degrees{}N \\ \cline{1-1} \cline{4-4}
D            &                      &                                  & 85\degrees{}E,10\degrees{}N \\ \hline
\mr{3}{*}{P} & \mr{3}{*}{Oval}      & \mr{3}{*}{As in (Ob) above}      & 120\degrees{}E,20\degrees{}N \\
             &                      &                                  & 150\degrees{}E,10\degrees{}N \\
             &                      &                                  & 130\degrees{}E,2\degrees{}S \\ \bottomrule
\end{tabular}
\label{tab: AP-SSTloc}
\end{center}
\end{table}

\begin{table}[htbp]
\caption{Magnitude (\ms{}) and location of peak zonal wind, and magnitude (\md{}) and centroid of mean precipitation for \ap{} simulations. Refer Fig. \ref{fig: AP-SSTloc} and table \ref{tab: AP-SSTloc} for shape and location of SST profiles.}
\begin{center}
\begin{tabular}{cccccccccc} \toprule
\mc{9}{c}{\textbf{Single heat source (at region S) simulations}} \\ \hline
\mr{2}{*}{Case} & \mc{2}{c}{SST}                        & \mc{4}{c}{Zonal wind}                           & \mc{3}{c}{Precipitation} \\ \cline{2-10}
                & Peak          & Shape                 & Peak  & Lon             & Lat           & Press & Mean  & Lon             & Lat \\ \midrule
\as{}1          & 29\degrees{}C & \mr{2}{*}{Circle (C)} & 23.64 & 70\degrees{}E   & 6\degrees{}N  & 175   & 8.81  & 90.2\degrees{}E & 21.9\degrees{}N \\
\as{}2          & 32\degrees{}C &                       & 33.97 & 80\degrees{}E   & 10\degrees{}N & 150   & 12.41 & 90.5\degrees{}E & 21.6\degrees{}N \\ \hline
\as{}3          & 29\degrees{}C & \mr{2}{*}{Oval (Oa)}  & 30.75 & 65\degrees{}E   & 6\degrees{}N  & 175   & 9.24  & 88.8\degrees{}E & 20.7\degrees{}N \\
\as{}4          & 32\degrees{}C &                       & 37.98 & 67.5\degrees{}E & 6\degrees{}N  & 150   & 14.74 & 89.4\degrees{}E & 21.4\degrees{}N \\ \hline
\as{}5          & 29\degrees{}C & \mr{2}{*}{Oval (Ob)}  & 35.58 & 45\degrees{}E   & 4\degrees{}N  & 175   & 10.17 & 86.3\degrees{}E & 20.5\degrees{}N \\
\as{}6          & 32\degrees{}C &                       & 42.56 & 40\degrees{}E   & 4\degrees{}N  & 150   & 15.06 & 88.6\degrees{}E & 20.1\degrees{}N \\ \hline
\mc{9}{c}{Uniform background temperature: 25\degrees{}C} \\ \midrule \midrule
\mc{9}{c}{\textbf{Multiple heat source simulations}} \\ \hline
\mr{2}{*}{Case} & \mr{2}{*}{SST region} & \mc{4}{c}{Zonal wind}                           & \mc{3}{c}{Precipitation} \\ \cline{3-9}
                &                       & Peak  & Lon             & Lat           & Press & Mean  & Lon             & Lat \\ \midrule
\am{}E1         & \mr{2}{*}{E}          & 15.54 & 52.5\degrees{}E & 2\degrees{}N  & 225   & 11.52 & 74.2\degrees{}E & 3.6\degrees{}S \\
\am{}E2         &                       & 17.58 & 37.5\degrees{}E & 6\degrees{}N  & 225   & 11.49 & 72.5\degrees{}E & 2.4\degrees{}S \\ \hline
\am{}N1         & \mr{2}{*}{B,C,D}      & 33.08 & 37.5\degrees{}E & 10\degrees{}N & 150   & 9.02  & 67.7\degrees{}E & 16.7\degrees{}N \\
\am{}N2         &                       & 30.41 & 57.5\degrees{}E & 6\degrees{}N  & 175   & 9.27  & 72.3\degrees{}E & 16.2\degrees{}N \\ \hline
\am{}NE1        & \mr{2}{*}{B,E}        & 26.73 & 37.5\degrees{}E & 8\degrees{}N  & 175   & 10.37 & 67.0\degrees{}E & 2.7\degrees{}N \\
\am{}NE2        &                       & 23.80 & 42.5\degrees{}E & 14\degrees{}N  & 150  & 7.51  & 67.3\degrees{}E & 3.1\degrees{}N \\ \hline
\am{}M1         & \mr{2}{*}{C,D,E}      & 30.89 & 55\degrees{}E   & 6\degrees{}N  & 125   & 11.40 & 70.8\degrees{}E & 5.4\degrees{}N \\
\am{}M2         &                       & 24.91 & 52.5\degrees{}E & 4\degrees{}N  & 150   & 11.62 & 71.6\degrees{}E & 2.6\degrees{}N \\ \hline
\am{}M3         & \mr{3}{*}{B,C,D,E}    & 24.31 & 40\degrees{}E   & 10\degrees{}N & 150   & 9.03  & 71.9\degrees{}E & 2.3\degrees{}N \\
\am{}M4         &                       & 36.02 & 35\degrees{}E   & 8\degrees{}N  & 150   & 9.03  & 68.7\degrees{}E & 5.4\degrees{}N \\
\am{}M5         &                       & 34.43 & 32.5\degrees{}E & 6\degrees{}N  & 150   & 10.86 & 69.2\degrees{}E & 4.9\degrees{}N \\ \hline
\am{}M6         & B,C,D,E,P             & 35.70 & 32.5\degrees{}E & 8\degrees{}N  & 150   & 10.22 & 68.8\degrees{}E & 3.1\degrees{}N \\ \midrule
\mc{5}{c}{Uniform background temperature: 22\degrees{}C} & \mc{4}{c}{Peak temperatures: $\leq$ 29\degrees{}C} \\ \bottomrule
\end{tabular}
\label{tab: AP-Umax,R}
\end{center}
\end{table}

\subsection{Single heat source simulations}\label{ap-single}

For single heat source simulations, the SST profiles imposed have circular and oval shapes all centered at 90\de{},20\dn{} which mimic the off-equatorial monsoonal heat source in the northern Bay of Bengal. The circular profiles (AP\_S1 and AP\_S2) have a diameter of 20\degrees{} while the oval shaped profiles have major axes of either 50\degrees{} (AP\_S3 and AP\_S4) or 90\degrees{} (AP\_S5 and AP\_S6). All oval profiles have a minor axis of 16\degrees{}. The simulations with oval-shaped SSTs serve to demonstrate the effect of the shape of the heating region on the jet. The peak SSTs are at 90\de{},20\dn{}. They go linearly down to the background temperature. However the profile with major axes of 90\degrees{} (AP\_S5 and AP\_S6) has a non-linear SST gradient which is used to study if the gradient has any qualitative change on the jet. The background temperature is 25\dc{}. In each set, the first simulation in each set has 29\dc{} peak SST while the second has 32\dc{} peak SST. For example, AP\_S3 has 29\dc{} and AP\_S4 has 32\dc{} SST peak on a uniform background temperature of 25\dc{}. 

\begin{figure}[htbp]
   \begin{center}
   \subfloat[Meridional cross-section]{\label{S6-yz}\includegraphics[trim = 0mm 15mm 80mm 60mm, clip, scale=0.35]{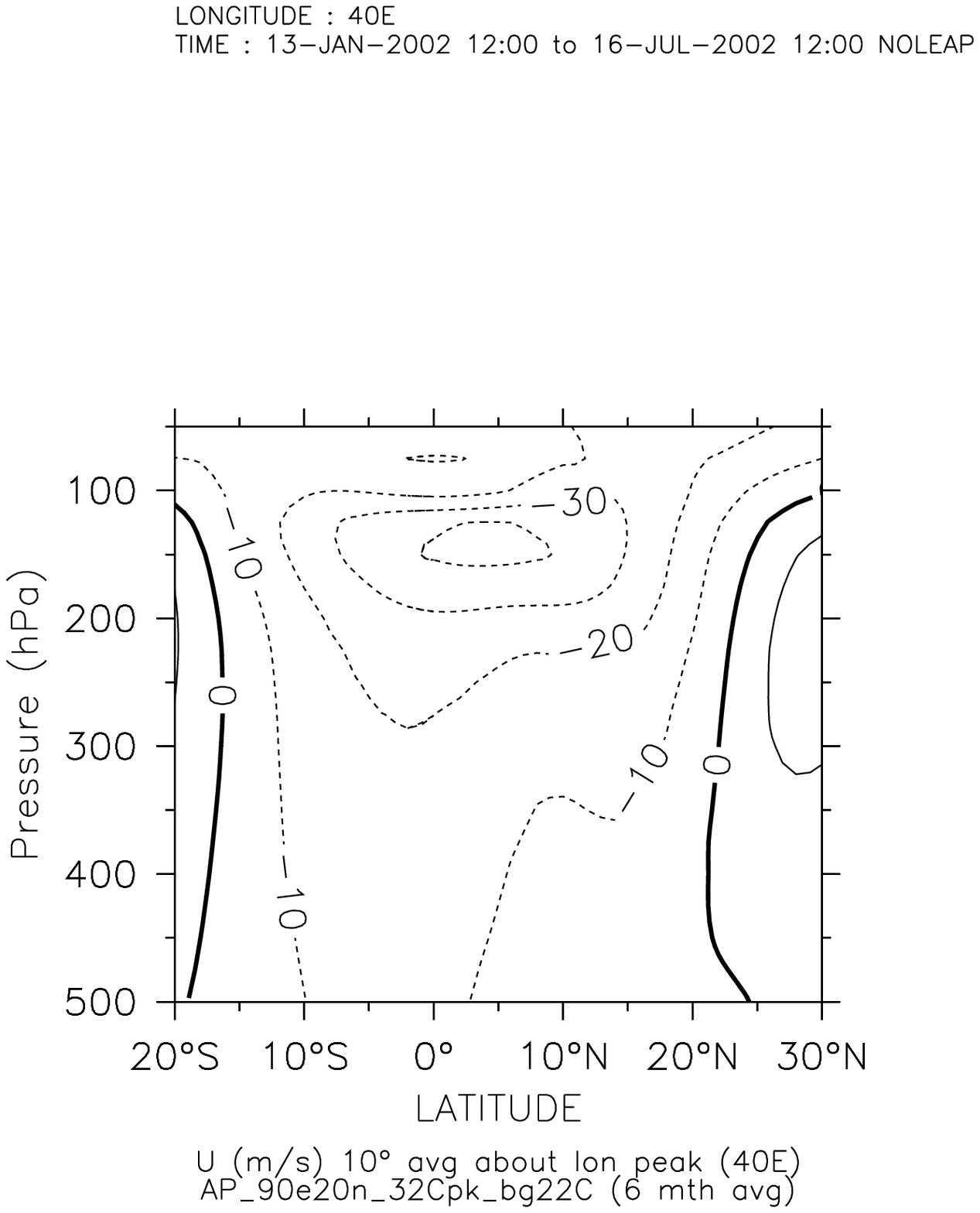}}
   \hspace{5mm}
   \subfloat[Zonal cross-section]{\label{S6-zx}\includegraphics[trim = 0mm 15mm 10mm 60mm, clip, scale=0.35]{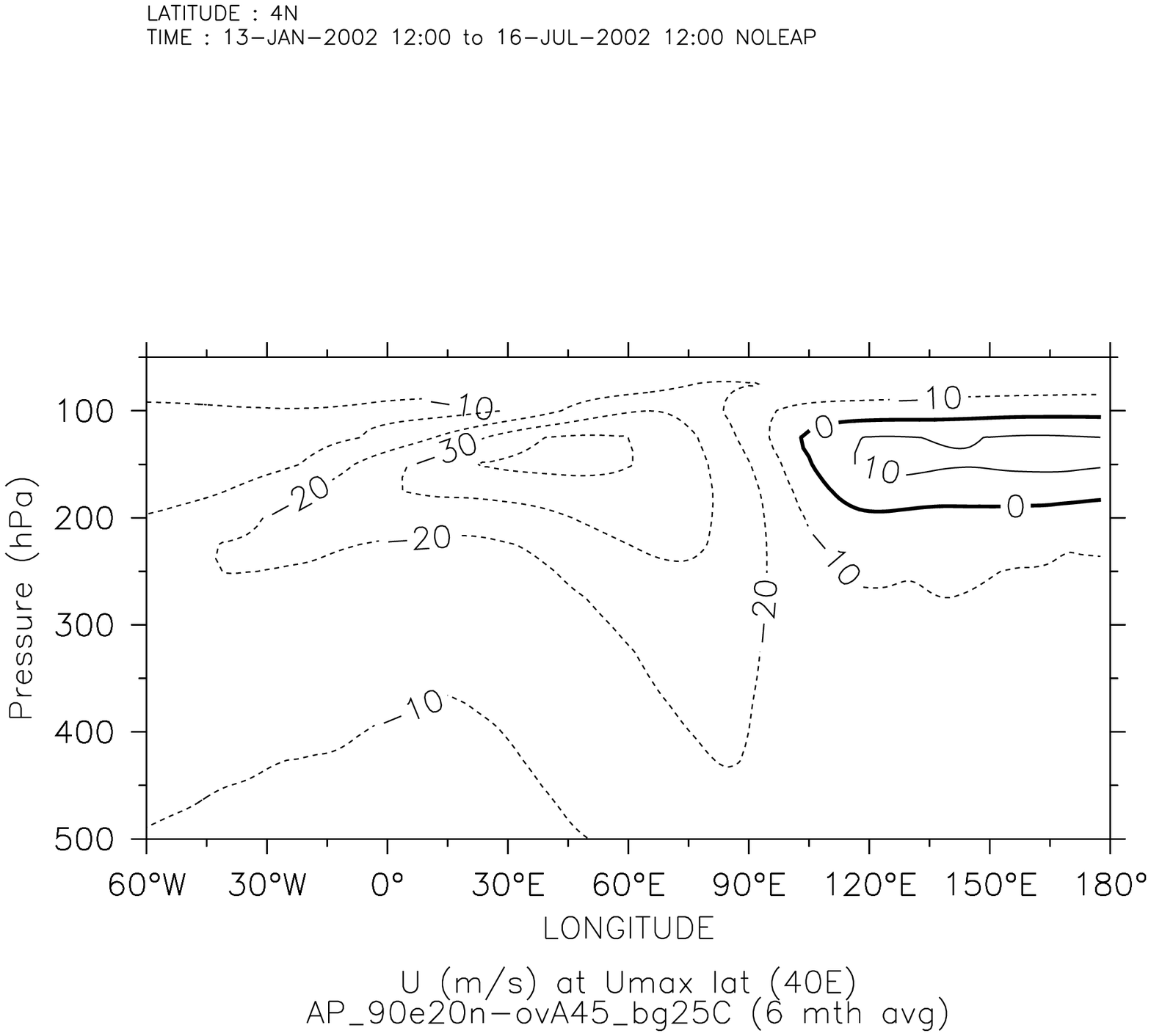}}
   \vskip 5mm
   \subfloat[Horizontal cross-section: (150 hPa)]{\label{S6-xy}\includegraphics[trim = 5mm 20mm 0mm 60mm, clip, scale=0.35]{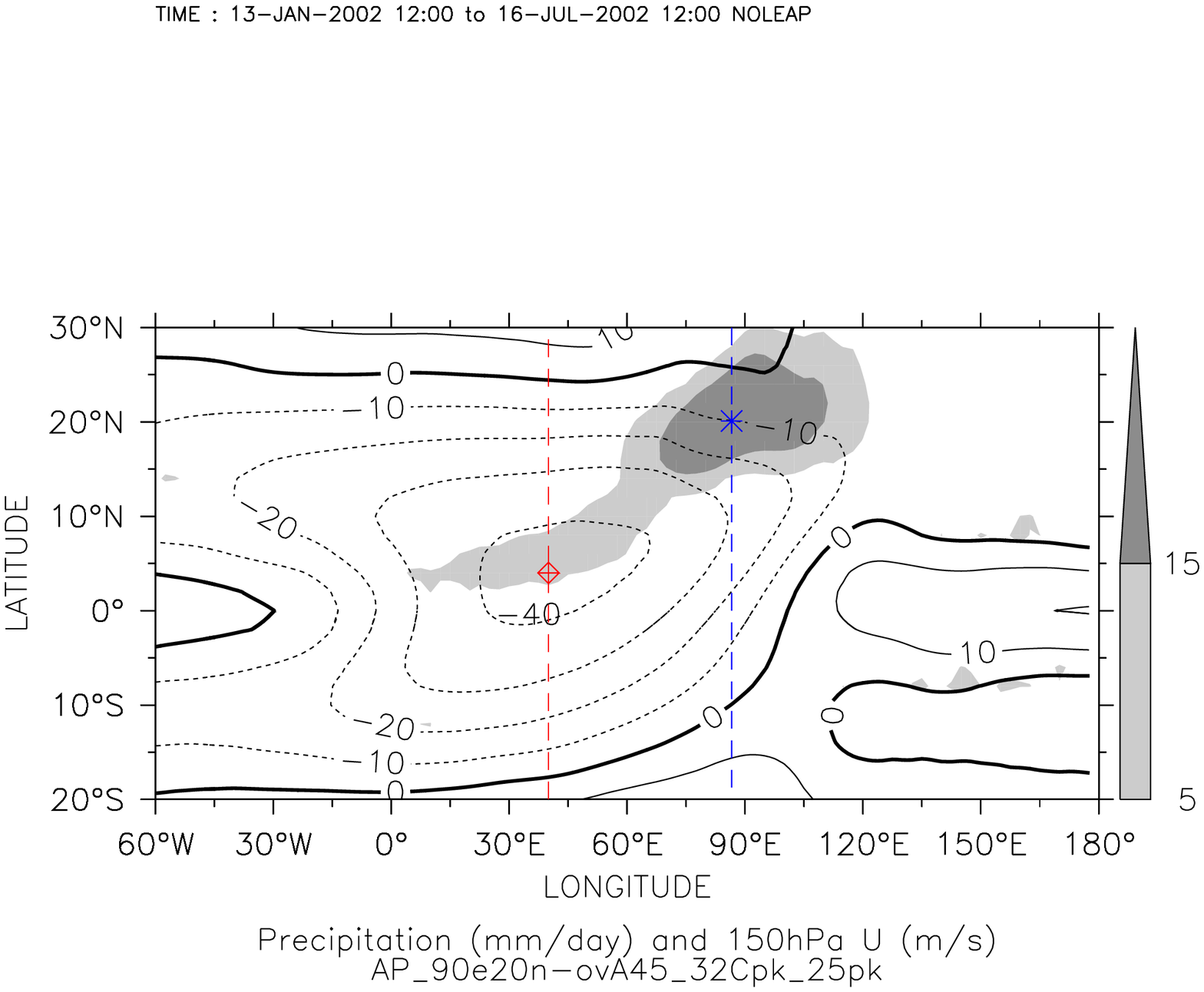}}
   \vskip 5mm
   \subfloat[Velocity vectors and geopotential height]{\label{S6-z3-xy}\includegraphics[trim = 5mm 5mm 10mm 70mm, clip, scale=0.35]{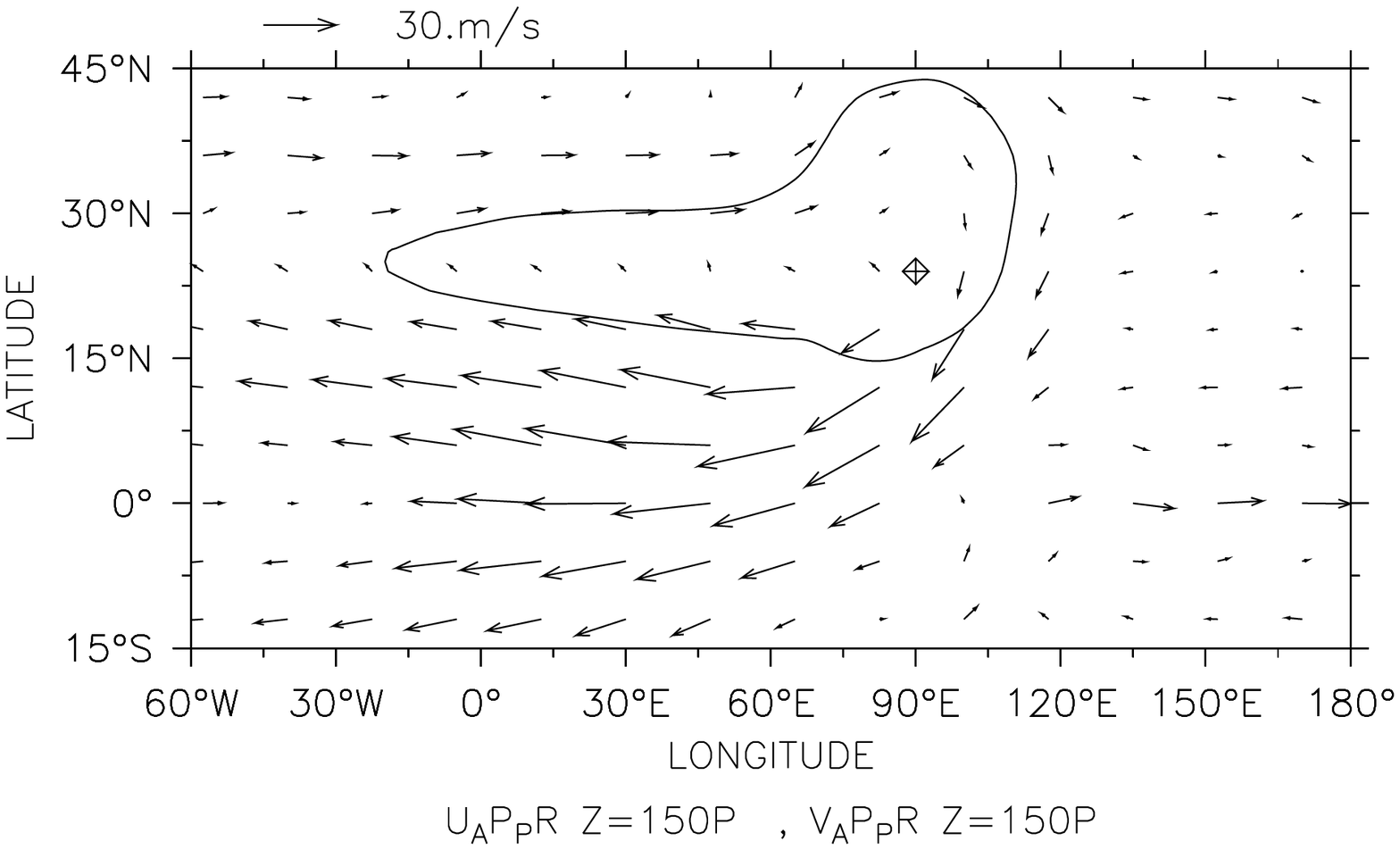}}
   \end{center}
   \caption[Zonal wind profile (meridional, zonal and horizontal), velocity vectors and geopotential height: \as{}6.]{Zonal wind (\ms{}) profile. Vertical sections are at the \protect\subref{S6-yz} longitude and \protect\subref{S6-zx} latitude where \um{} is attained; \protect\subref{S6-xy} horizontal cross-section at pressure level where \um{} is attained, `cross-diamond' is location of \um{}, `star' is location of \pc{}, precipitation contours (\md{}) are shaded; \protect\subref{S6-z3-xy} velocity vectors and geopotential height ($\times10^{3}$m) at 150 hPa; `cross-diamond' is location of peak geopotential height; contour value is 14.09 which is 99.5\% of peak: \as{}6 (Aqua-planet with single off-equatorial heat source in region S shown in Fig. \ref{fig: AP-SSTloc}).}
   \label{fig: S6-xyz}
\end{figure}

The meridional and zonal cross-sections (Figs. \ref{S6-yz},\subref*{S6-zx}) are shown for AP\_S6 simulation. Meridionally, the jet is not symmetric. The major difference with \rn{} and \rp{} simulations lies in the vertical shape of the jet. With single off-equatorial heat source the jet has a pole to equator tilt towards higher pressure levels. This is opposite to \rn{} and \rp{} (Figs. \ref{nc-yz},\subref*{cc-yz}). Below 500 hPa, there is also a region of weak low-level jet which is the \ap{} counterpart of the Somali jet (not shown). Further, the low-level westerly was observed to be either very weak or almost non-existent for those simulations with reduced spatial extent of heating and/or lesser overall heating rates, for example AP\_S2 and AP\_S5.

To the east of the jet, there is an upper-level westerly intrusion. A hint of this westerly is also observed in the \rn{} and \rp{} simulations (Fig. \ref{fig: mix_jul_Umax_yz-zx}, right panel). In Fig. \ref{S6-xy}, the precipitation region is demarcated by the 5\md{} and above shaded contours. It can be seen that precipitation (also refer Table \ref{tab: AP-Umax,R}) for the AP\_S6 case has very high rates which is unrealistic. For comparatively low heating rates, a baroclinic structure similar to \rn{} and \rp{} simulations does not exist. This only suggests that for such cases, the easterlies developed are more of an add-on to the equatorial easterlies developed with uniform background SSTs. However even low heating more than doubles the peak zonal winds generated in comparison to only uniform SST everywhere. This shows that even relatively higher zonal easterly wind speeds may not create vertical structures that have any resemblance with \rn{} and \rp{} cases. So one cannot understand the structure of the TEJ with a single heat source. In all the single heat-source simulations the easterlies were present around the entire tropics. In each of the circular and oval sets, increased heating due to increased SSTs also increases the zonal wind speeds. Points A and B in Fig. \ref{fig: mix-Pc-Umax-loc} are for AP\_S2 and AP\_S6.

The first thing to note is that compared to the change in mean precipitation magnitudes, the zonal separation differs significantly. The precipitation centroid also varies by less than $\sim$5\degrees{}. The maximum separation is for the 90\degrees{} major axis oval-shaped SST (AP\_S6) while the least is for the circular SST profile with 32\dc{} peak (AP\_S2). This demonstrates that more zonally constricted heating reduces the zonal separation between \pc{} and \um{}. When the spatial extent of heating is less, the zonal wind has relatively rapid zonal acceleration and a slower zonal deceleration. Intense convergence in a relatively small region and subsequent south-easterly and then almost easterly movement of the air mass will cause zonally rapid acceleration and slow deceleration. \cite{mishra-87} attributed the southward movement of the jet to the spherical geometry of the earth. Non-linearity was also responsible for causing the jet to shift towards the equator (\cite{mishra-93}). Increased spatial extent of heating causes increased acceleration length as is also evident from Table \ref{tab: AP-Umax,R}. It appears that spatially extensive heating imparts greater energy to the mean tropical flows, since higher zonal winds exist over longer zonal lengths. This is an important point and will be stressed upon in the discussion where multiple heat sources are introduced. The jet is thus to the south-west of the peak heating which is a departure from \rn{} and \rp{} simulations.

The location of the geopotential contour is very near to the region of maximum precipitation. This is clearly seen in Fig. \ref{S6-z3-xy}. Between 70\de{}-110\de{} the meridional velocities are also significant. This is a major departure from \rp{} simulations where the contour is very much to the west of the region of precipitation.

Additional simulations were also done with SST peaks at 90\de{},10\dn{}. The uniform background temperatures were 20\dc{} and 25\dc{}. The magnitudes of the SST peaks ranged from 23\dc{} to 37\dc{}. For heating at 10\dn{}, zonal wind speeds remained approximately $\sim$20\ms{} and did not show any increase even with higher heating rates. In contrast, for heating at 20\dn{}, zonal winds showed a monotonic increase with heating. The wind speeds increased from $\sim$20\ms{} to $\sim$50\ms{} as heating increased. Thus heating in lower latitudes does not result in increase in jet zonal velocities.

\subsection{Simulations with multiple heat sources resembling \cc{} simulation} \label{ap-mh}

The major finding of the previous section is that a single off-equatorial heat source is inadequate in explaining the structure of either the \rn{} TEJ or the full AGCM TEJ. Hence it is necessary to study the impact of combinations of heat sources. Fig. \ref{fig: AP-SSTloc} shows the regions where precipitation exists in \cc{} simulation. An additional advantage that \cc{} simulation has more precipitating regions than \rn{} in the region of interest and so it is more beneficial to understand these heat sources in determining the location and strength of the TEJ. If these \ap{} simulations have a jet structure resembling the TEJ in \cc{} or \nglo{} simulation in location as well as in magnitude, then it is an important step forward. This is because it can be then claimed that the location of the TEJ as largely a construct of the location of heating and not on any complicated land-ocean interactions. This section is an attempt to understand the interactions between the various heating and how they interact to produce jet-like structures.

The names AP\_N1 to AP\_M6 are for these cases. The locations and peak SSTs are given in Tables \ref{tab: AP-SSTloc}, \ref{tab: AP-Umax,R} (also refer Fig. \ref{fig: AP-SSTloc}). As explained before, in regions of overlap the maximum SSTs are chosen which removes any kinks in the profiles. The uniform background temperature is 22\dc{}. Though it would have been more realistic to incorporate a 25\dc{} background temperature, it was thought that 22\dc{} would enable us to clearly decipher the role the different zones play as elsewhere precipitation would be less while at the same time 22\dc{} background temperature would allow for minimal convection to occur. The peak SSTs again do not exceed 29\dc{} and in most cases are less than that. In all cases, SSTs linearly increase to the peak value. Although only a few simulations of each type have been listed, more such simulations have been conducted with different SST peaks in the same locations in order to check the validity of the assertions made. They are not presented for want of space.

\subsubsection{Simulations without near-Equatorial heating}

\begin{figure}[htbp]
   \begin{center}
   \subfloat[\am{}N1 (150 hPa)]{\label{N1-xy}\includegraphics[trim = 5mm 20mm 55mm 60mm, clip, scale=0.35]{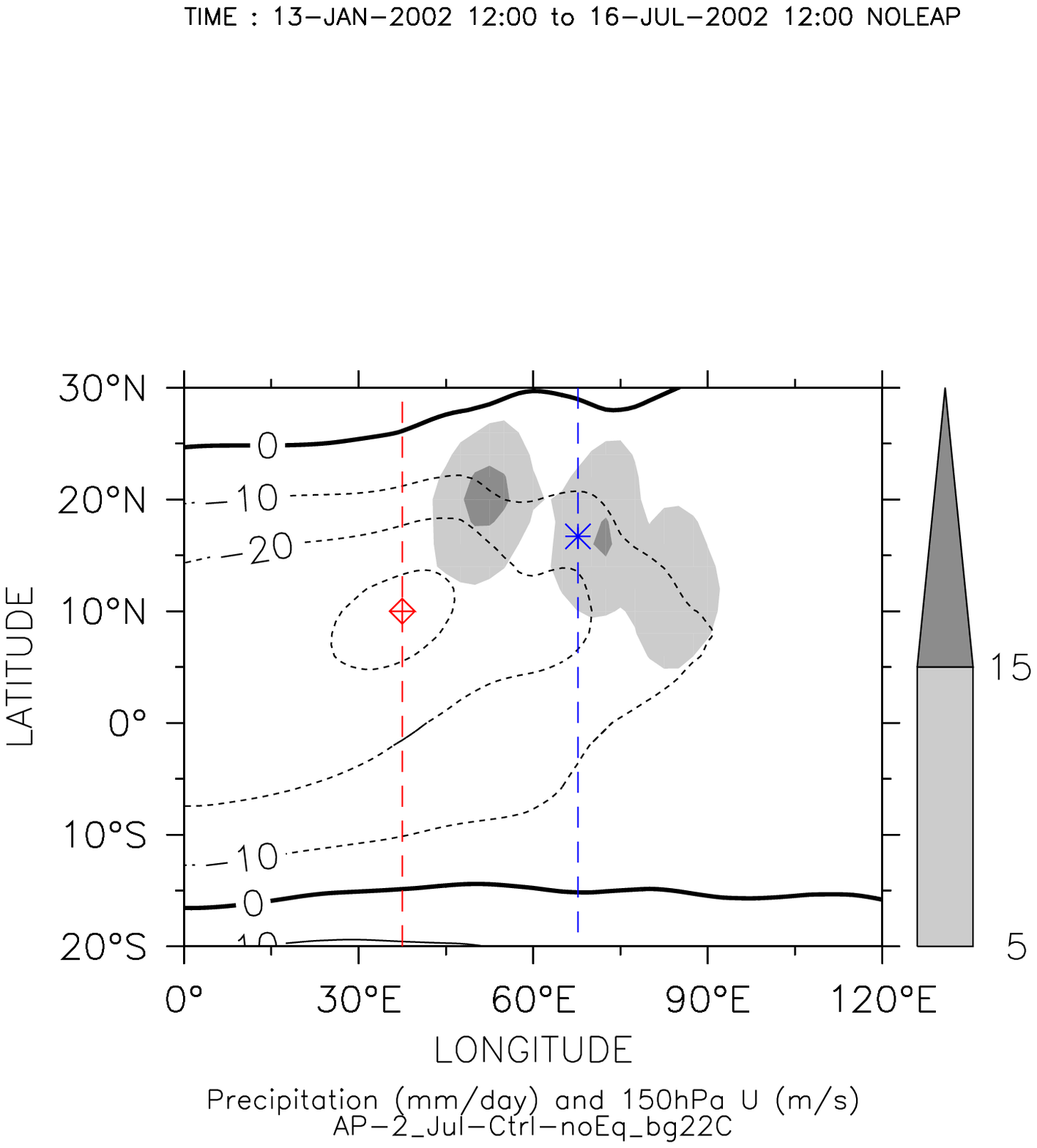}}
   \hspace{5mm}
   \subfloat[\am{}N2 (175 hPa)]{\label{N2-xy}\includegraphics[trim = 5mm 20mm 55mm 60mm, clip, scale=0.35]{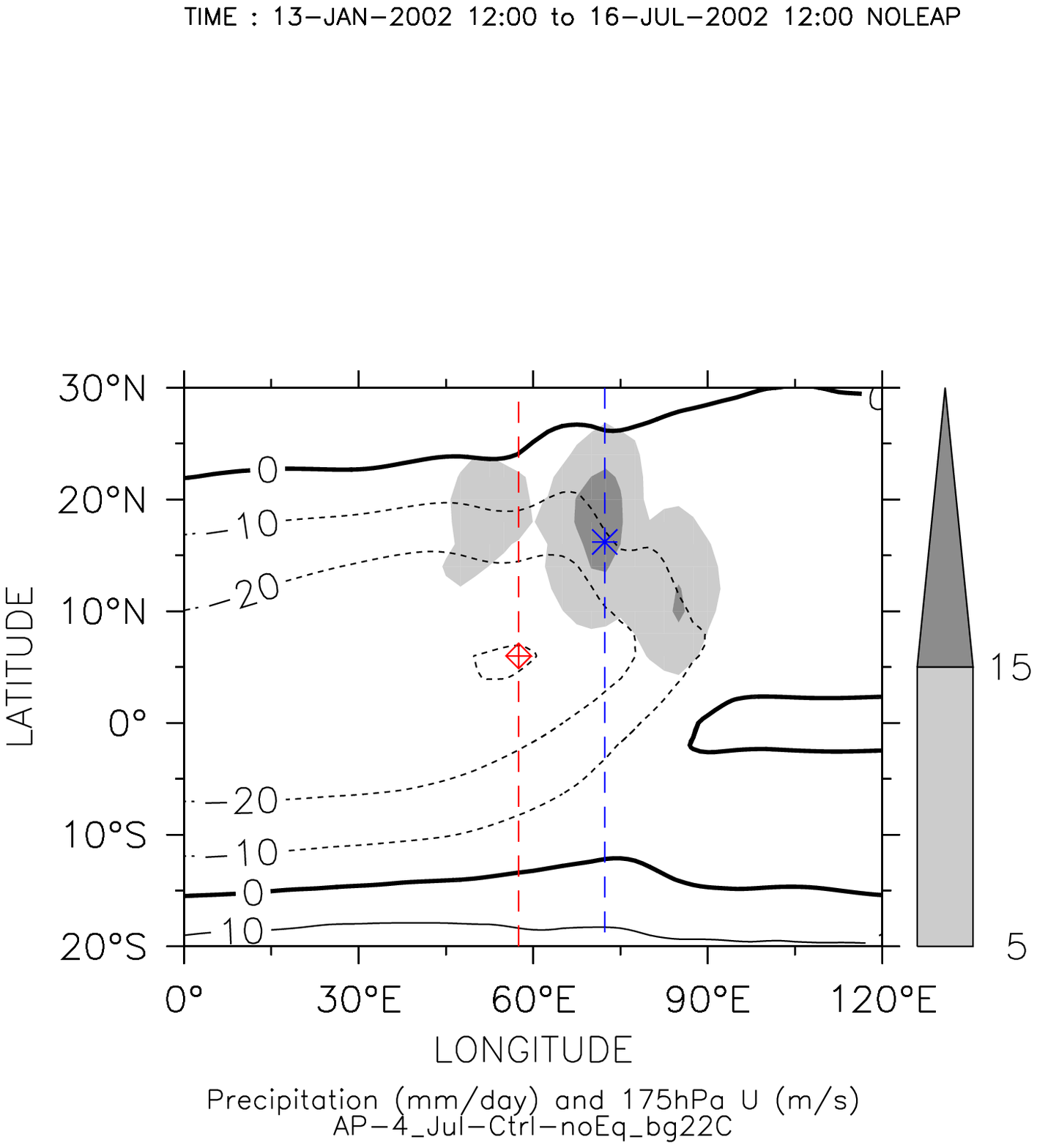}}
   \end{center}
   \caption[Horizontal zonal wind profile at pressure level where \um{} is attained: \am{}N1 and \am{}N2.]{Shift in location of \um{} in response to locations of heating. Zonal wind (\ms{}) profile at pressure level where maximum zonal wind is attained. Precipitation contours (\md{}) are shaded, `cross-diamond' indicates location of \um{}, `star' is location of \pc{}. Aqua-planet with multiple off-equatorial heat sources in regions B, C and D shown in Fig. \ref{fig: AP-SSTloc}.}
   \label{fig: N12_Umax_xy}
\end{figure}

These cases are AP\_N1 and AP\_N2 which have heating in regions B, C and D as shown in Fig. \ref{fig: AP-SSTloc}. This set has no equatorial heating. Since the heating is mostly off-equatorial, the meridional and zonal structures are similar to those with single heat source at 90\de{},20\dn{} and hence are not shown. The increased heating caused a low-level westerly in both cases. The interesting point to note from Table \ref{tab: AP-Umax,R} and Fig. \ref{fig: N12_Umax_xy} is the $\sim$25\degrees{} westward shift of \um{} in the 1$^{st}$ simulation (Fig. \ref{N1-xy}) in comparison to the 2$^{nd}$ simulation (Fig. \ref{N1-xy}). The zonal separation between \pc{} and \um{} (Fig. \ref{fig: mix-Pc-Umax-loc}) also shows a similar difference. The precipitation region in these two simulations shows that in the second case, precipitation region B is less than 10\md{}. This shows the importance of the heating that is closest to the peak zonal wind location. When the westernmost heating is below a certain threshold, the jet is influenced by the next closest heating, in this case the one in region C in Fig. \ref{fig: AP-SSTloc}. Simulations with heating only in regions C and D have also been conducted (not shown). In these cases, the zonal separation between \pc{} and \um{} is about 15\degrees{} less compared to AP\_N1 and AP\_N2. In other words, additional heating at location B increases this separation. This could be one of the reasons why \cc{} and \nglo{} simulations show a westward shift in the TEJ relative to \rn{}. In the \rp{} simulations, there is significant precipitation in the Saudi Arabian region.

\subsubsection{Importance of near-equatorial heating} \label{ap-mh-eq}

Referring to Fig. \ref{cc-jul-R-xy} one can see that in the \cc{} simulation, during July, there is a broad precipitation zone just south of the Equator between 50\de{}-100\de{} and a zone just north of this zonal tongue centered at 60\de{}. Our goal is to induce similar precipitation patterns by imposing SST peaks in this region. Two such simulations have been conducted which are labelled as AP\_E1 and AP\_E2. The details are given in Tables \ref{tab: AP-SSTloc}, \ref{tab: AP-Umax,R} and Fig. \ref{fig: AP-SSTloc}. Heating is present only in region E. Whenever there is an overlap, the maximum temperature is used which ensures that there are no kinks in the SST profiles. However the background SST is now 22\dc{}. The rationale for this choice will be explained in the next subsection where multiple-heat source simulations are discussed. Both rectangular and circular SST profiles have been imposed. The peak SSTs do not exceed 29\dc{} and are made to increase linearly to the peak value.

\begin{figure}[htbp]
   \begin{center}
   \subfloat[AP\_E2: Meridional section]{\label{E2-yz}\includegraphics[trim = 0mm 15mm 80mm 60mm, clip, scale=0.35]{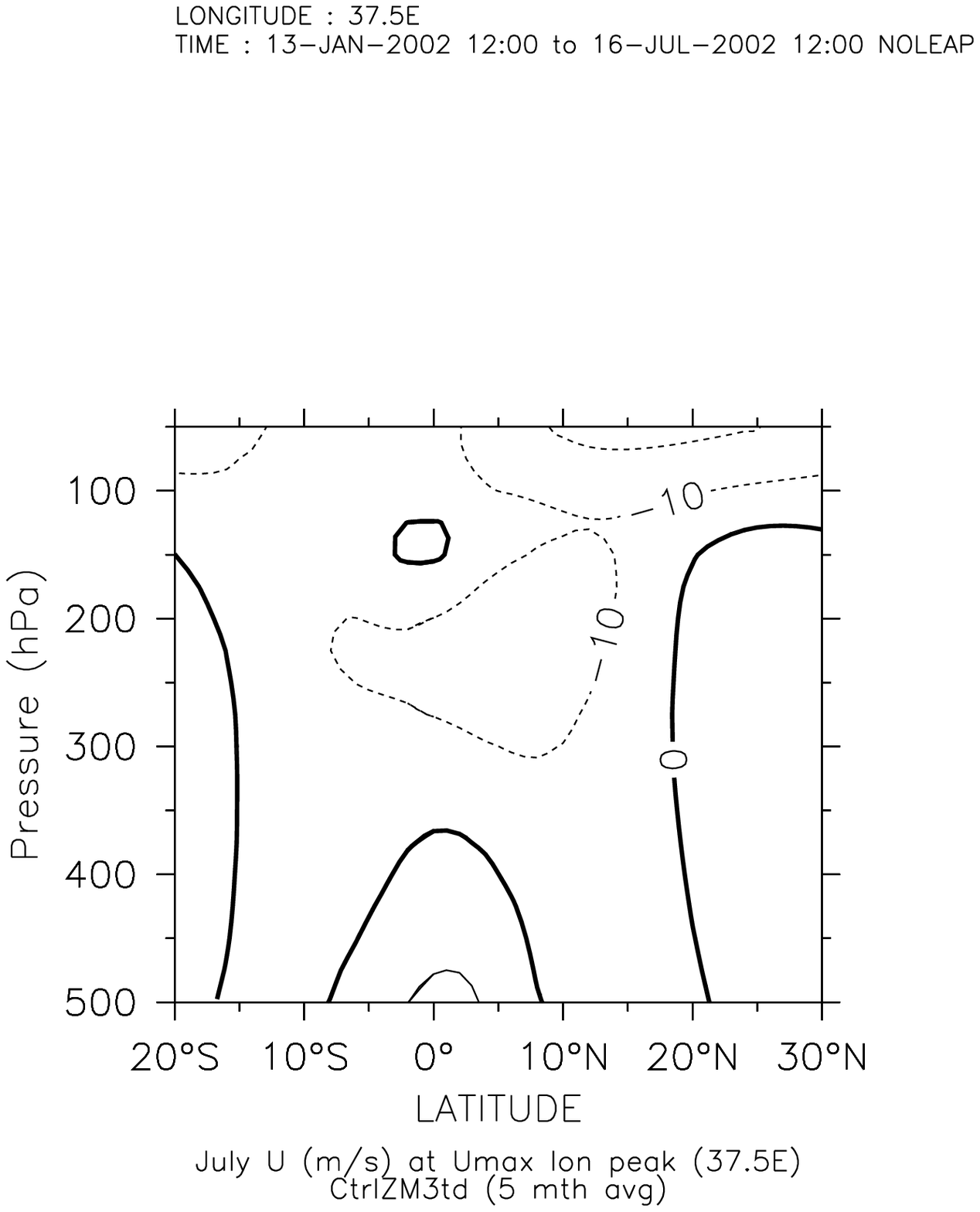}}
   \subfloat[AP\_E2: Horizontal section: (225 hPa)]{\label{E2-xy}\includegraphics[trim = 5mm 20mm 0mm 60mm, clip, scale=0.35]{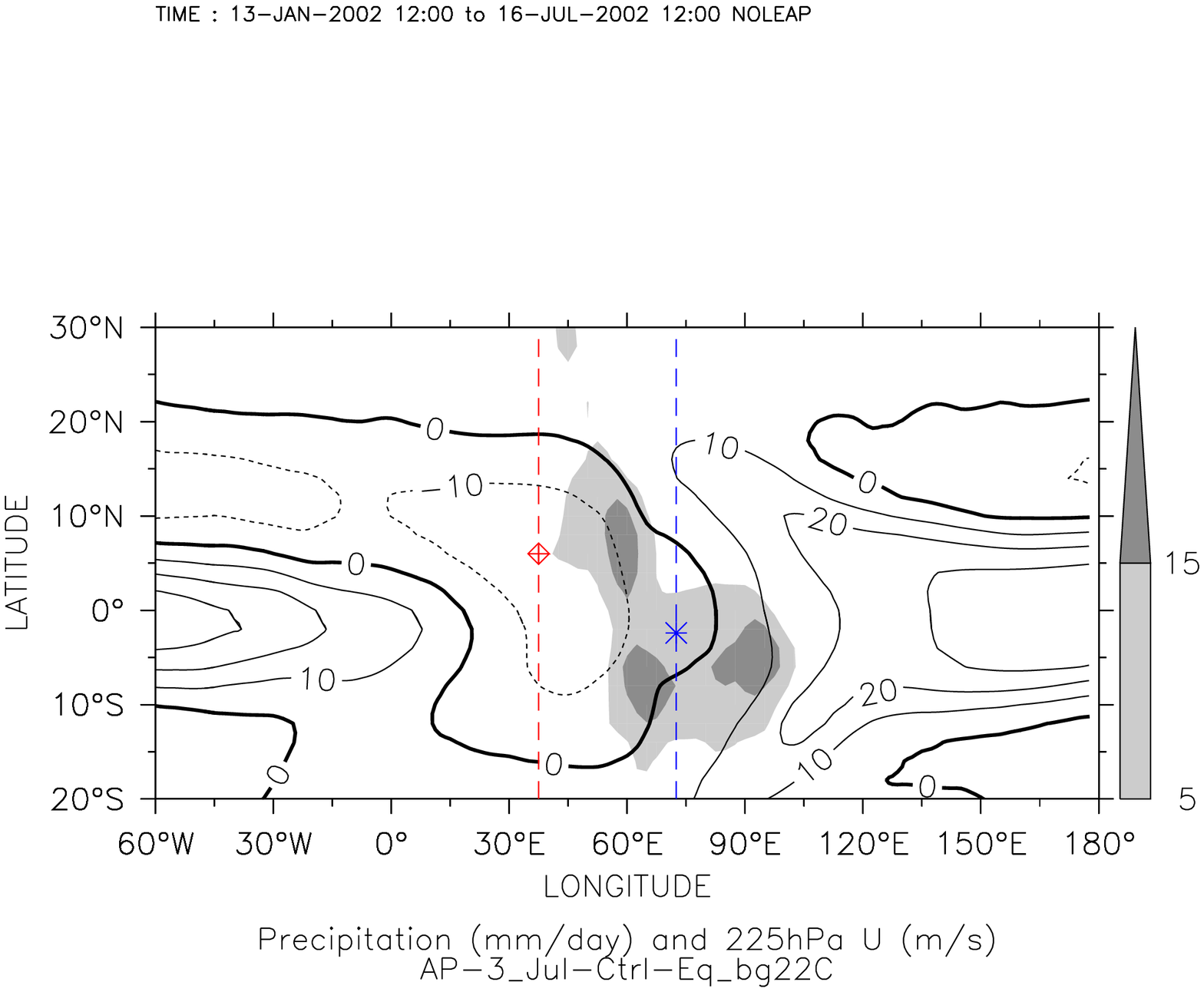}}
   \end{center}
   \caption{Zonal wind (\ms{}) profile. \protect\subref{E2-yz} meridional section is at the longitude where \um{} is attained; \protect\subref{NE2-xy} horizontal cross-section at pressure level where \um{} is attained, `cross-diamond' is location of \um{}, `star' is location of \pc{}, precipitation contours (\md{}) are shaded: Aqua-planet with multiple heat sources in region E shown in Fig. \ref{fig: AP-SSTloc}.}
   \label{fig: E2-xyz}
\end{figure}

From Table \ref{tab: AP-Umax,R} it can be seen that the zonal wind speeds barely qualify as a jet. The speeds are hardly more than one-and-a-half times the easterly obtained for a uniform background simulation. The zonal wind structure is shown in Fig. \ref{fig: E2-xyz}. The location of peak zonal wind is at a lower height (below 200 hPa). The most important observation is the stark contrast in the meridional structure vis-a-vis the off-equatorial heating. From the meridional section (Fig. \ref{E2-yz}), it is seen that the equator to pole tilt is towards higher pressure levels as in \cc{} and \rn{}. A low-level westerly of comparable strength extending below 500 hPa was also observed. Thus with just equatorial heating similar to \rp{}, the vertical structure becomes baroclinic. Compared to \rn{} and \rp{} there is a significant westerly to the east of the upper-level easterly. This signifies the reduced zonal extent of the easterly flows in comparison to single heat-source cases.

But the horizontal structure (Fig. \ref{E2-xy}) immediately shows that the zonal wind structure is actually nowhere near the \rp{} TEJ. The peak zonal wind was observed to be to the west of 90\dw{} and hence very far from \pc{}. This far-off value has not been tabulated since the region chosen for locating the peak zonal wind is 0-90\de{}. In spite of the maximum heating being in the south of the equator, the easterly is in the northern hemisphere. All these show that near-equatorial heating is necessary for imparting a baroclinic structure somewhat resembling reality. But stand-alone near-equatorial heating is by itself insufficient to generate a jet structure that resembles the TEJ.

\begin{figure}[htbp]
   \begin{center}
   \subfloat[AP\_NE1: Meridional section]{\label{NE2-yz}\includegraphics[trim = 0mm 15mm 80mm 60mm, clip, scale=0.35]{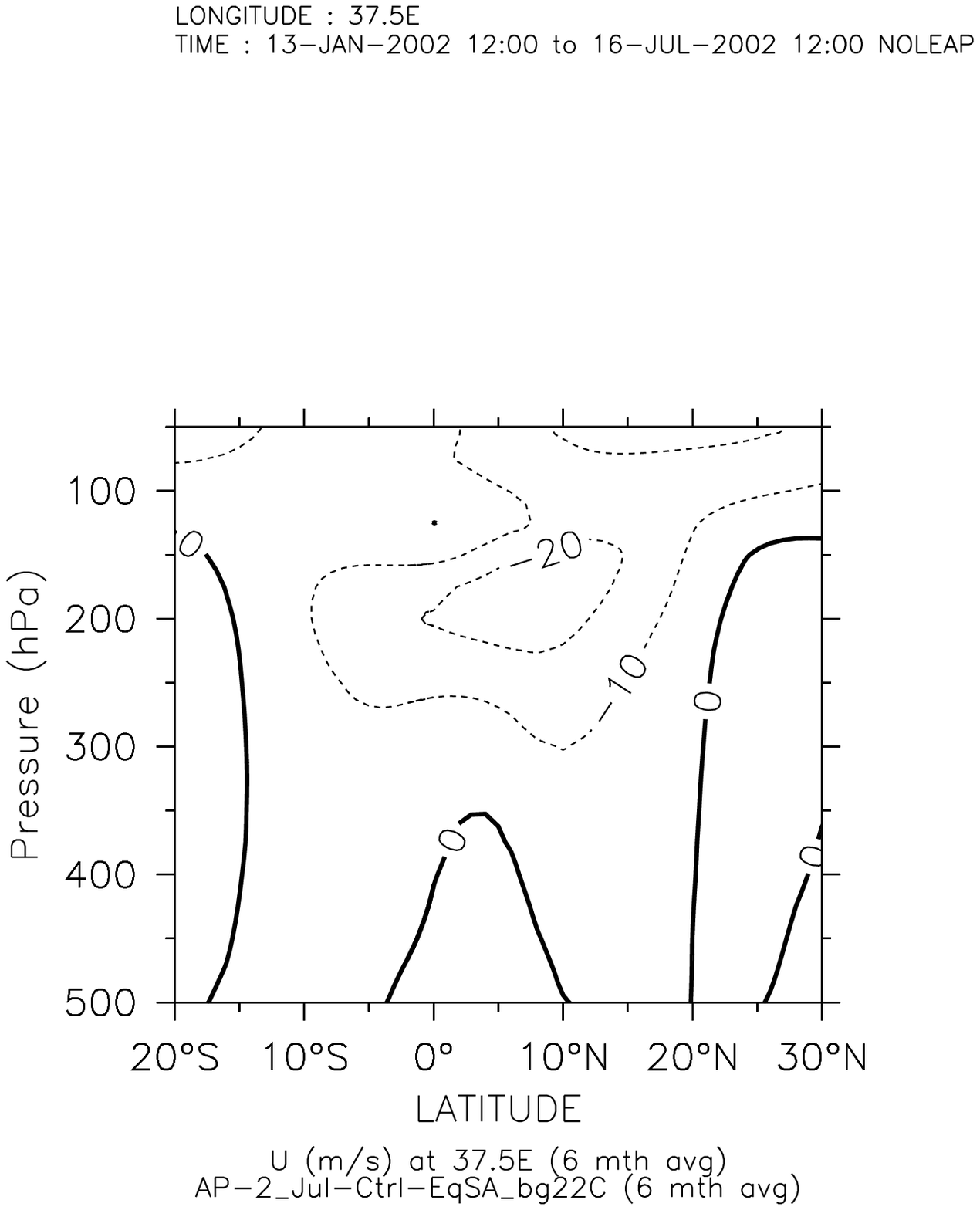}}
   \subfloat[AP\_NE1: Horizontal section: (175 hPa)]{\label{NE2-xy}\includegraphics[trim = 5mm 20mm 0mm 60mm, clip, scale=0.35]{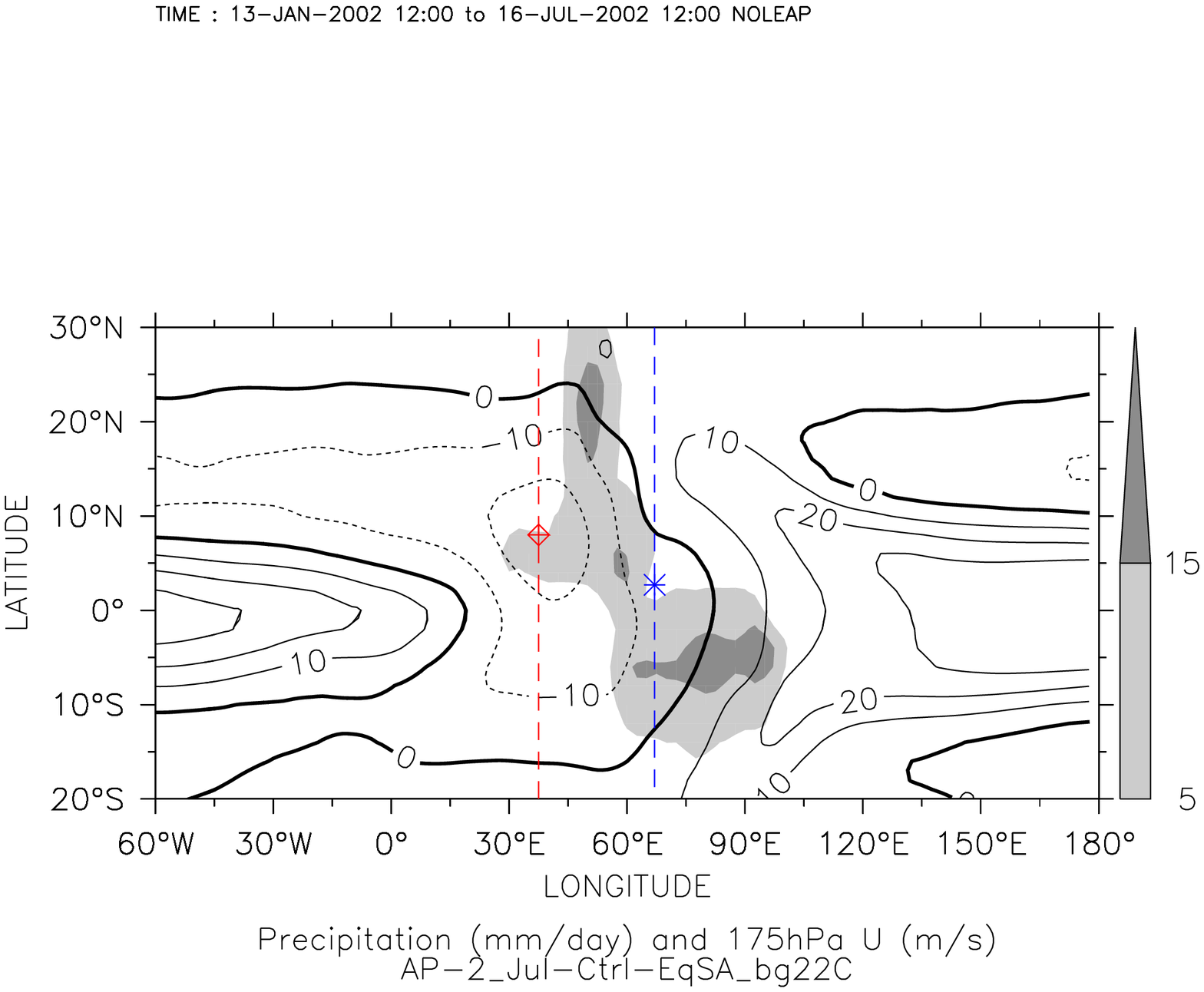}}
   \vskip 5mm
   \subfloat[AP\_NE2: Meridional section]{\label{NE3-yz}\includegraphics[trim = 0mm 15mm 80mm 65mm, clip, scale=0.35]{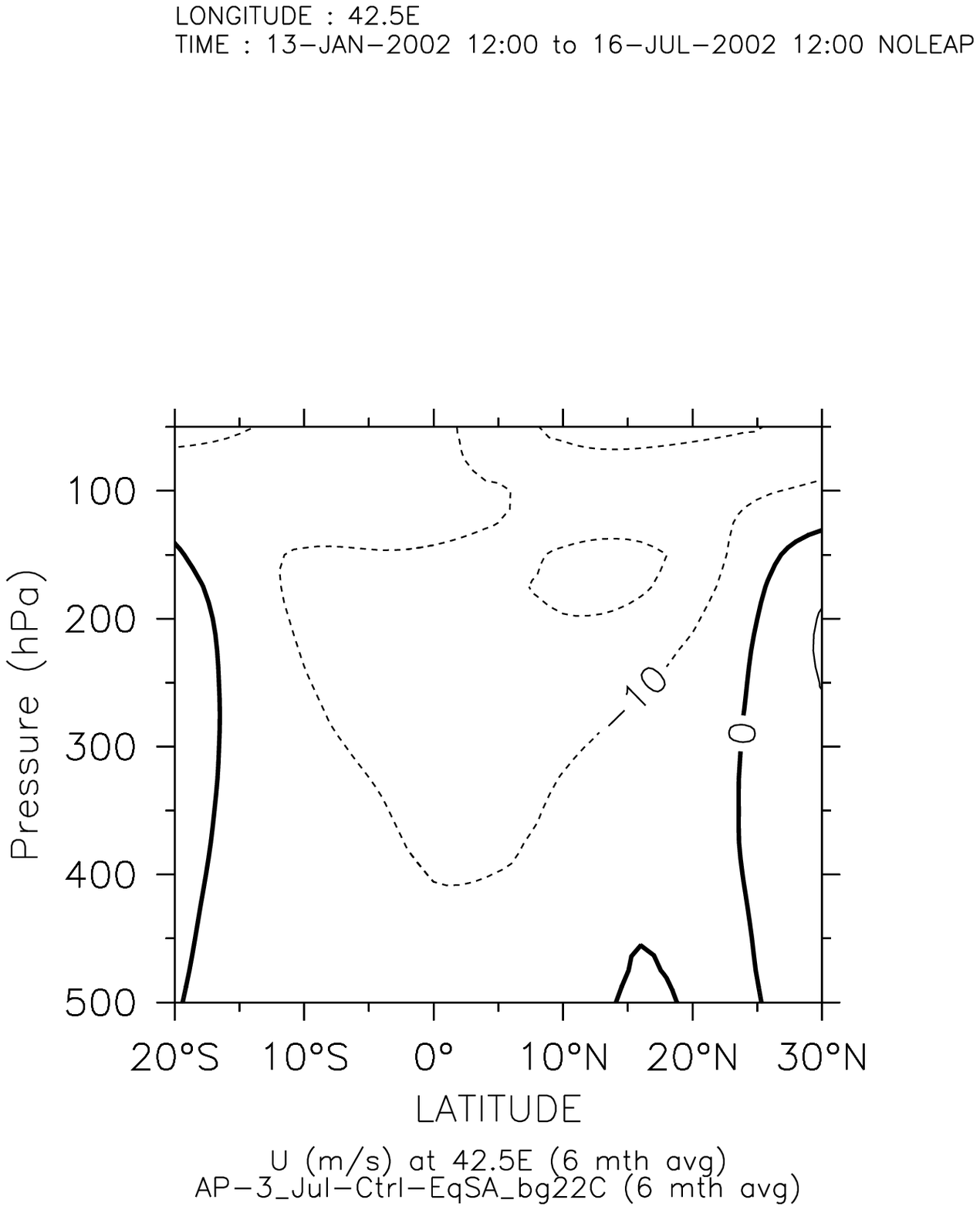}}
   \subfloat[AP\_NE2: Horizontal section: (150 hPa)]{\label{NE3-xy}\includegraphics[trim = 5mm 20mm 0mm 60mm, clip, scale=0.35]{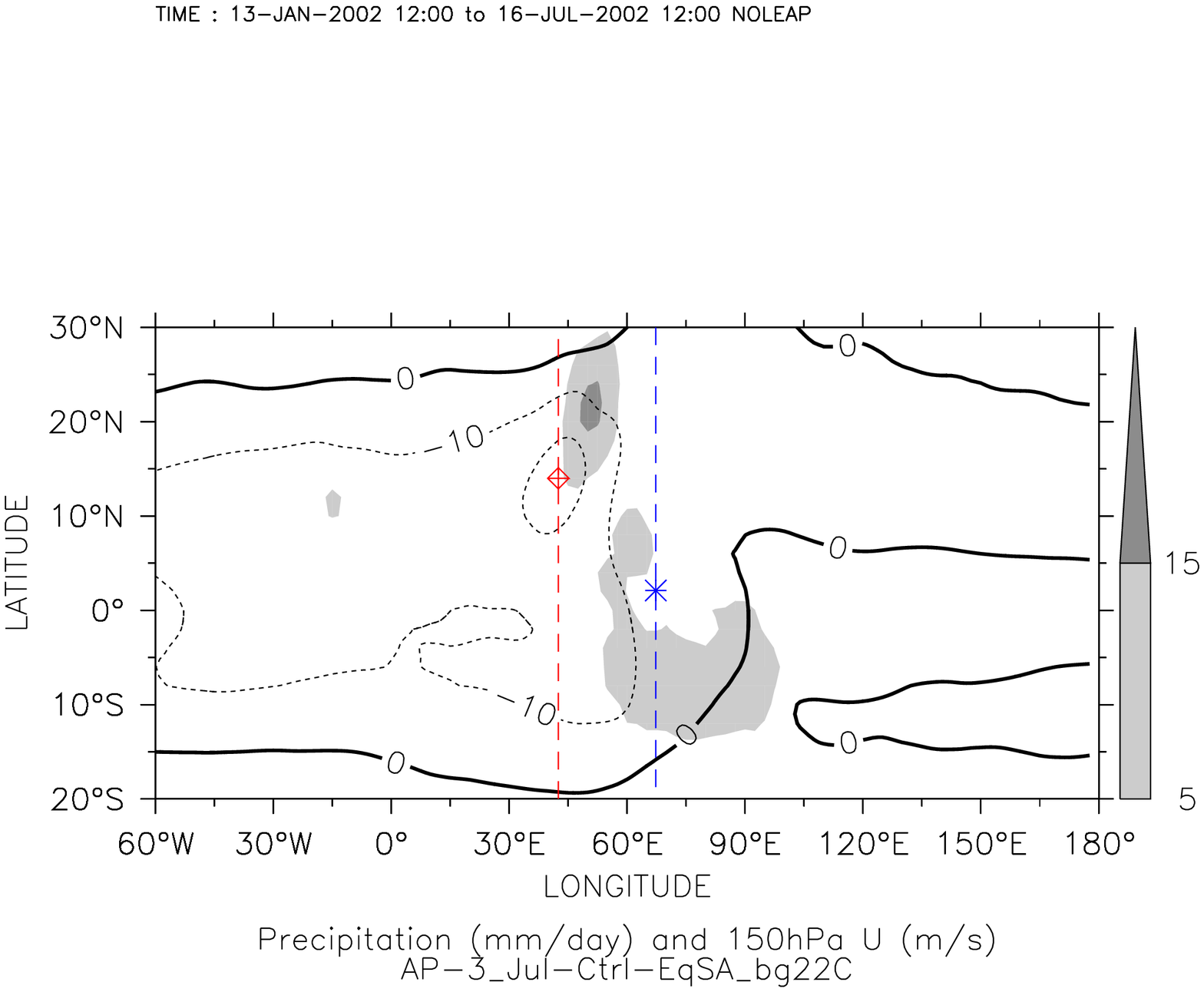}}
   \end{center}
   \caption[Zonal wind profile (meridional, zonal and horizontal)]{Zonal wind (\ms{}) profile. \protect\subref{NE2-yz},\protect\subref{NE3-yz} meridional sections are at the longitude where \um{} is attained; \protect\subref{NE2-xy},\protect\subref{NE3-xy} horizontal cross-section at pressure level where \um{} is attained, `cross-diamond' is location of \um{}, `star' is location of \pc{}; precipitation contours (\md{}) are shaded: Aqua-planet with multiple heat sources in regions B and E shown in Fig. \ref{fig: AP-SSTloc}.}
   \label{fig: NE2-xyz}
\end{figure}

\subsubsection{Interplay between heating in near-Equator and Saudi Arabian regions} \label{ap-mh-noeq}

It is interesting to study the effect of heating in region B in addition to region E discussed above. These are the AP\_NE1 and AP\_NE2 simulations. It can be seen from Tables \ref{tab: mix-Umax,R} and \ref{tab: AP-Umax,R}, that in these simulations the location of \um{} is closer to \cc{} simulation.

The meridional structure (Fig. \ref{NE2-yz}) is similar to the equatorial heating case. There is also the equator to pole tilt that is observed in \rp{} and \rp{} cases. The meridional structures show the familiar equator to pole tilt (as in \rn{}) till the precipitation in region B (Saudi Arabia) does not increase beyond a threshold. This increase can be inferred from Fig. \ref{NE3-yz}. Even then the depth is moderated by the equatorial heating effects. The change in tilt is due to precipitation being more in region B as can be seen from Fig. \ref{NE3-xy} in comparison to Fig. \ref{NE2-xy}. Here the equatorial heating in the latter is quite low and yet heating in B is not enough to increase the tilt in comparison to similar heating rates in the single heat source at 20\dn{}).

This shows that heating near the equator is essential in creating a meridional structure resembling \rn{} and \rp{}.

\subsubsection{Effect of all heat sources except Pacific ocean warm pool region} \label{ap-mh-nopo}

The discussion on the influence of and near-equatorial and off-equatorial heating can now be used to combine all of them and study how close the net effect is on the \cc{} simulation. These are the AP\_M1 to AP\_M5 simulations (refer Table \ref{tab: AP-Umax,R}). In this effort, once again the role of the easternmost heating on the jet structure and location is studied. The only difference between AP\_M1 and AP\_M2, and AP\_M3 to AP\_M5 simulations is the presence of heating in region B in the latter set. Heating in region D is expected to have minimal role to play as it is not the easternmost source.

The first and foremost difference, as observed from Table \ref{tab: AP-Umax,R}, is in the zonal location of \um{}. Precipitation in region B, as expected, causes a 15\degrees{}-20\degrees{} westward shift in the jet location while the location of \pc{} is hardly changed. This is clearly seen from Fig. \ref{fig: mix-Pc-Umax-loc} where points G and H are for AP\_M1 and AP\_M5 simulations. The zonal separation without heating in region B in the former causes the location of \um{} to be much closer to \pc{}. The jet zonal velocities are generally greater and this is because of heating in regions B,C and D (see section \ref{ap-mh-noeq}).

\begin{figure}[htbp]
   \begin{center}
   \subfloat[Meridional section]{\label{M5-yz}\includegraphics[trim = 0mm 15mm 80mm 60mm, clip, scale=0.35]{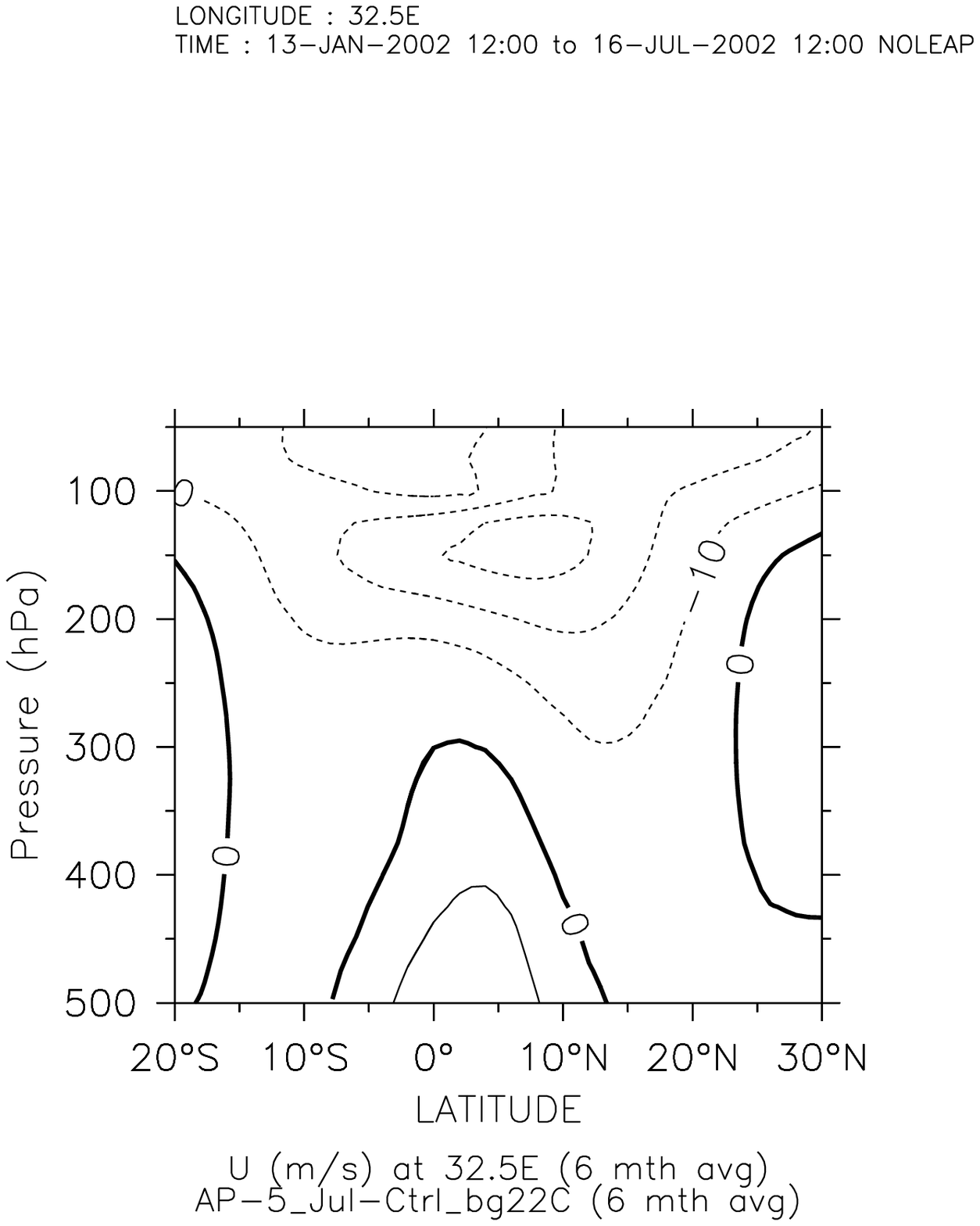}}
   \hspace{5mm}
   \subfloat[Zonal section]{\label{M5-zx}\includegraphics[trim = 0mm 15mm 10mm 60mm, clip, scale=0.35]{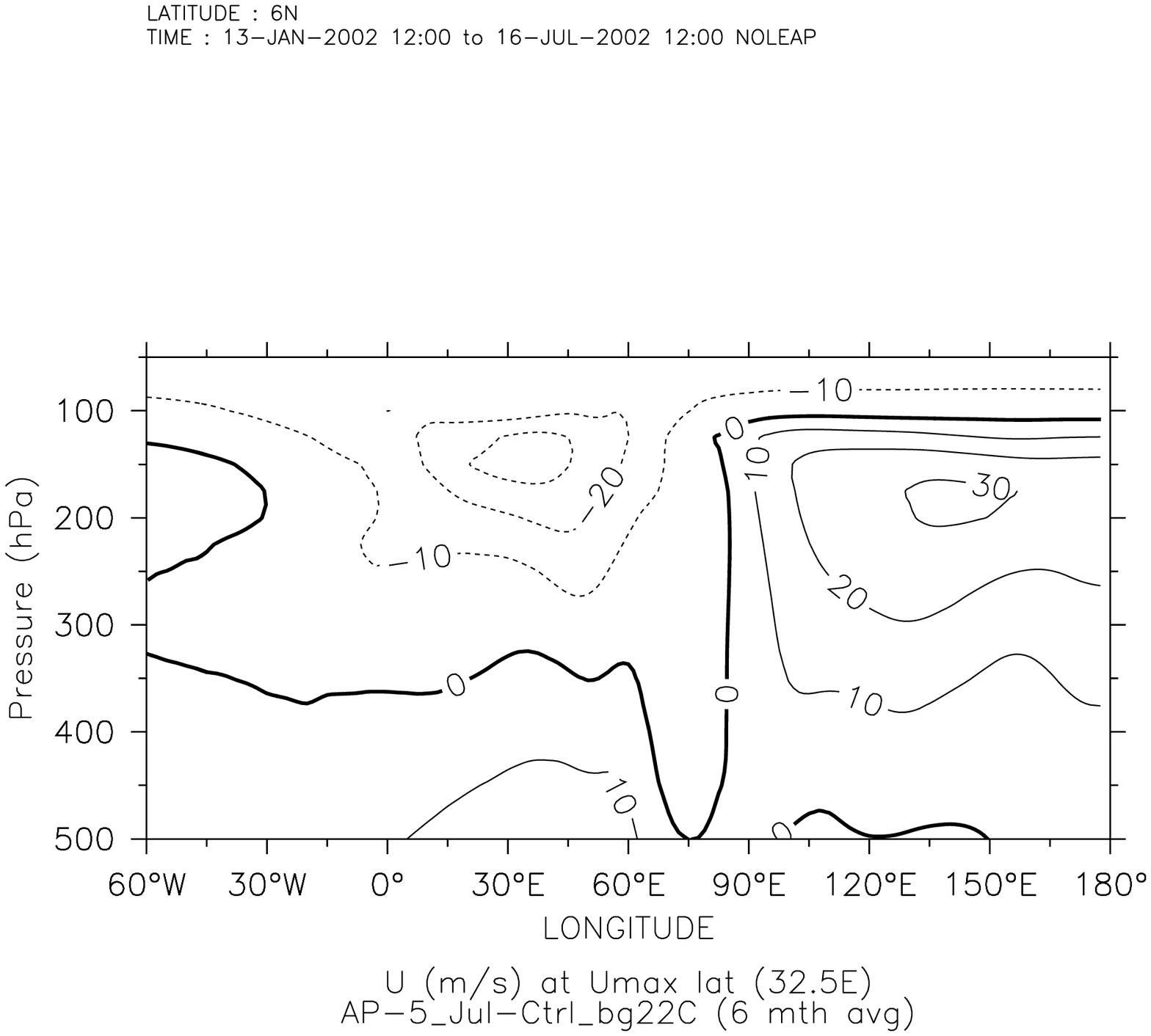}}
   \vskip 5mm
   \subfloat[Horizontal section (150 hPa)]{\label{M5-xy}\includegraphics[trim = 5mm 20mm 0mm 60mm, clip, scale=0.35]{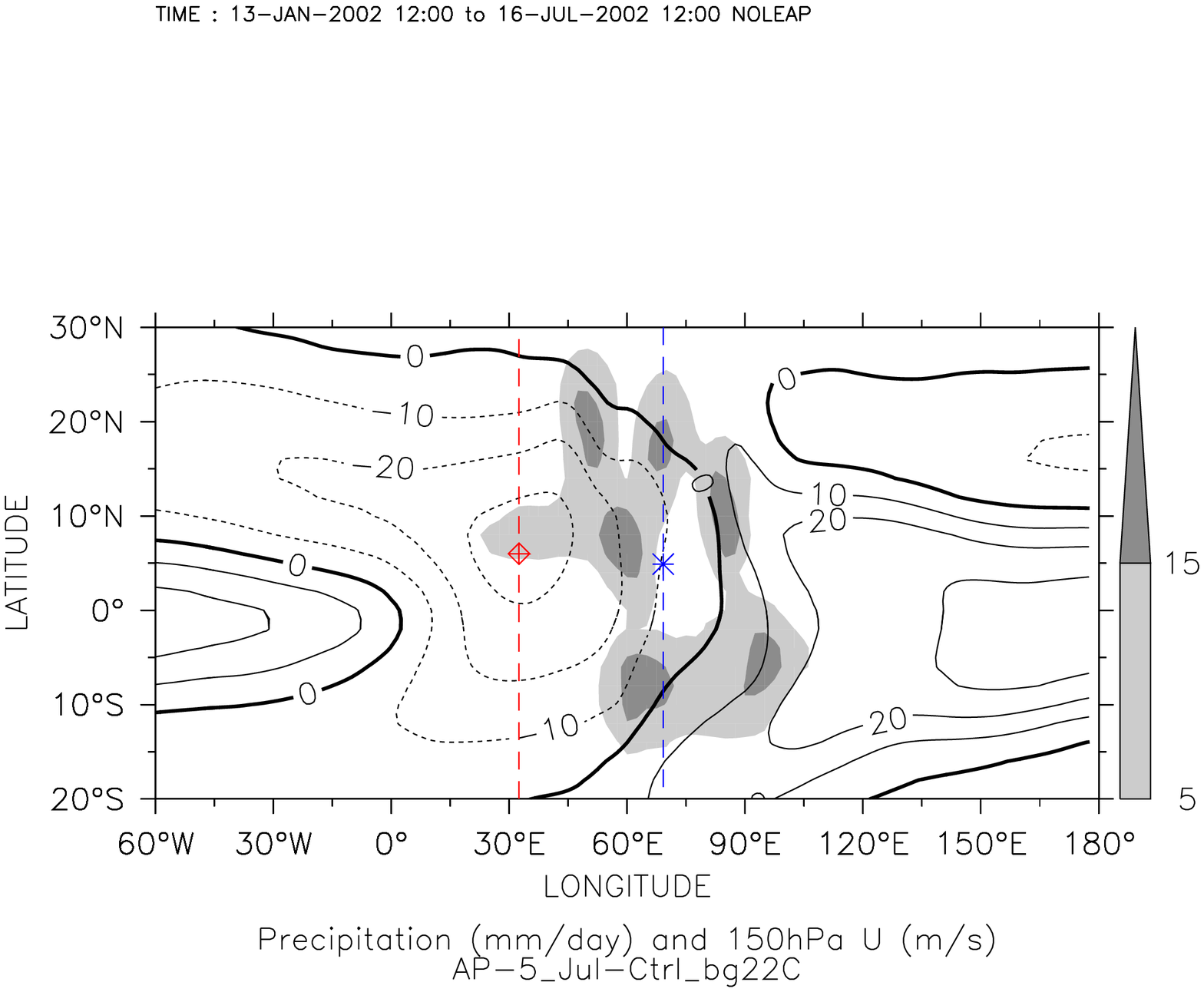}}
   \vskip 5mm
   \subfloat[Velocity vectors and geopotential height (150 hPa)]{\label{M5-z3-xy}\includegraphics[trim = 5mm 5mm 10mm 70mm, clip, scale=0.35]{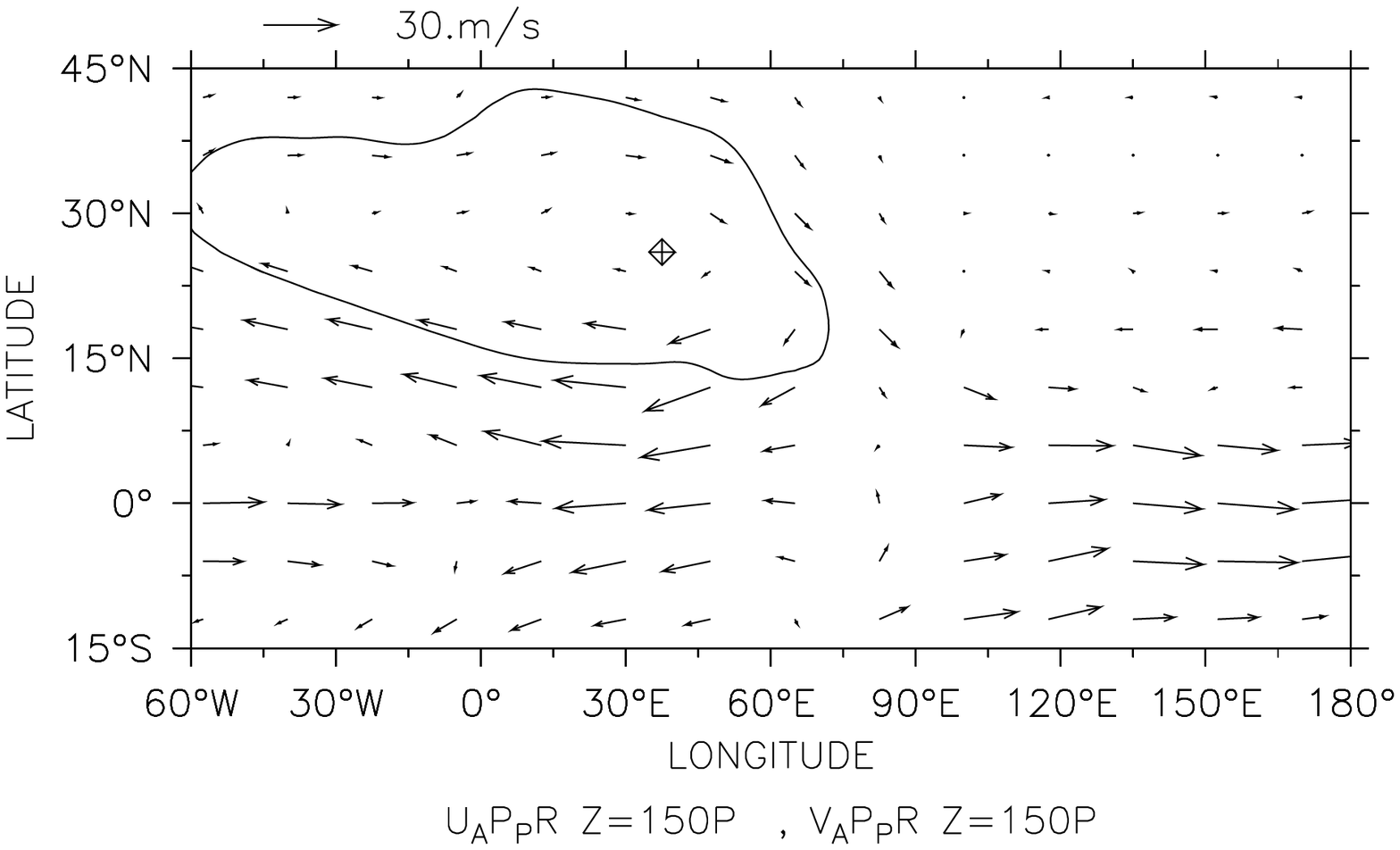}}
   \end{center}
   \caption[Zonal wind profile (meridional, zonal and horizontal), horizontal velocity vectors and geopotential height: \am{}M5.]{Zonal wind (\ms{}) profile. Vertical sections are at the \protect\subref{M5-yz} longitude and \protect\subref{M5-yz} latitude where \um{} is attained; \protect\subref{M5-xy} horizontal cross-section at pressure level where \um{} is attained, `cross-diamond' is location of peak zonal wind, `star' is location of \pc{}, precipitation contours (\md{}) are shaded; \protect\subref{M5-z3-xy} velocity vectors and geopotential height ($\times10^{3}$m) at 150 hPa, `cross-diamond' is location of peak geopotential height, contour value is 13.79 which is 99.5\% of peak. \am{}M5 (Aqua-planet with multiple heat sources in regions B, C, D and E shown in Fig. \ref{fig: AP-SSTloc}).}
   \label{fig: M5-xyz}
\end{figure}

In all cases, the other zonal wind structures are similar in all cases (except for the difference in location of \um{} when heating in region B is present). Hence in Fig. \ref{fig: M5-xyz} only AP\_M5 simulation is shown. The upper-level meridional structure of the jet (Fig. \ref{M5-yz}) resembles \rp{} simulations (Fig. \ref{fig: mix_jul_Umax_yz-zx}, left panel). The equator to pole tilt is always towards higher pressure levels. This demonstrates that in \ap{} configuration, precipitation patterns similar to \rp{} simulations result in meridional shapes similar to full AGCM simulations. The zonal jet lengths are still less than \rp{} simulations.The eastward extent of the upper-level westerly intrusion is also unrealistic. This discrepancy will be addressed in section \ref{ap-mh-po} below.

The relatively weaker and more westward location of peak zonal wind speed in AP\_M3 simulation compared to AP\_M5 is because the precipitation in region B in AP\_M3 was much lower in comparison to AP\_M5 (not shown). This once again underscores the influence of heating in region B (Saudi Arabia) in distorting the location of the jet in the \cc{} simulation. In spite of lower heating in region B in AP\_M3 simulation, the meridional structure has the familiar equator to pole tilt (not shown). Except for the westerly to the east of 90\de{}, the overall TEJ simulated has a structure similar to \cc{} (Figs. \ref{cc-xy} and \ref{cc-yz},\subref*{cc-zx}).

The location of the geopotential high (Fig. \ref{M5-z3-xy}) is now bears resemblance to \cc{} simulation (Fig. \ref{cc-z3-xy}). Although this peak in AP\_M5 is shifted more eastwards in comparison, from Tables \ref{tab: mix-Umax,R} and \ref{tab: AP-Umax,R} it is seen that the location of \um{} as well as \pc{} is also westwards in comparison. On closer inspection, from Fig. \ref{fig: mix-Pc-Umax-loc} (points J and H) it can be observed that the zonal and meridional separation between \pc{} and \um{} almost the same. Thus in the \ap{} simulations if all the SST profiles had been shifted westwards by $\sim$10\degrees{}, then the mosaic of precipitation, zonal wind and geopotential heights would have shifted westwards by about the same amount. Then there would be more similarity with \rp{} simulations.

It is also interesting to observe that the jet velocities increase by only $\sim$5\ms{} when compared with simulations with just an additional source at region B. In the previous single heat source simulations, the jet velocities increased when the intensity of heating increased both spatially and magnitude wise. Here, additional the heat source at 20\dn{} does not significantly increase the jet strength. This shows that when multiple, well-distributed heat sources resembling reality are present, it is the intensity of heating and not the number of heating zones that determines the zonal wind strength.

\subsubsection{Importance of heating in the Pacific Ocean warm pool} \label{ap-mh-po}

\begin{figure}[htbp]
   \begin{center}
   \subfloat[Zonal cross-section]{\label{M6-zx}\includegraphics[trim = 5mm 15mm 10mm 60mm, clip, scale=0.35]{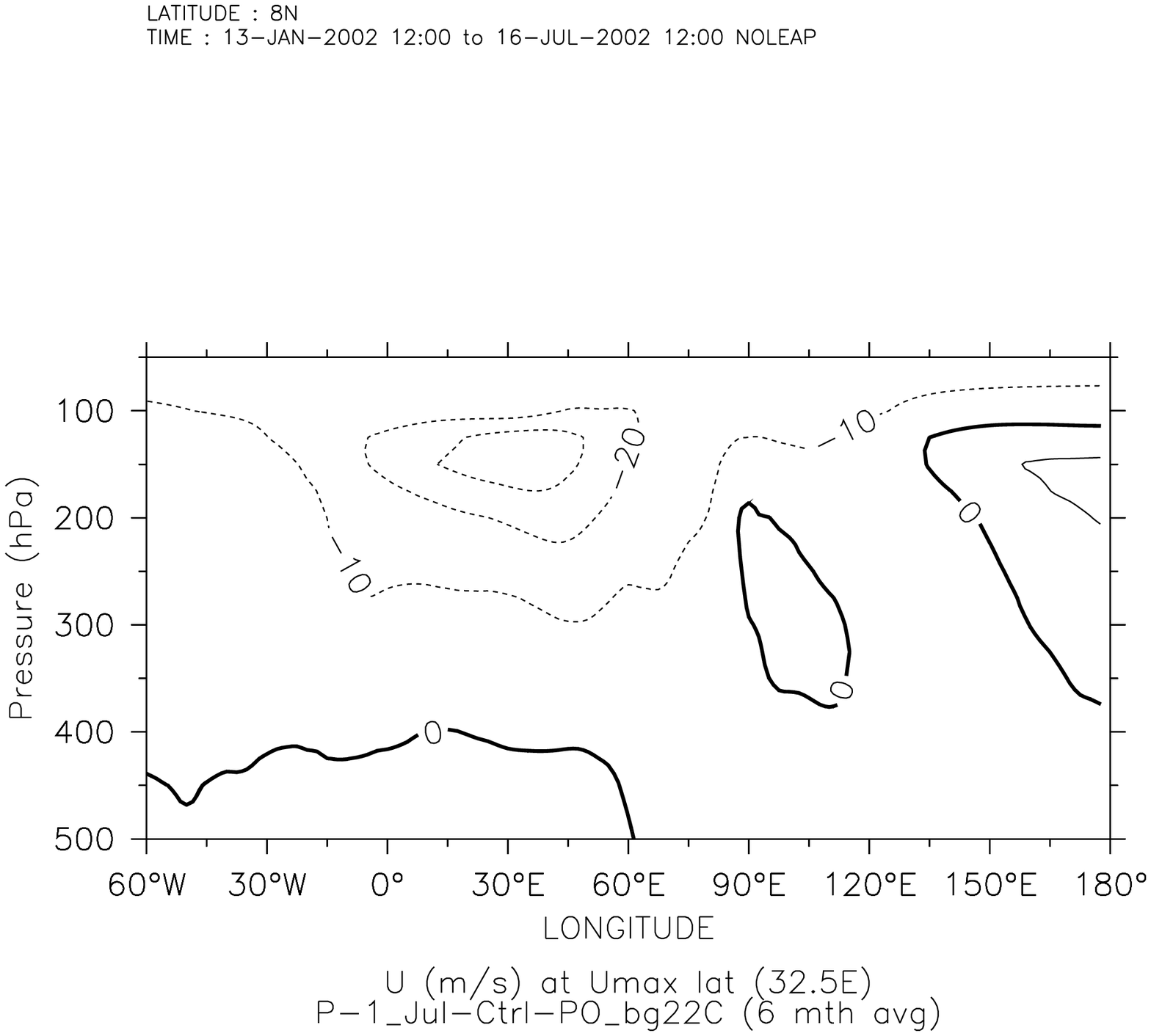}}
   \vskip 5mm
   \subfloat[Horizontal cross-section (150 hPa)]{\label{M6-xy}\includegraphics[trim = 5mm 20mm 0mm 60mm, clip, scale=0.35]{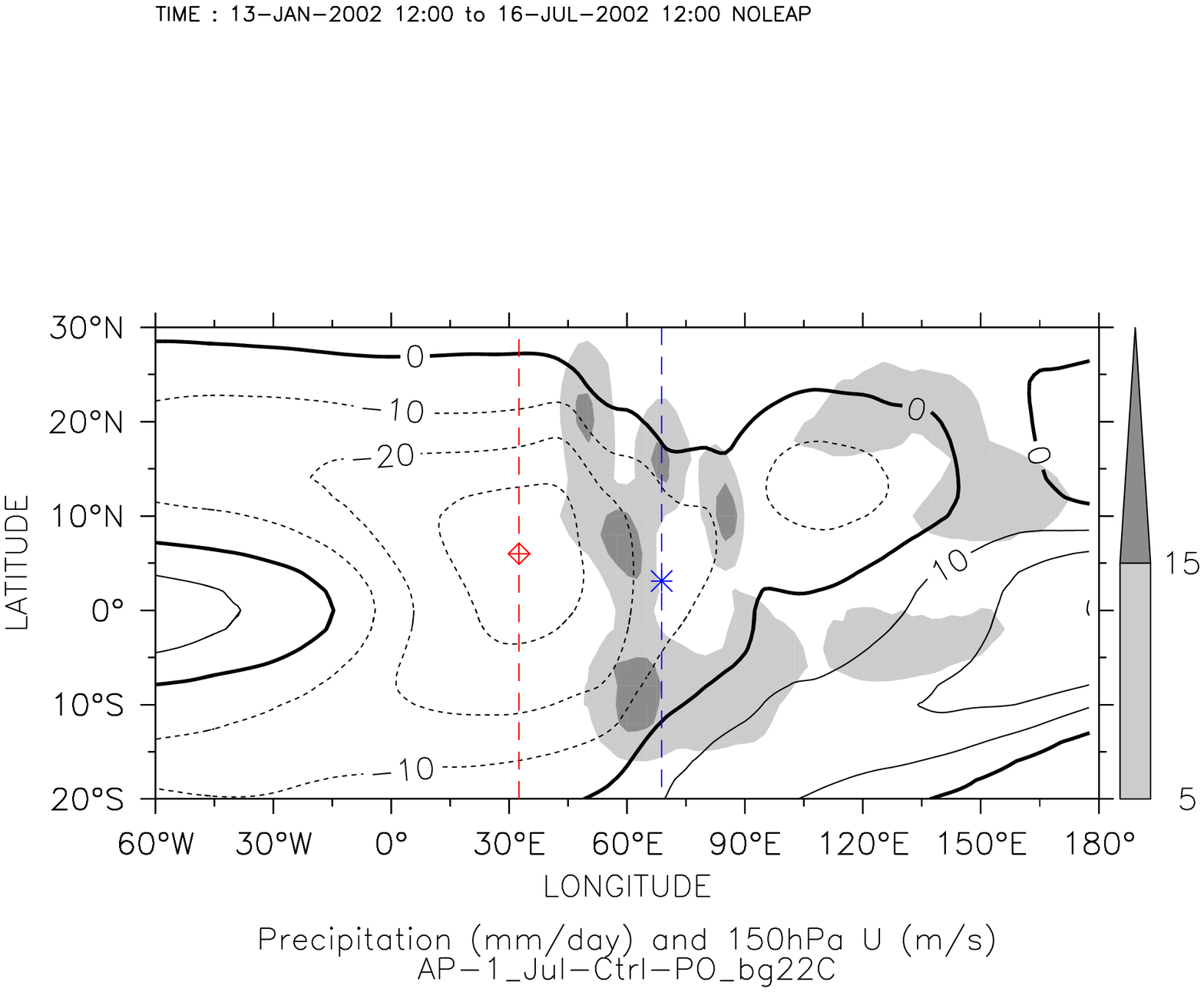}}
   \end{center}
   \caption[Zonal and horizontal profiles of zonal wind: \am{}M6.]{Zonal wind (\ms{}) profile. \protect\subref{M6-zx} zonal wind at the latitude where \um{} is attained, \protect\subref{M6-xy} horizontal cross-section at pressure level where \um{} is attained, `cross-diamond' is location of \um{}, `star' is location of \pc{}, precipitation contours (\md{}) are shaded. \am{}M6 (Aquaplanet with multiple heat sources in all regions, except S, shown in Fig. \ref{fig: AP-SSTloc}).}
   \label{fig: M6-xyz}
\end{figure}

In all the above multiple heat source simulations, the greatest mismatch in zonal wind structure with the \rp{} simulations is the westerly that was always present to the west of 90\de{}. This westerly was prominently reflected in all the zonal cross-sections. From Fig. \ref{fig: rean-cam3-Jul-R-xy} it is observed that in \rn{} there is significant precipitation occurring east of 100\de{}. Although in this region, in both \cc{} and \nglo{}, the precipitation is less than observation, there was significant precipitation in excess of 5\md{} in the Pacific warm pool which had not been incorporated in the \ap{} configurations. Hence an additionale experiment, AP\_M6, incorporating additional heating in region P as shown in Fig. \ref{fig: AP-SSTloc}, was conducted. The ratio of mean precipitation between 110\de{}-180\degrees{} and 20\ds{}-30\dn{} for AP\_M6 and \cc{} is $\sim$0.95. Everywhere else the maximum SSTs were same as AP\_M5. The results are shown in Fig. \ref{fig: M6-xyz}. The horizontal section (Fig. \ref{M6-xy}) now shows the extension of the easterly beyond 90\de{} and this is also reflected in the zonal section (Fig. \ref{M6-zx}). Thus the acceleration length of the TEJ is now closer to \rp{} and \rn{}. The meridional section is similar to AP\_M5 simulation and hence is not shown. The location and magnitudes of peak zonal wind in AP\_M5 and AP\_M6 are almost the same. Thus heating in the Pacific warm pool is necessary for the eastward extent of the TEJ.

\section{Conclusions}\label{con}

The July climatological structure of the \tej{} in observations has been studied and compared with the simulations by an AGCM. The TEJ in \rn{} in July 1998 and 2002 had significant zonal shifts. Reduced precipitation in the Indian region in 2002 made the jet have its maxima in the southern Indian peninsula as compared to 1988 when high rainfall in the Indian region resulted in the TEJ having its maxima over the Arabian sea region.

The TEJ in the \cc{} simulation has errors in the spatial location; otherwise the simulated TEJ bears similarities with observations. Removing orography left the spatial location and structure is practically unchanged. This leads us to conclude that the TEJ is not directly influenced by orography. The primary reason for the shift in the simulated TEJ was because the location of precipitation in both \cc{} and \nglo{} is westwards when compared to \rn{}.

Additional experiments were conducted to check if the TEJ is primarily influenced by latent heating. Changing the deep-convective relaxation time scale both in \cc{} and \nglo{} simulations confirmed this. In these new simulations the precipitation is more accurately simulated and most importantly anomalous precipitation in Saudi Arabia no longer occurred. The jet followed the shift in precipitation and relocated to the correct climatological position. The absence of orography once again had no impact on the location of the jet. This conclusively proves that the TEJ is independent of orography. Changing the default value of deep convective time-scale also demonstrated the secondary role of orography. Both \cct{} and \nglot{} have very similar precipitation patterns and hence the TEJ in both is correctly simulated.

To understand why the TEJ was shifted westward in the AGCM, \ap{} experiments were conducted.

\begin{enumerate}[1]
\item The simulation of TEJ in an \ap{} configuration of the AGCM shows that orography and land-sea interactions are not as important as latent heat release.
\item The total acceleration length is cirrelated to the zonal extent of the heating.
\item A heat source at 20\dn{} appears more to be robust in generating wind speeds that may be referred to as a jet while equatorial heating alone does not generate TEJ.
\item Equatorial heating is necessary to generate a strong low-level westerly that imparts the vertical baroclinic structure to the TEJ. However it is insufficient in generating true TEJ horizontal structure.
\item Equatorial heating is essential to create meridional structures seen in observations and full AGCM simulations. Greater poleward depth is possible only if equatorial heating is present.
\item The longitudinal location of peak zonal wind is influenced by the off-equatorial heating that is closest to it. It has been demonstrated that rainfall in Saudi Arabia causes the extreme westward shift of the TEJ in full AGCM simulations in comparison to observations.
\item Heating in the Pacific warm pool is essential to cause eastward extension and increased acceleration length of the TEJ.
\end{enumerate}

When all the important heat sources are incorporated in the \ap{} configuration, many observed features of the TEJ were simulated. Thus \ap{} simulations play an important role in understanding the role of heat sources in the absence of any influence of land and orography.

\bibliographystyle{plainnat}
\bibliography{references}   

\end{document}